\documentclass{emulateapj}
\usepackage{graphicx}

\shorttitle{Variability of Double-Peaked Emission Lines}
\shortauthors{Gezari et al.}

\newcommand {\aplt} {\ {\raise-.5ex\hbox{$\buildrel<\over\sim$}}\ }

\newcommand{\kms}{km s$^{-1}$~}
\newcommand{\msun}{$M_\sun$}

\newcommand{\mbh}{$M_{\rm BH}$}

\newcommand{\eg}{e.g.,}
\newcommand{\ie}{i.e.,}
\newcommand{\etal}{et al.~}
\newcommand{\rg}{$r_{g}$}

\begin{document}

\title{Long-Term Profile Variability of Double Peaked Emission Lines\\ in Active Galactic Nuclei}

\author{S. Gezari}
\affil{California Institute of Technology, Pasadena, CA; suvi@srl.caltech.edu}

\author{ J. P. Halpern}
\affil{Department of Astronomy, Columbia University, New York, NY}

\and 

\author{M. Eracleous} \affil{Department of Astronomy and Astrophysics,
The Pennsylvania State University, University Park, PA}

\begin{abstract}
We present up to two decades of spectroscopic monitoring of the
double-peaked broad H$\alpha$ emission lines of seven broad-line radio
galaxies.  These extremely broad, double-peaked lines are successfully
modeled by emission from gas in the outer accretion disk, and their
profiles vary on timescales of months to years.  This variability is a
valuable probe for understanding the structure and dynamics of the
accretion disk.  We characterize the long-term broad-line profile
variability in detail, and compare the evolution of the line profiles
with predictions from a few simple physical models for dynamical
processes in the accretion disk.  We find no evidence for persistent,
periodic variability that would be indicative of a precessing
elliptical disk, or a circular disk with a long-lived, single-armed
spiral or warp.  We do find transient, periodic variability on the
orbital timescale, possibly resulting from shocks induced by tidal
perturbations, and variability in the wings of the profile which
suggests changes in the emissivity of the inner accretion flow.
Dramatic but localized profile changes are observed during flares and
high-states of broad-line flux.  In 3C 332, these changes can be
explained by a slow, smooth, secular change in disk illumination.  In
Arp 102B and 3C 390.3, however, a simple disk model cannot explain the
asymmetric profile changes, suggesting a need for more sophisticated
models.  We also observe sharp, transient features that appear in the
blue peak of the objects, which require a well-organized velocity
field.
\end{abstract}


\section{Introduction \label{intro}}
In the standard model for active galactic nuclei (AGNs), the central
engine is powered by an accretion disk feeding a supermassive black
hole.  The first direct dynamical evidence  for the presence of the
nuclear accretion disk assumed to be powering AGNs emerged in a class
of broad emission-line AGNs, known as ``double-peaked emitters," with
extremely broad (FWHM = 10,000--20,000 km s$^{-1}$) double-peaked
Balmer emission lines (Eracleous \& Halpern 1994) with line profiles
reminiscent of the double-peaked lines known to originate in the
accretion disks of cataclysmic variable stars (e.g., Horne \& Marsh
1986).  The detailed shape of the ``disk-like'' line profiles are well
fitted by emission from photoionized gas in a circular or elliptical
accretion disk around a central black hole (Chen et al.~1989; Chen \&
Halpern 1989; Eracleous et al.~1995).  Ultraviolet spectroscopy of
several double-peaked emitters obtained with {\it HST}  (Halpern \etal
1996; Eracleous \etal 2004) revealed that the Mg~II $\lambda$2800 line
also has a double-peaked profile similar to the optical H$\alpha$ and
H$\beta$ line profiles and appears to share the same origin as the
Balmer lines, namely, a photoionized accretion disk.  In 1/3 of the
objects, the Ly$\alpha$ line is weak and narrow, which is a prediction
of the photoionized accretion disk model (Dumont \& Collin-Souffrin
1990a,b,c; Rokaki \etal 1992).  In the disk model, Ly$\alpha$ photons
in the dense inner regions of the accretion disk are trapped by
scattering and destroyed by collisional de-excitation, resulting in a
suppressed  emissivity of Ly$\alpha$ in the inner disk where the
rotational velocity of the gas is highest.   In 2/3 of the objects,
the higher-ionization UV lines and Ly$\alpha$ are strong, with broad,
single-peaked profiles, which is a prediction of the accretion disk +
wind model, in which a highly ionized radiatively accelerated
accretion disk wind produces broad, single-peaked profiles with
logarithmic wings (Murray \& Chiang 1997).

Alternative models have been proposed for the origin of the
double-peaked line emission, including:   a bipolar outflow, binary
black holes with separate broad line regions, an anisotropically
illuminated spherical distribution of clouds, and  bloated stars with
cometary tails.  These models  have been progressively ruled out by
inconsistencies with observational  results and by basic physical
arguments.  Eracleous \etal (1997) were able to reject the binary
supermassive black holes scenario favored by Gaskell (1983, 1996) for
three double-peaked emitters (Arp 102B, 3C 390.3, 3C 332), based on
long-term monitoring of their broad double-peaked Balmer lines (see
also Halpern \& Filippenko 1988).   For $\gtrsim$ 20 yr their line
profiles showed no evidence for  the systematic drift of the radial
velocity of the peaks expected for a binary system.  The stationary
peaks also rule out the possibility that the broad peaks are
associated with individual clouds or bloated stars, since the peaks
would then move on the dynamical timescale of the broad line region,
which is not observed.   A collection of clouds on random orbits (Goad
\& Wanders 1996) is not physically plausible since the collisional
timescale between clouds is on the order of a few dynamical times of
the broad line region,  not allowing the configuration of clouds to
last long enough to produce the persistent  double-peaked emission.
In at least one double-peaked emitter, 3C 390.3,  a bipolar outflow
has been definitively ruled out by reverberation mapping,  which
showed no time lag between the response of the wings of the light-echo
of the broad H$\alpha$, H$\beta$ line profiles, as well as the broad
Ly$\alpha$ and  UV C IV line profiles, which would be indicative of a
radial outflow (Dietrich \etal 1998;  O'Brien \etal 1998; Shapavalova
\etal 2001; Sergeev \etal 2002).

\subsection{Demography of Double-Peaked Emitters}\label{intro-sec1}
The first double-peaked emission line was discovered in the H$\beta$
profile of the broad-line radio galaxy (BLRG) 3C 390.3 (Lynds 1968;
Burbidge \& Burbidge 1971; Osterbrock, Koski, \& Phillips 1975), and
its large width and distinct red and blue peaks were attributed to
motion in an accretion disk (Oke 1987; Perez \etal 1988).  Subsequent
discoveries of double-peaked H$\alpha$ lines in the BLRGs Arp 102B and
3C 332 had profiles that were successfully modeled as emission from
photoionized gas in the outer regions of  a circular relativistic
Keplerian accretion disk around a central black hole (Chen \etal 1989,
Chen \& Halpern 1989; Halpern 1990).  Motivated by the fact that at
that time the three known double-peaked emitters were hosted by BLRGs,
Eracleous \& Halpern (1994, 2003) conducted a spectroscopic survey of
moderate redshift ($z < 0.4$) radio-loud AGN, and discovered 21 new
double-peaked broad emission lines (20\% of the sample), of which 13
of the profiles could be fitted with a simple circular disk model.
Careful starlight subtraction of ground-based spectra, and narrow-slit
\textsl{HST} spectroscopy have revealed the presence of double-peaked
emission lines in a large fraction of nearby low-ionization nuclear
emission line regions (LINERs), the most common type of AGN, including
NGC 1097, (Storchi-Bergmann \etal 1993), M 81 (Bower \etal 1996), NGC
4203 (Shields \etal 2000), NGC 4450 (Ho \etal 2000), and NGC 4579
(Barth \etal 2001).  The Sloan Digital Sky Survey (SDSS; York 2000)
spectroscopic AGN sample ($z <$ 0.332) detected 116 additional
double-peaked emitters which made up 4\% of the entire sample.  Of
these, 76\% are radio-quiet, 12\% are classified as LINERs, 40\% fall
in the parameter space of a circular accretion disk profile, and 60\%
require some form of asymmetry in the disk (Strateva \etal 2003).

\subsection{Accretion Disk Model}\label{intro-sec2-ssec1}
The circular relativistic Keplerian accretion disk model from Chen \&
Halpern (1989) is described by the following parameters: $i$, the
inclination of the disk; $\xi_{i}$ and $\xi_{o}$, the inner and outer
radii of the line-emitting portion of the disk in units of
gravitational radii (\rg $= GM/c^{2}$); $q$, the emissivity power-law
index ($\epsilon \propto \xi^{-q}$); and $\sigma$, the local
broadening in \kms.  The local broadening parameter is added to smooth
out the peaks of the profile, which otherwise would be cuspy due to
the sharp inner and outer boundary to the line-emitting region (see
Chen \etal 1989).  Another version of the disk model uses a continuous
emissivity power law with a break radius, instead of local broadening,
to match the smooth peaks of the profile.  In this model, there is no
outer boundary to the line-emitting region, but simply a
characteristic break radius at which the slope of the power-law
changes.  The emissivity power-law index depends on the geometry of
the continuum source.  At radii much larger than the size of the
continuum source, the flux illuminating the disk drops off as
$r^{-3}$, and thus the emissivity, which is proportional to the
illuminating flux (Collin-Souffrin \& Dumont 1989), also has a
power-law index, $q$, equal to 3 (Chen \& Halpern 1989).  In the disk
model,  double-peaked emission lines will arise in the extreme case in
which the line-emitting portion of the disk,  $\xi_{i} \leq \xi \leq
\xi_{o}$, is relatively small (\ie~a ring),  resulting in displaced
broad peaks with a signature blueshifted peak higher than the
redshifted peak due to Doppler boosting, and a gravitational redshift
of the wings of the line.  The disk model produces an excellent fit to
the peaks and wings of the broad H$\alpha$ and H$\beta$ lines, but
there is often a central broad emission-line component at lower
velocities that is not accounted for by the disk model.  This excess
emission may originate from the ``standard''  BLR associated with the
more common single-peaked broad emission lines.  Eracleous \etal
(1995) adapted the circular accretion disk model to elliptical disks
in order to fit the profiles of double-peaked emitters with a red peak
stronger than the blue peak (not possible in a circular disk, in which
the blue peak is always Doppler boosted to be stronger than the red
peak).  This model introduces eccentricity and phase angle parameters
to the circular model described above, and $\xi$ then refers to the
pericenter distance of the elliptical orbits.

The symmetric, and often logarithmic, profiles of broad single-peaked
optical and UV emission lines in standard broad-line AGNs can be
reproduced with many velocity fields and geometric configurations of
the broad-line region (BLR), including both rotational and radial
motion (see review by Mathews \& Capriotti 1985).   Because of this
degeneracy, the kinematics and structure of the BLR remains enigmatic.
Disks are a promising candidate for the BLR structure (Shields 1977),
and they are also invoked to explain many other properties of AGNs.
Reverberation experiments (see review by Peterson 2001) measure the
light-crossing time of the BLR in nearby Seyfert 1 galaxies as 1--2
lt-weeks (\eg Maoz \etal 1990,1991; Netzer 1990),  approximately the
size expected for an accretion disk around a supermassive black hole.
A photoionized accretion disk can  produce ``standard'' single-peaked
broad emission lines  if either the disk is large (Rokaki \etal 1992;
Jackson, Penston, \& Perez 1991), is oriented face-on (Corbin 1997),
or  emits a wind (Murray \& Chiang 1997).  Given the frequency of
``disk-like'' emitters discovered in nearby galaxies, broad emission
lines originating from the nuclear accretion disk may be a universal
phenomenon in AGNs.

\subsection{Testing Models with Long-Term Profile Variability}\label{intro-sec3}

Spectroscopic monitoring of double-peaked emitters has revealed  that
a ubiquitous property of the double-peaked broad emission lines is
variability of the profile shape on the timescale of months to years
(Miller \& Peterson 1990; Romano \etal 1998).   This slow, systematic
variability of the line profile is on the timescale of dynamical
changes in an accretion disk, and has been shown to be unrelated to
the shorter timescale variability ($\sim$ days) seen in the overall
flux in the line, due to reverberation of the  variable ionizing
continuum.  Patterns in the variability of the broad Balmer lines are
often a gradual change and reversal of the relative strengths of the
blueshifted  and redshifted peaks.   Newman \etal (1997) successfully
modeled the periodic oscillation in the relative heights of the blue
and red peaks of the broad line in Arp 102B over 4 yr as an
inhomogeneity, or bright spot, orbiting within the line-emitting
accretion disk.  From the orbital period of the bright spot (2.2 yr)
and its velocity, they were able to solve for the mass of the black
hole from  the orbital radius of the spot  (1.5 x 10$^{16}$ cm) and
the inclination angle determined from the profile fit.  This yielded a
central black hole mass of 2.2$^{+0.2}_{-0.7}$ x 10$^{8}$ \msun.

Periodic variability of the red and blue peak strengths has also been
attributed to a precessing elliptical disk (NGC 1097: Storchi-Bergmann
\etal 1995, 1997) a precessing single-armed spiral (3C 332, 3C 390.3:
Gilbert \etal 1999; NGC 1097: Storchi-Bergmann \etal 2003), and a
precessing warp in the disk.   Elliptical disks and spiral waves have
been detected in the accretion disks of cataclysmic variables
(Steeghs, Harlaftis, \& Horne 1997; Baptista \& Catalan 2000), and a
radiation induced warp has been detected in the large-scale disk of
the AGN NGC 4258 (Maloney, Begelman, \& Pringle 1996). Asymmetries in
an AGN accretion disk, such as eccentricity, a spiral arm, or a warp,
will be induced by the tidal effects of a binary companion in orbit
around the primary black hole, or by the close passage of a massive
star cluster (Chakrabarti \& Wiita 1993, 1994).   If the disk is
self-gravitating, a spiral wave can emerge from a growing density
perturbation, as is  observed in the massive disks of young stellar
objects (Adams, Ruden, \& Shu 1989; Laughlin \& Korchagin 1996).
Elliptical disks have also been proposed to be a natural consequence
of the tidal disruption of a star by a supermassive black hole, when
the tidal disruption debris settles into bound eccentric orbits around
the central black hole (\eg~Rees 1988; Syer \& Clarke 1992).   The
tidal disruption event scenario is a favored interpretation of the
dramatic, sudden appearance of double-peaked emission lines in several
LINERs (NGC 1097, M 81) and the BLRG Pictor A (Halpern \& Eracleous
1994; Sulentic \etal 1995).   Storchi-Bergmann \etal (2003)
successfully fitted the detailed variability of NGC 1097 with a
logarithmic one-armed spiral with a pattern period physically
consistent with the central black hole mass estimated from the
\mbh-$\sigma$ relation.  Spiral waves are a physically desirable
model.  They can play an important role in accretion disks, because
they provide a mechanism for transporting angular momentum outward in
the disk, allowing the gas to flow inwards towards the central black
hole.

Long-term profile variability is thus a useful tool for extracting
information about the structure and dynamics of the accretion disk
most likely producing the double-peaked emission lines.  In this
paper, we present up to two decade's worth of spectroscopic
observations of a sample of double-peaked emitters with the goal of
testing the robustness of the accretion disk model, and models for
physical processes in the accretion disk as well as alternative models
outside the disk paradigm, by directly comparing the observed
evolution of the line profiles with the model predictions.  Confirming
or ruling out any of these models brings significant progress in
developing a model for the origin and motion  of the broad-line gas in
the ``special case" of double-peaked emitters, as well  as in the
standard broad-line regions of normal AGNs.
 
\section{Spectroscopic Monitoring Program \label{ch2-sec1}}

\subsection{Observations \label{obs-sec1}}

A long-term observing program was initiated in 1991 to systematically
take moderate resolution (4--6 \AA) optical spectra of 20 known
double-peaked emitters, selected for having apparent magnitudes bright
enough to be observed with 2 m class telescopes in a reasonable amount
of exposure time.  The majority of the galaxies in our monitoring
program are BLRGs from the Eracleous \& Halpern (1994, 2003) survey of
moderate-redshift radio-loud AGNs.  Included are objects with clearly
double-peaked Balmer lines as well as objects whose Balmer lines are
very asymmetric with a displaced broad peak and shoulders (\eg~3C 227
and Mkn 668).  The spectra were obtained with the Lick 3m, the KPNO
2.1m, the MDM 2.4m, the CTIO 1.5m telescope, and recently with the
9.2m Hobby-Eberly telescope.  Here we present the monitoring of seven
BLRGs: Arp 102B, 3C 390.3, 3C 332, PKS 0235+023, Mkn 668, 3C 227, and
3C 382. The remaining seven BLRGs are presented in Lewis \& Eracleous
(2006).  The broad H$\alpha$ profiles of 12 of the 14 BLRGs in our
monitoring program  have been successfully fit at one time with a
circular or elliptical disk profile in the literature.  Here we
present an elliptical disk profile fit for Mkn 668, leaving only 3C
227 with a profile that is too complex to fit with a single disk
component.  Table \ref{table:dp} lists the BLRGs presented in this
paper, along with their apparent visual magnitude, redshift, and
Galactic reddening.

\begin{figure} [tbp]
\begin{center}
\includegraphics[width=0.9\linewidth]{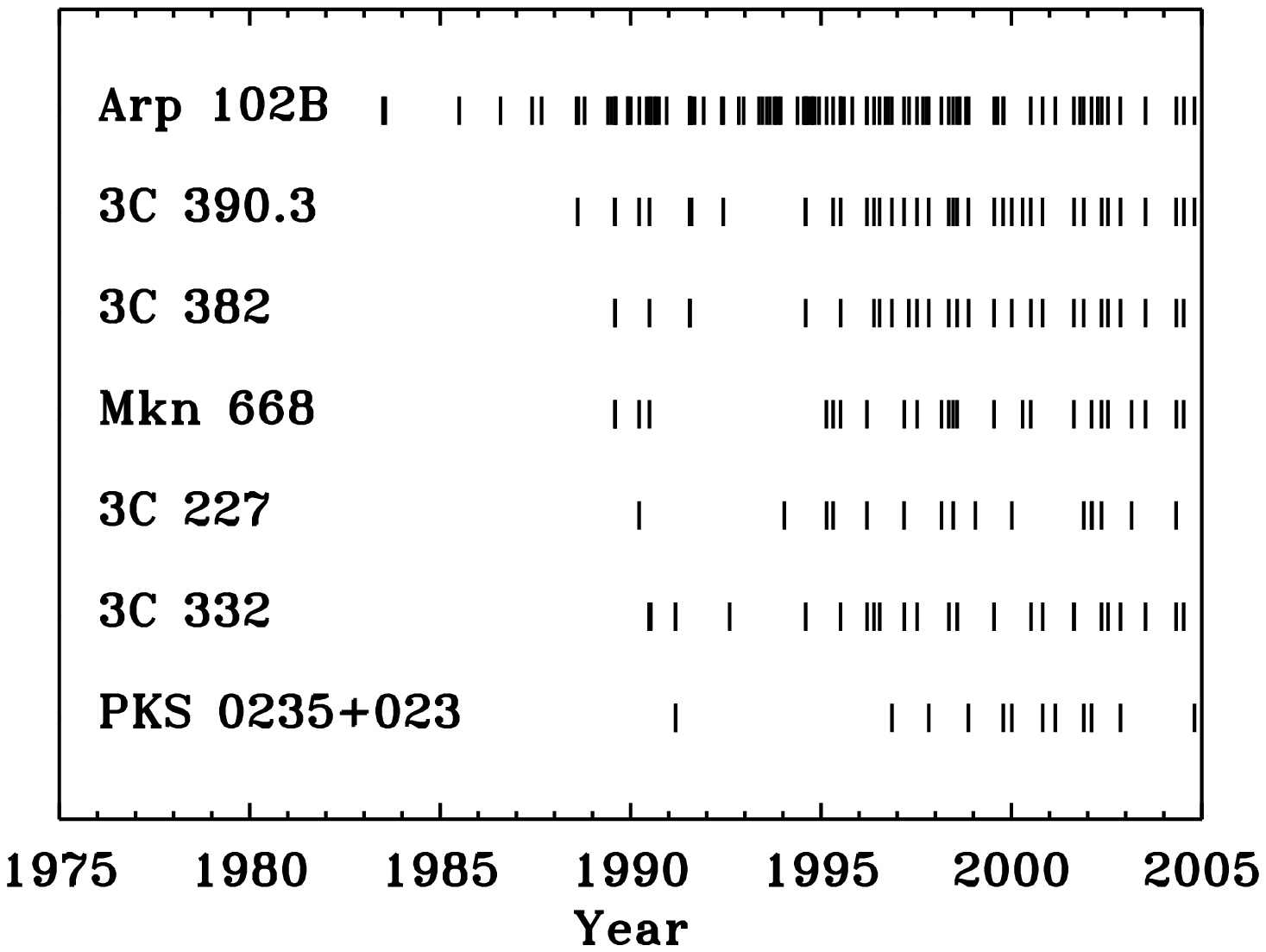}
\caption[Temporal coverage of monitoring program.]{Temporal coverage of the spectroscopic monitoring program 
for the seven double-peaked BLRGs in our program.  Each tick mark represents
a separate observation. \label{ch2:fig:dates}}
\end{center}
\end{figure}

In order to sufficiently sample the long-term variability of the line
profile, where significant changes typically occur on the timescale of
months, when possible, 3--4 observations of the objects were taken per
year.  The log of spectroscopic observations for the objects is listed
in Tables \ref{table:arp}--\ref{table:382}, including the telescope
and exposure time.    Figure \ref{ch2:fig:dates} shows  a histogram of
the temporal spectroscopic coverage of the BLRGs presented in this
paper.

\subsection{Data Reduction \label{obs-sec2}}

The spectra obtained before 1999 were reduced with standard IRAF
routines for bias correction, flat fielding, sky subtraction,
cosmic-ray rejection, and spectrum extraction.  The calibration of
one-dimensional spectra (wavelength calibration, atmospheric
extinction and absorption band correction, and flux calibration) was
carried out with custom-made programs developed by our group
(described in Eracleous \& Halpern 1994).  Spectra obtained after 1999
were reduced with IDL routines written for our observing program,
which include scaling and shifting of the atmospheric bands for
optimal correction, and the removal of bad pixels based on outliers
from the mean local spatial profile.   Spectrophotometric standard
stars (BD +17$\degr$4708, BD +26$\degr$2606, HD 19445, HD 84937: Oke
\& Gunn 1983; Fukugita \etal 1996) were observed to correct for
atmospheric absorption features and for construction of the
sensitivity function.   The extraction window was chosen to include
90\% or 95\% of the total enclosed flux, depending on the dominance of
galaxy starlight.  For our study we isolated the broad H$\alpha$ line,
since  it is the strongest double-peaked emission line, and its high
signal-to-noise makes it possible to study complex profile variability
in detail.

The continuum near the broad H$\alpha$ line was modeled by the
least-squares fit of the linear combination of a stellar template and
a straight line (Eracleous \& Halpern 1994).  The stellar template was
constructed from a scaled spectrum of a normal galaxy [NGC 4339 (E0),
NGC 5576 (E3), NGC 7332 (S0 pec sp), NGC 7457 (SA(rs)0), or UGC 555
(S0/a)], which was chosen from the fit that gave the smoothest
residuals at the strong  Na I D stellar absorption feature near 5900
\AA.  Correct scaling of the stellar template is important for removal
of the contribution of stellar absorption lines to the broad lines.
However, the removal of  H$\alpha$ absorption may not be perfect,
since Na I D and H$\alpha$ could come from different stars, and there
might be some gas-phase absorption in Na I D.  The spectra were
corrected for Galactic extinction, and shifted to their rest
wavelengths using the mean redshift measured from parabolic fits to
the peaks of the strong narrow emission lines in the spectrum.

In order to isolate the broad component of the H$\alpha$ emission
lines, the narrow  emission lines superimposed on the broad H$\alpha$
line (narrow H$\alpha$, [O I] $\lambda$6300, $\lambda$6363, [N II]
$\lambda\lambda$6548,6583, and [S II] $\lambda\lambda$6716,6731) were
fitted with Voigt profiles plus a quadratic function approximating the
broad ``continuum'' below it and then subtracted from the spectrum.  A
Voigt profile fits the wings of the line profile more accurately than
a single Gaussian.  Due to the uncertainty in the shape of the broad
line below the narrow lines, the following constraints were put on the
Voigt profile fits: 1) the widths of the lines were fixed to the width
of the isolated narrow [O I] $\lambda$6300 line (or narrow H$\alpha$
line for objects where  [O I] is too weak), and 2) the [O I] and [N
II] forbidden-line doublets were fixed to have a total intensity ratio
of 3:1.  In the case of Mkn 668,  the multiple-Voigt profile fit to
the H$\alpha$ + [N II] narrow-line complex underestimates  the flux in
the narrow H$\alpha$ line for some of the epochs.  In order to improve
the fits, an additional constraint for [N II]$\lambda$6583/H$\alpha$=
0.97 was added, in order to match the narrow-line ratios of the
best-fit profiles.  The continuum-subtracted spectra were then
normalized by the flux in the narrow [O I] $\lambda$6300 (or narrow
H$\alpha$) line for relative flux calibration.  The narrow emission
lines are used for relative flux calibration since the light-crossing
time of the narrow-line region is long($\sim$ 10$^{2}$--10$^{3}$ yr),
and thus the narrow-line fluxes should not vary on the timescale of
the monitoring program.  Absolute flux calibration is not critical to
this study, since we are most concerned with measuring the changes in
the {\it shape} of the line profile that reflect dynamical changes in
the broad-line gas and not the absolute flux of the broad line in
response to the continuum.  Since the narrow-line region is
predominantly unresolved in the spectra,   the line ratios should be
constant for all the observations, regardless of the slit-width or
seeing conditions of each observation.  The scatter in the narrow-line
ratios therefore reflects the systematic error introduced by the
narrow-line fitting, and is typically on the order of 10\%.  The
velocity scale for the spectra shifted to their rest wavelengths was
determined from the relativistic Doppler formula:

\begin{equation}
\frac{v_{r}}{c} = \frac{\lambda_{\rm obs}^{2} - \lambda_{\rm
rest}^{2}}{\lambda_{\rm obs}^{2} + \lambda_{\rm rest}^{2}}.
\label{ch2:eq:vel}
\end{equation}

The narrow-line--subtracted, narrow-line--flux-calibrated broad
H$\alpha$ profiles of the objects for all the spectra in our
monitoring program are shown in the Appendix.  In the case of 3C 382,
the [O I] $\lambda$6300 narrow line was too weak to use as a template
for line fitting and flux calibration, and the scatter in the narrow
H$\alpha$+[N II] multiple-Voigt profile fit was too large to use the
narrow H$\alpha$ line as a reliable flux calibrator.  Therefore, the
profiles were neither narrow-line--subtracted nor  flux-calibrated,
and the profiles are shown normalized by the flux in the regions of
the broad H$\alpha$ line not contaminated by the narrow lines.

\subsection{Characterization of Variability \label{obs-sec3}}

We characterize in detail the variability of the line profiles on
timescales longer than the light crossing time of the broad-line
region ($\tau_{LT} = r_{BLR}/c$ $\sim$ weeks).  Variability on such
long timescales (months to years) is unlikely to be the result of
reverberation of a variable ionizing continuum (cf.~Blandford \& McKee
1982; Peterson 1993).  Rather, it is caused by changes in the
structure, velocity field, or emissivity of the line-emitting gas.
Simultaneously to changes in the line profile there may also be slow
changes in the integrated flux of the broad lines caused by long-term
variations in the flux of the ionizing continuum.

Mean and root mean square (rms) profiles were created from the broad
H$\alpha$ profiles normalized by their total broad-line flux.  The
profiles are normalized by the broad-line flux in order to remove the
first-order multiplicative response of the profile shape due to
changes in the line flux.  The rms profile thus measures deviations of
the shape of the profile from the mean profile shape.  Sharp features
occur in the rms spectra at the location of the narrow emission lines
due to systematic errors in the narrow-line removal.  The velocities
of the red and blue peaks of the mean spectrum were measured from the
flux-weighted centroid of the top 10\% of the peaks.  The red and blue
peaks of the individual profiles were characterized by measuring the
first three moments of the top 10\% of the peak: $M_{0}$, the average
flux [$P$(R), $P$(B)]; $M_{1}$,  the flux-weighted velocity centroid
[$V$(R), $V$(B)];  and when the signal-to-noise was sufficient,
$M_{2}$,  the dispersion [$\sigma$(R), $\sigma$(B)].  The moments of
the peaks were measured instead of using a Gaussian fit, in order to
make no assumptions about the shape of the peaks, which are often
sharp and asymmetric.  Only the top 10\% of the peaks were used to
measure the moments of the peaks in order to avoid confusion with the
neighboring broad profile shape. The wings of the profile were
characterized by the full width at quarter-maximum (FWQM) and the
shift at quarter-maximum (QMS) measured from the central velocity at
quarter-maximum.  The flux-weighted centroid at quarter-maximum (QMC),
the total integrated flux, and the integrated flux in fixed velocity
intervals were also measured.  The statistical errors in the
measurement of the average flux, centroid, and velocity dispersion of
the peaks were calculated from the signal-to-noise in the continuum at
a rest wavelength of 6900--6930 \AA~(this wavelength range was chosen
so that it was covered by all the spectra in the study).   The noise
per pixel of the continuum in all the spectra is better than
$\sigma_{pix} \aplt 0.01$, which corresponds to a statistical error in
the average flux and centroid of the peaks of less than 5\%.   Peak
velocities and velocity dispersions  are only given for peaks that do
not have narrow lines superimposed on them, as the narrow-line
subtraction is not accurate enough to reveal the true shape of the
peak below it, and thus does not allow for a useful measure of
velocity and width.  The 1~$\sigma$ error in measurements that include
regions superimposed by the narrow lines, including the total
broad-line flux,  is calculated from adding in quadrature the
1$\sigma$ error in the flux of the least-squares Voigt profile fit
and the systematic error in the relative flux calibration measured
from the scatter in the narrow-line ratios (which dominates the
error).  In order to facilitate future studies of these objects, all
of the measurements for the objects are reported in Tables
\ref{tbl:arp:ha_props} -- \ref{tbl:227:ha_props}.

\section{Observed Trends}

\subsection{Arp 102B \label{res-sec1}}

\begin{figure}[tbp]
\begin{center}
\includegraphics[width=0.9\linewidth]{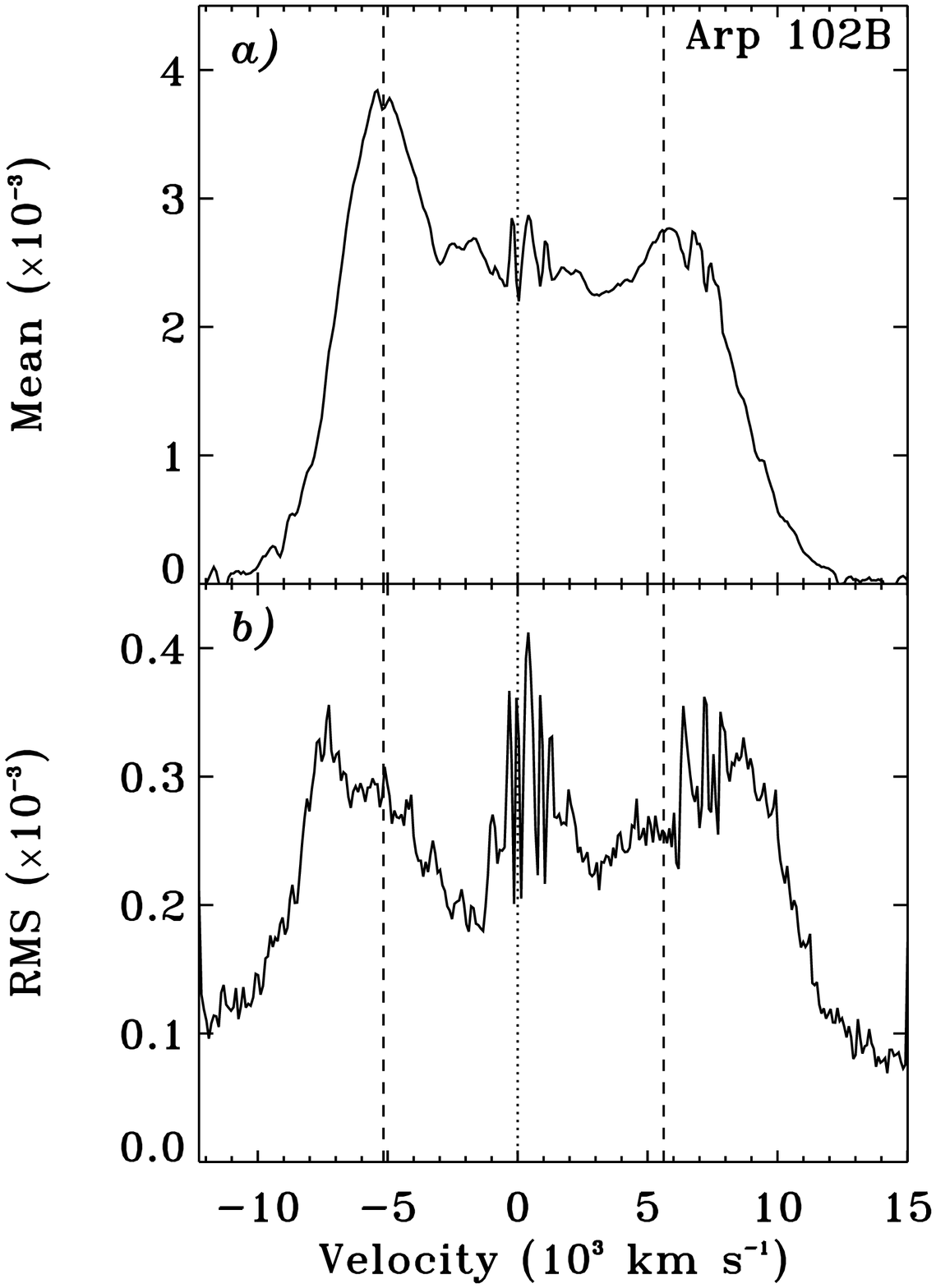}
\caption[Arp 102B: Mean and rms profile.]{Arp 102B: (a) Mean and (b) rms profile of the broad H$\alpha$ line in units of $f_{\nu}$ normalized by the total broad-line flux, with dashed lines indicating 
the velocities of the red and blue peaks of the mean profile, measured from the flux-weighted velocity
centroid of the top 10\% of the peaks.  \label{arp:fig:meanrms}}
\end{center}
\end{figure}

\begin{figure} [tbp]\centering

\includegraphics[width=0.75\linewidth]{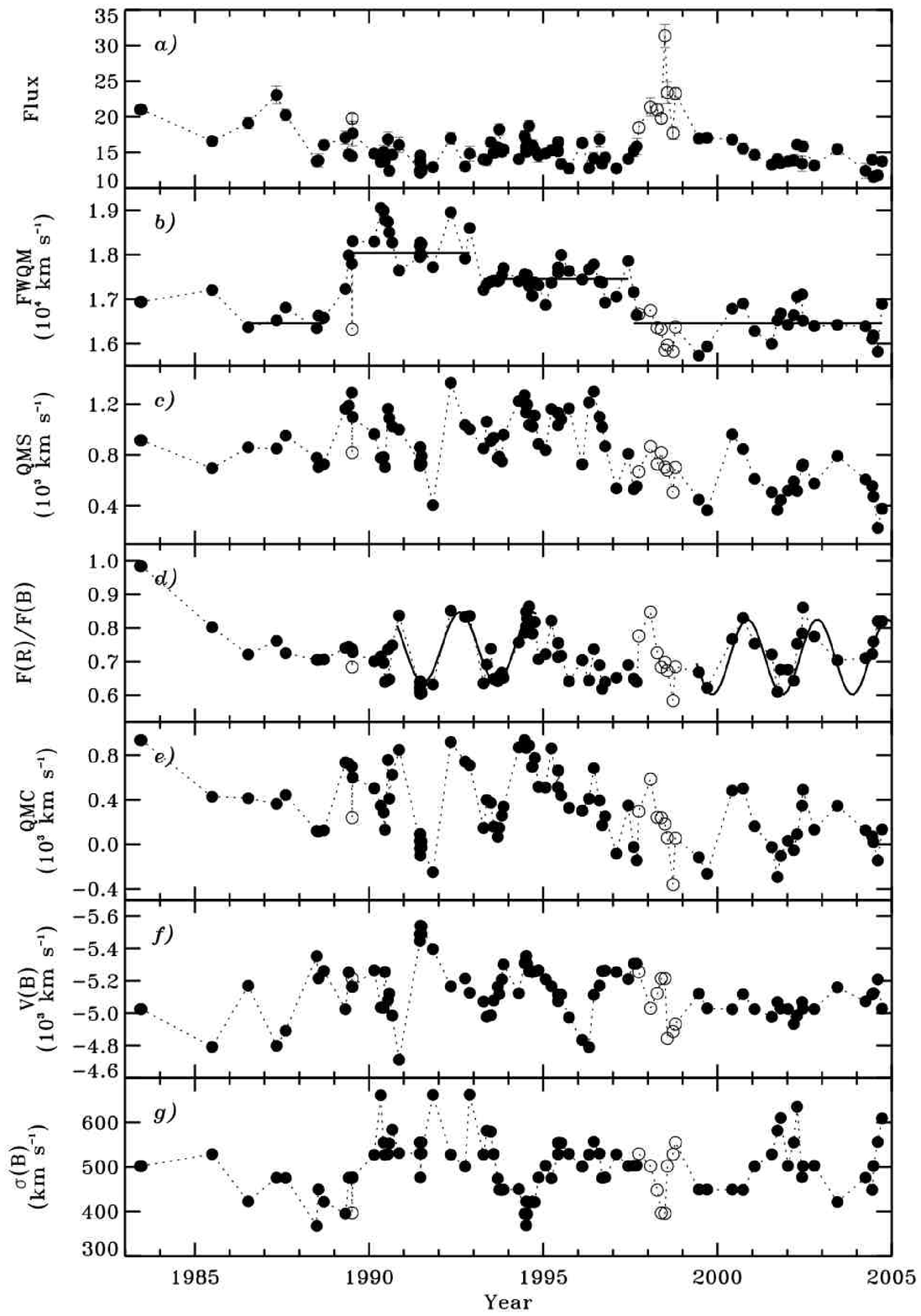}
\caption[Arp 102B: Characterization of the profiles.]{Arp 102B: (a) Total broad-line flux calibrated relative to the narrow 
[O I] $\lambda$6300 line, (b) full width at quarter-maximum, (c) shift at quarter-maximum, 
(d) ratio of the average flux in fixed intervals in the red (+4900 to +6400 \kms) and blue ($-$4500 to $-$5900 \kms) peaks, (e) flux-weighted centroid at quarter-maximum,
(f) flux-weighted velocity centroid and (g) dispersion of top 10\% of the blue peak of the broad H$\alpha$ profiles over the entire duration of our monitoring program.
Solid lines show the step-like variation of the FWQM that are reproduced by model disk profiles.
The least-squares fit sine curves to the oscillations of $F$(R)/$F$(B) are plotted with solid lines. 
The times of a rapid flare in total broad-line flux 1989, and a high-state of total broad-line flux in 1989 during
which a large amplitude flare occurred, are plotted with open circles.
\label{arp:fig:tp_arp_3}}
\end{figure}

The mean spectrum of Arp 102B (Figure \ref{arp:fig:meanrms}a) has redshifted and blueshifted peaks with 
$V$(R) = +5600 and $V$(B) = $-$5200 km s$^{-1}$, respectively.  The rms spectrum (Figure \ref{arp:fig:meanrms}b) indicates strong variability in the shape of the profile near the peaks
of the mean spectrum, with the strongest variability in the wings of the profile.  Figure \ref{arp:fig:tp_arp_3}
shows the a) flux calibrated relative to narrow [O I]$\lambda$ 6300, b) FWQM, c) QMS, d) $F$(R)/$F$(B), e) QMC, f) $V$(B), and g) $\sigma$(B) of the broad H$\alpha$ line over the entire duration of the monitoring program.   The $F$(R)/$F$(B) ratio is measured from the average flux in the velocity ranges corresponding to the top 10\% of the red peak (+4900 to +6400 km s$^{-1}$) and blue peak ($-$4500 to $-$5900 km s$^{-1}$) of the mean spectrum.  Fortunately, the velocity range for the red peak does not overlap with the location of the narrow [S II] lines.
 
{\it Flares in broad-line flux.} -- A rapid flare in the
broad-line flux occurred in 1989, with a duration of $\sim$ 10 days, and 
an increase in flux of 35\%.  In addition, there was a year-long period of a high state 
in flux in 1998,
during which a large amplitude flare occurred with a duration of $<$ 123 days, 
and an increase in flux of 60\%.  Both flares are also seen in the continuum flux near
H$\beta$, and thus are not the result of systematic effects in the relative flux calibration due to changes in seeing or slit width.  The times of the flares and the high-state of
flux are plotted with open circles in Figure \ref{arp:fig:tp_arp_3}.  
During both
flares, the FWQM decreased to $<$ 1.65 $\times$ 10$^{4}$ km s$^{-1}$ and the width
of the blue peak narrowed to $\sigma$(B) = 400 \kms.  During the rapid flare, the QMS and QMC shifted by $\sim$ $-$500 \kms, as a result of a sudden decrease in the strength of the red wing during the flare.   

{\it Systematic variations in FWQM.} -- The FWQM varies in a step-like
manner, with a rapid increase from 1.67 $\times$ 10$^{4}$ to 1.80 $\times$ 10$^{4}$ km s$^{-1}$
from 1988.71 to 1989.50, and a drop to 1.75 $\times$ 10$^{4}$ km s$^{-1}$ in
1993, and then a drop to 1.67 $\times$ 10$^{4}$ km s$^{-1}$ in 1998.  The
changes in the FWQM reflect systematic changes in the wings of the profile.  Along with changes in the width of the profile, the QMS also shifts
from a mean of +1000 km s$^{-1}$ to a mean of +600 km s$^{-1}$ after 1997, revealing a weakening of the red wing relative to the blue wing.

{\it Red peak higher than blue peak.} -- The shape of the profile in 1983 is
quite distinct from the  later spectra, with a strong, broad red peak
and $F$(R)/$F$(B) close to 1.  In fact, spectra of the broad H$\alpha$ line 
reported by Miller \&
Peterson (1990) taken in 1982 have $F$(R)/$F$(B) $>$ 1.

{\it Oscillation of $F$(R)/$F$(B).} --  $F$(R)/$F$(B) oscillates from 1991 to 1994, which
was reported by Newman \etal (1997) (and
measured from slightly different wavelength intervals), and this oscillation of $F$(R)/$F$(B) reappears in
1999.  The least-squares fit of the $F$(R)/$F$(B) ratio to a sine curve with
weights set to 1/$\sigma_{stat}^{2}$ for the times of oscillation from 1990 -- 1994 and 1999 -- 2005 
results in a period of 2.073 $\pm$ 0.018 and 2.010 $\pm$ 0.006 yr, respectively.  Our period for the first
oscillation is within the errors of Newman et al.'s (1997) measurement of 2.16 $\pm$ 0.07 yr.  The oscillation of $F$(R)/$F$(B) is also evident in the QMC, which has the same pattern of oscillation.
During the oscillation of $F$(R)/$F$(B), the blue peak has a minimum
velocity during a maximum of $F$(R)/$F$(B) in 1990.86 (when the blue peak
is weak), and a maximum velocity during a minimum of $F$(R)/$F$(B) in
1991.47 (when the blue peak is strong). The width of the blue peak
spikes during the $F$(R)/$F$(B) oscillation  at times when the very top of
the blue peak appears to be suppressed in strength, and does not
always  occur during a maximum of $F$(R)/$F$(B).

{\it Systematic drift of the blue peak velocity.} -- The velocity of
the blue peak varies by a maximum amplitude of 500 \kms from the mean.   
The velocity systematically decreases from 1994.87 to 1996.31,
from $-$5300 to $-$4800 km s$^{-1}$, and then increases back to $-$5300 km
s$^{-1}$ in 1996.69.

{\it Systematic narrowing of the blue peak.} -- The width of the blue peak
systematically narrows to a minimum of 400
km s$^{-1}$ from 1994.46 to 1994.75, when the shape of the blue peak is visibly sharper.

\subsection{3C 390.3 \label{ch2-sec4}}

\begin{figure} [tbp]
\begin{center}
\includegraphics[width=0.9\linewidth]{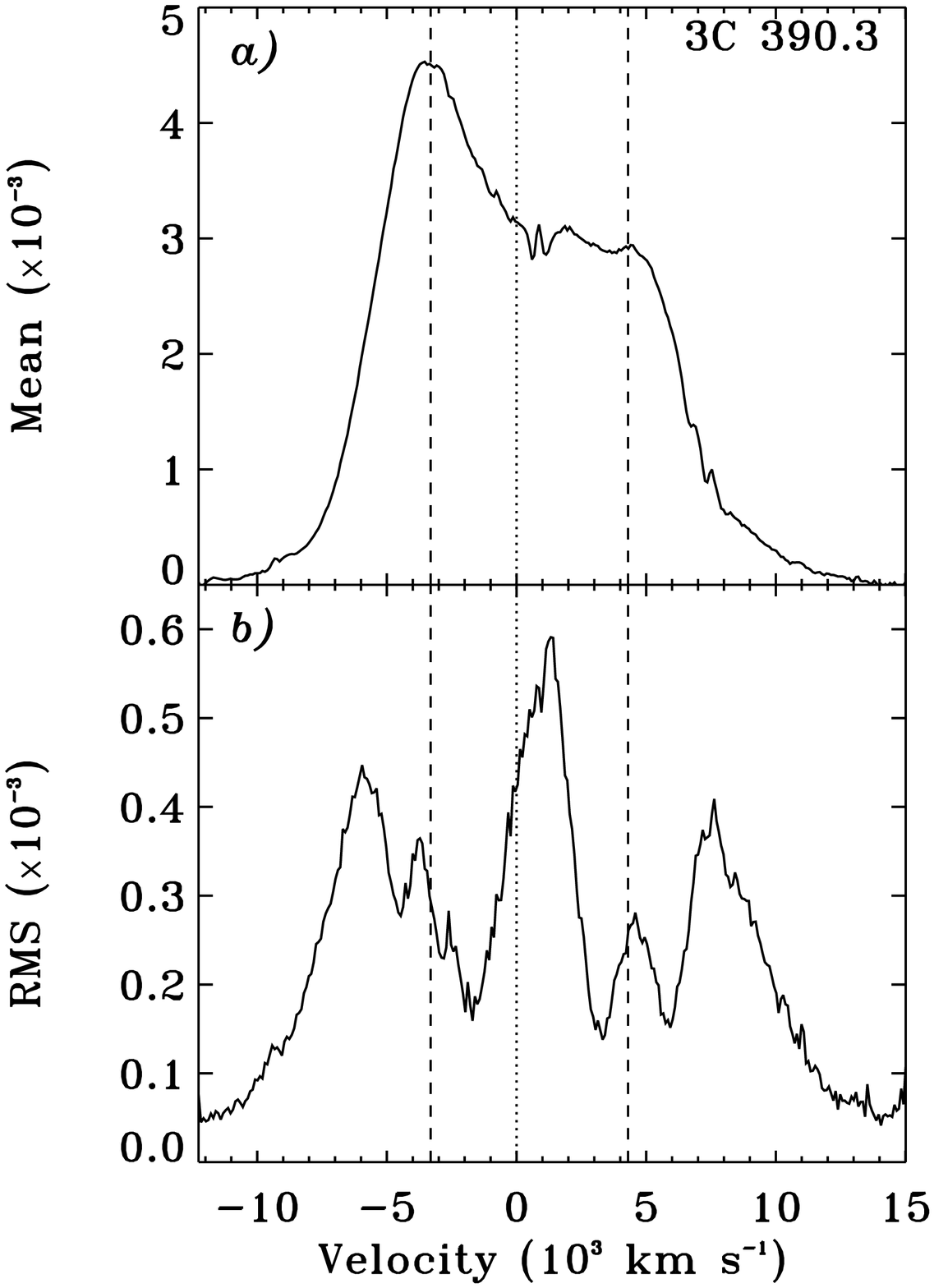}
\caption[3C 390.3: Mean and rms profile.]{3C 390.3: (a) Mean and (b) rms profile of the broad H$\alpha$ line in units of $f_{\nu}$ normalized by the total broad-line flux, with dashed lines indicating 
the velocities of the red and blue peaks of the mean profile, measured from the flux-weighted velocity
centroid of the top 10\% of the peaks.
\label{390:fig:meanrms}}
\end{center}
\end{figure}

\begin{figure}[tbp]\centering
\includegraphics[width=0.8\linewidth]{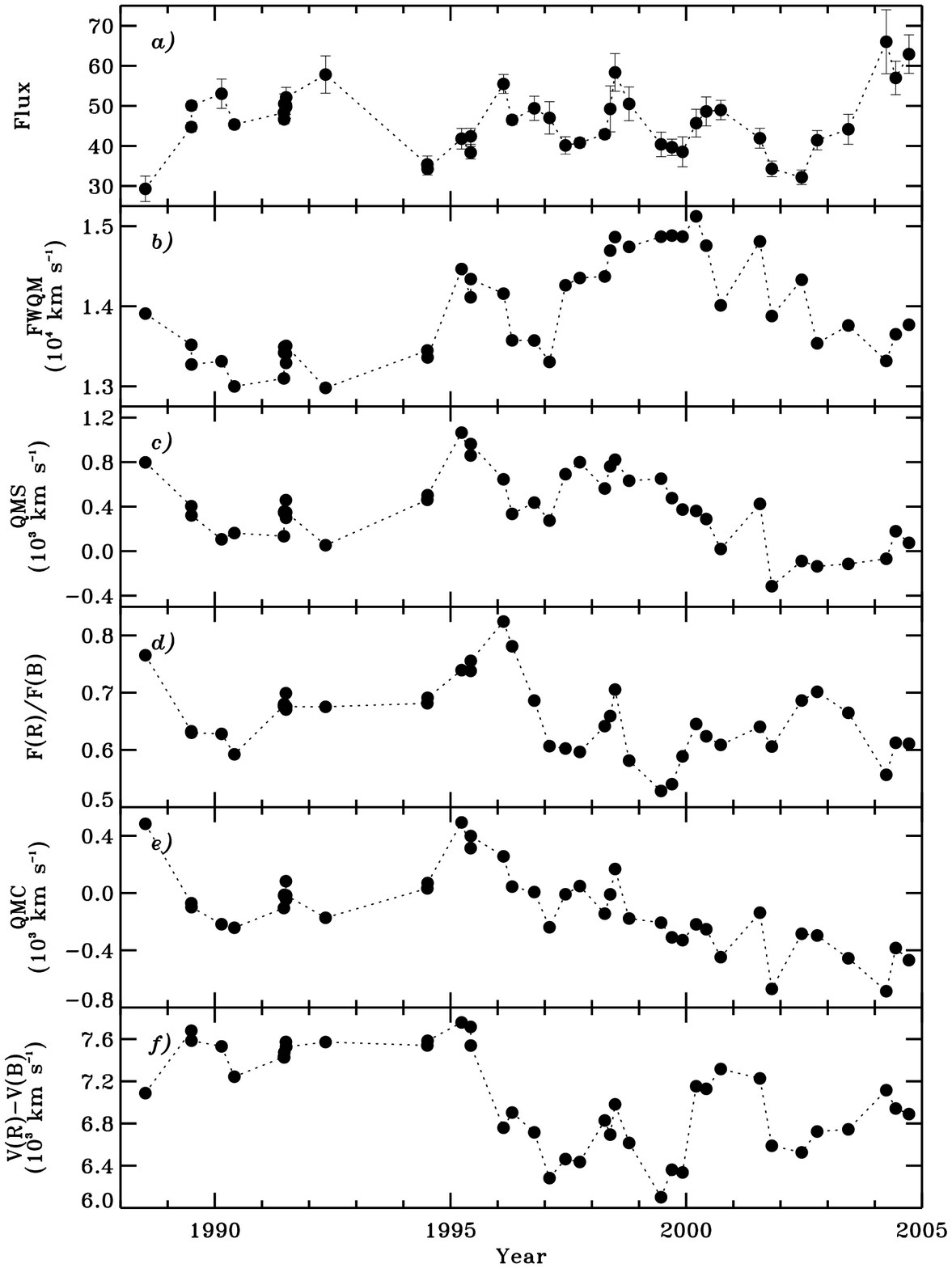}
\caption[3C 390.3: Characterization of the profiles.]{3C 390.3: (a) Total broad-line flux calibrated relative to the narrow [O I] $\lambda$6300 line, (b) full width at quarter-maximum, (c) shift at quarter-maximum, 
(d) ratio of the average flux in fixed intervals in the red (+3300 to +5900 \kms) and blue ($-$1700 to $-$4500 \kms) peaks, (e) flux-weighted centroid at quarter-maximum,
and (f) peak velocity separation of the broad H$\alpha$ profiles over the entire duration of our monitoring program.
\label{390:fig:tp_390_3}}
\end{figure}

\begin{figure} [tbp]\centering
\includegraphics[width=0.9\linewidth]{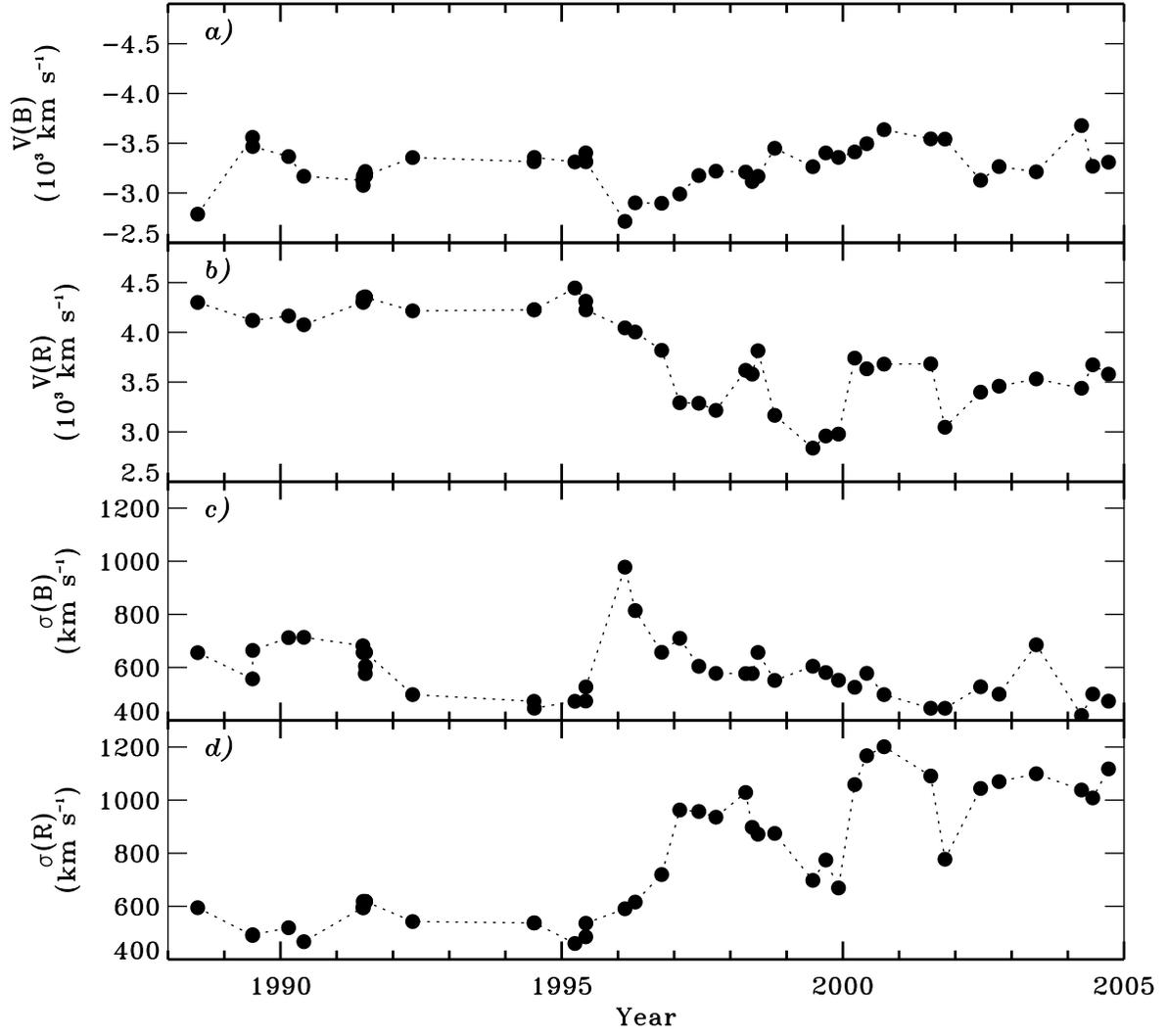}
\caption[3C 390.3: Characterization of the red and blue peaks.]{3C 390.3: (a, b) Flux-weighted velocity centroid and (c, d) dispersion of the top 10\% of the
red and blue peaks of the broad H$\alpha$ profiles over the entire duration of
our monitoring program.
\label{390:fig:velsig}}
\end{figure}

\begin{figure} [tbp]\centering
\includegraphics[width=0.9\linewidth]{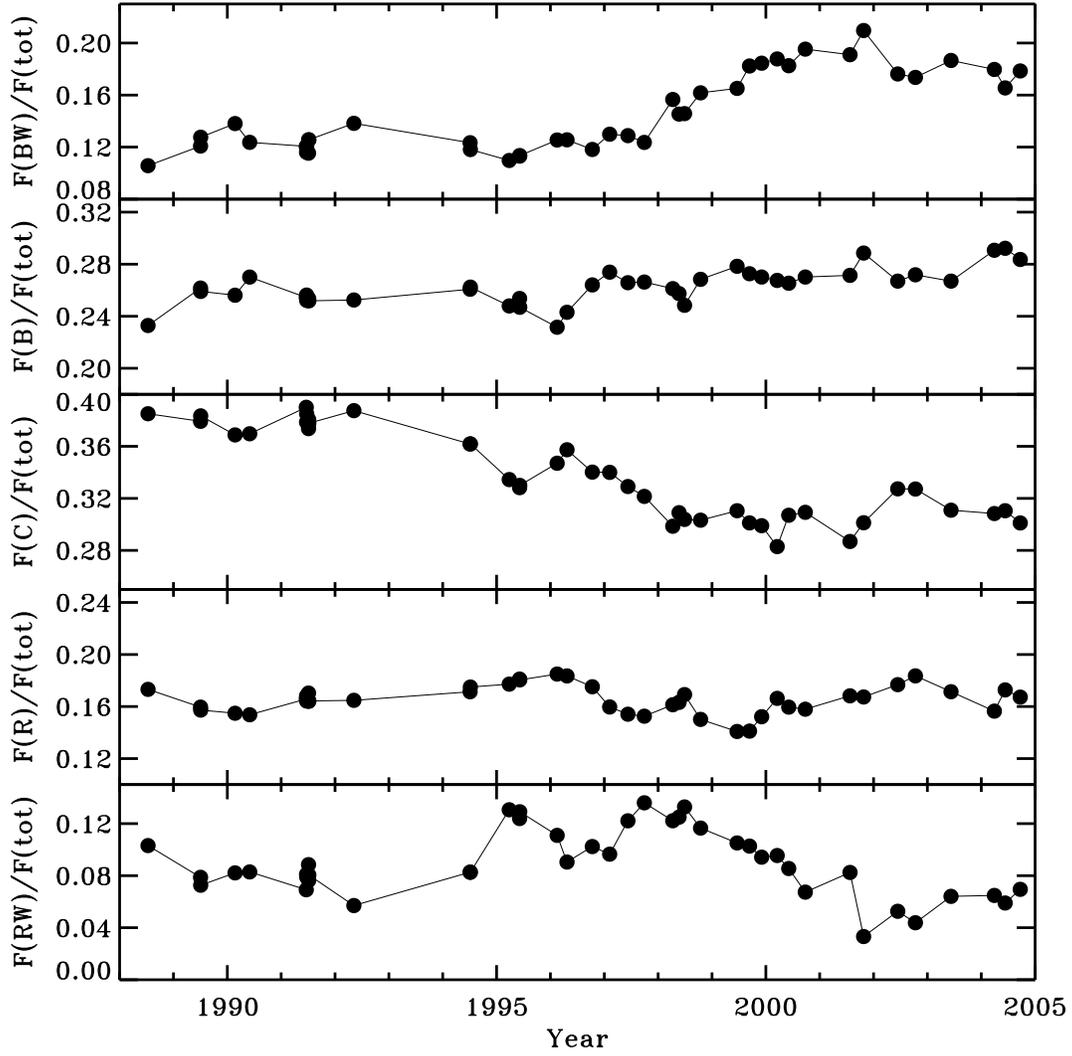}
\caption[3C 390.3: Normalized integrated flux in different parts of the profile.]{3C 390.3: 
Normalized integrated flux in fixed intervals in the blue wing [$F$(BW): $-$4500 to $-$12,300 \kms], 
blue peak [$F$(B): $-$1700 to $-$4500 \kms], central peak [$F$(C): $-$1700 to +15,000 \kms], 
red peak [$F$(R) :+3300 to +5900 \kms], and red wing [$F$(RW): +5900 to +15,000 \kms]
of the  broad H$\alpha$ profiles over
the entire duration of our monitoring program.
\label{390:fig:regions}}
\end{figure}

The mean spectrum of 3C 390.3 (Figure \ref{390:fig:meanrms}a) has a strong blue peak
with $V$(R) = $-$3300 \kms and a red peak on a flat shelf, with $V$(R) = +4300 \kms.
The rms profile (Figure \ref{390:fig:meanrms}b) shows five
distinct regions of strong profile variability in
the red central part of the profile, peaking at a velocity of +1300 \kms,
near the velocities of the red peak and the blue peak of the mean
profile, and in the wings of the profile.  
It should be noted that there is a spike in the rms profile
at $-$2500 \kms that corresponds to the strong $B$-band atmospheric
absorption feature at 6870 \AA, which is not always divided out
perfectly by the atmospheric standard stars in the reduction of the
spectra.  Figure \ref{390:fig:tp_390_3} shows the a) flux calibrated relative to narrow [O I] $\lambda$6300, b) FWQM,  c) QMS, d) $F$(R)/$F$(B), e) QMC, and f) $V$(R)-$V$(B) of the broad
H$\alpha$ line over the entire duration of the monitoring program.  Figure \ref{390:fig:velsig} shows the a)
$V$(B), b) $V$(R), c) $\sigma$(R), and d) $\sigma$(B) of the peaks of the profiles.
The $F$(R)/$F$(B) ratio is measured from the average
flux in the velocity ranges corresponding to the top 10\% of the red peak (+3300 to +5900 km s$^{-1}$) and blue peak ($-$1700 to $-$4500 km s$^{-1}$) of the mean spectrum.
Figure \ref{390:fig:regions} shows the integrated flux in
the five coherent variable regions of the profile determined from the
rms spectrum, relative to the total flux of the broad line: the blue
wing ($-$4500 to $-$12,300 km s$^{-1}$), blue peak ($-$1700 to $-$4500 km s$^{-1}$), 
central peak ($-$1700 to +3300 km s$^{-1}$), red peak (+3300 to +5900 km s$^{-1}$), 
and red wing (+5900 to +15,000 km s$^{-1}$).   The dramatic changes
in the shape of the profile are also demonstrated in the individual profiles plotted in
comparison with the mean profile scaled to the total flux of the line
(see Appendix).

Figure \ref{390:fig:vz91} shows $F$(B)/$F$(R) defined by Veilleux \&
Zheng (1991) (hereafter VZ91) and the velocity of the blue peak over
40 yr, with data points from 1974 to 1990 from VZ91 shown with open
circles, and data from our monitoring program shown with filled
circles.  The definition of $F$(R) from VZ91 is actually centered on
the red {\it central} peak located at +1800 \kms, and $F$(B) includes
the blue peak and blue wing of the profile.  The quasi-periodic
oscillation that they reported does not continue after 1990, and the
$F$(B)/$F$(R) ratio continues to rise due to the decreasing strength
of the central peak (see Figure \ref{390:fig:regions}).  The linear
decrease of the velocity of the blue peak from $-$4700 \kms in 1968 to
$-$3200 \kms in 1988 was reported by Gaskell (1996), and used to fit a
double-line spectroscopic binary model with a period of 300 yr and a
binary mass of 6.6 x 10$^{9}$ \msun.  The linear trend in the blue
peak velocity  does not continue after 1988,  and the velocity levels
off at $-$3200 \kms,  which was also noted by Shapovalova \etal (2001)
from their monitoring of the blue peak velocity  of the H$\beta$
profile from 1995 to 2000.
 
Shapovalova \etal (2001) reported an anticorrelation between the peak
velocity separation in the H$\beta$ profile (measured from difference
profiles, in which the minimum-activity average spectrum was
subtracted from spectra {\it not} normalized by their broad-line flux)
and the total broad line flux, as well as an anticorrelation with the
flux measured in the continuum at 5125 \AA.  We do not see such a
relationship between peak velocity separation and total flux in our
monitoring of H$\alpha$; however we measure the peak velocities from
the narrow-line--subtracted spectra, and have peak velocity
separations that range from (6 -- 8) $\times$ 10$^{3}$ km s$^{-1}$,
while their measurements of $V$(R)-$V$(B) are in the range (8 -- 15)
$\times$ 10$^{3}$ km s$^{-1}$.  It seems that their measurement of the
peak velocity separation is affected when producing the difference
profiles.

{\it Fluctuation of $F$(R)/$F$(B).} -- A fluctuation of $F$(R)/$F$(B)
is observed from 1994 to 1998 over 4 yr, with an amplitude of 0.11.
During the maximum of $F$(R)/$F$(B) in 1996.12, the velocity of the
blue peak drops by +600 \kms, and the  width of the blue peak jumps by
90\%.  After the maximum of $F$(R)/$F$(B), the width of the blue peak
steadily decreases, and the blue peak velocity increases.

{\it Flattening of the red peak.} -- The dramatic changes in the width
and velocity of the red peak starting in 1997 are a consequence of the
shape of the red peak flattening and the red wing forming into a sharp
shelf in 2000 (see individual spectra in the Appendix).

{\it Blueward drift of QMS and QMC.} -- After reaching a peak velocity
in 1995, the QMS and QMC drift by $\sim$ $-$1000 \kms, reflecting the
increase in strength of the blue peak and blue wing of the profile
relative to the red wing.

{\it Transient sharp features in the blue peak.} -- Transient sharp
spikes in the blue peak (shown in Figure \ref{390:fig:spike1})  appear
in 1989.50, persisting for $>$ 2 yr with a FWHM $\simeq$ 750 km
s$^{-1}$ and a velocity of -4300 km s$^{-1}$, and again in 1998.27
with a FWHM $\simeq$ 700 km s$^{-1}$ that drifts from $-$3000 to
$-$2600 km s$^{-1}$ over a period of 1.5 months.  Although the
velocity of the spike in 1998 is quite close to the location of the
atmospheric $B$-band feature at $-$2500 km s$^{-1}$, it is a real
feature that is also detected in Shapovalova \etal's (2001) monitoring
of the H$\beta$ profile of 3C 390.3 taken with a different telescope
from 1998.15 to 1998.54.

{\it Stronger blue peak during high state of flux.} -- The total flux
of the broad line is in a high state in 2004 and the profile has a
stronger and sharper blue peak than in the mean profile.  However, it
also appears that the blue peak is getting systematically sharper and
stronger after 2000 (Figure \ref{390:fig:pfits2}), thus this behavior
may be unrelated to the increase in flux.

\subsection{3C 332 \label{ch2-sec5}}

The mean spectrum of 3C 332 (Figure \ref{332:fig:meanrms}a) has
well-separated red and blue peaks with $V$(R) = +8300 \kms and $V$(B)
= $-$6400 km s$^{-1}$, respectively.  The rms spectrum (Figure
\ref{332:fig:meanrms}a) is strongest in the red wing, and also shows
profile variability in the blue wing.  Figure \ref{332:fig:tp_332_3}
shows the a) flux calibrated relative to narrow H$\alpha$, b) FWHM, c)
HMS, d) $F$(R)/$F$(B), e) HMC, and f) $V$(R)--$V$(B) of the broad
H$\alpha$ line over the entire duration of the monitoring program.
Figure \ref{332:fig:velsig} shows the velocity of the red and blue
peaks of the profiles.  The width, shift, and centroid is measured at
half-maximum for this object, since the quarter-maximum is often
located too far out in the wings of the spectrum to be measured
reliably.  The $F$(R)/$F$(B) ratio is measured from the average flux
in the velocity ranges corresponding to the top 10\% of the red peak
(+7200 to +10,000 km s$^{-1}$) and blue peak ($-$5300 to $-$8000 km
s$^{-1}$) of the mean spectrum.

{\it Maximum in $F$(R)/$F$(B).} -- $F$(R)/$F$(B) rises to a maximum in
1995.43,  at which point the velocity of the blue peak shifts to a
lower velocity, and the velocity of the red peak shifts to a higher
velocity.  During this maximum, the HMC rises to a maximum red
velocity of +2500 \kms, and then returns back to +500 \kms in 1998,
where it remains constant.

{\it Changes in profile shape during high state of flux.} --  The FWHM
and $V$(R)--$V$(B) begin to decrease when the flux increases in 1998.
Figure \ref{332:fig:corrmatrix} shows strong negative  correlations
between flux and the peak velocity separation, FWHM, and the
integrated flux in the red wing (+10,000 to +12,800 \kms)  and the
blue wing ($-$8000 to $-$10,800 \kms), normalized by the total flux.
These negative correlations indicate that at higher fluxes, the peaks
shift closer together, and the wings of the profile are weaker.  There
also appears to be a bluer HMS during the high-state of flux,
indicating a weakening of the red wing relative to the blue wing.

\subsection{PKS 0235+023 \label{ch2-sec6}}

The mean spectrum of PKS 0235+023 (Figure \ref{0235:fig:meanrms}a) has
a narrow blue peak with $V$(B) = $-$4700 \kms, and red peak with
$V$(R) = +2700 \kms.  The rms profile (Figure \ref{0235:fig:meanrms}b)
shows variability in the shape of the profile blueward of the blue
peak, and redward of the red peak of the mean spectrum.  Figure
\ref{0235:fig:tp_0235_3} shows the a) flux calibrated relative to
narrow H$\alpha$, b) FWQM, c) QMS, d) $F$(R)/$F$(B), e) QMC, and f)
$V$(B) of the broad H$\alpha$ line over the entire duration of the
monitoring program.  The $F$(R)/$F$(B) ratio is defined from the
integrated flux in the regions of maximum variability in the rms
spectrum near the red peak (+2300 to +5200 \kms) and blue peak
($-$3900 to $-$6900 \kms) of the mean spectrum.

{\it Dramatic change in profile shape after 1991.} -- The drop in
$F$(R)/$F$(B) by 50\%, the drift in the velocity of the blue peak by
+500 \kms, and the drift in velocity of the QMS and QMC by $\sim$
$-$2000 \kms,  reveal a dramatic change in the profile shape between
the 1991 profile and the profiles after 1996, when the entire profile
appears to get skewed towards the blue.

{\it Linear increase of $V$(B)}. -- The blue peak velocity increases
roughly linearly from $-$3600 \kms in 1991 to $-$5000 \kms in 2005.

{\it Minimum in FWQM.} -- The minimum in the FWQM of the profile from
1998--1999 is the consequence of a stronger blue peak as well as an
almost complete disappearance of the red peak and red wing of the
profile.

\subsection{Mkn 668 \label{ch2-sec7}}

The mean profile of Mkn 668 (Figure \ref{668:fig:meanrms}a) has a
strong central red peak with $V$(R) = +1700 \kms, and a broad base
with a blue shelf.  The rms spectrum (Figure \ref{668:fig:meanrms}b)
shows strong variability redward and blueward of the central red peak
of the mean profile.  Figure \ref{668:fig:tp_668_3} shows the a) flux
calibrated relative to narrow H$\alpha$, b) FWQM, c) QMS, d)
$F$(R)/$F$(B), e) QMC, and f) $V$(R) of the broad H$\alpha$ line over
the entire duration of the monitoring program.  The $F$(R)/$F$(B)
ratio is measured from the integrated flux in the regions of strong
variability measured by the rms spectrum, redward of the central peak
(+3200 to +5600 \kms) and blueward of the central peak ($-$800 to
$-$3200 \kms) of the mean profile.

{\it Fluctuation of $F$(R)/$F$(B).} -- The $F$(R)/$F$(B) ratio
fluctuates, with a maximum of 2.5 in 1990.42, decreasing to a minimum
of 1.5 from 1996 to 2001, and then increasing to a maximum of 2.3 in
2002.  These changes in $F$(R)/$F$(B) are reflective of the dramatic
changes in the shape and peak velocity of the central peak of the
profile.  The velocity of the central peak shifts from a maximum of
+2500 \kms during the maximum of $F$(R)/$F$(B), and to a minimum of
$\lesssim$ +1700 \kms during the minimum of $F$(R)/$F$(B),  while the
shape of the profile changes from a skewed to a more symmetric shape.

{\it Dramatic changes in $V$(R).} -- The historical velocity curve of
the central red peak is shown in Figure \ref{668:fig:vels} with data
points in 1974 from Osterbrock \& Cohen (1979), in 1982 from Gaskell
(1983), and data points from Marziani \etal (1993) from 1985 to 1991
measured from the H$\beta$ profile.  The central peak was at a maximum
velocity of +2700 \kms from 1974 to1982, before shifting to a lower
velocity of +1800 \kms in 1985.  When the peak shifts to smaller
velocities, the peak velocity measurement of the H$\alpha$ profile is
affected by systematic errors in the narrow-line subtraction.  Due to
the uncertainty in the exact shape of the broad-line underneath the
narrow H$\alpha$+[N II] narrow-line complex, upper limits are shown
for our measurement of $V$(R) when the measured velocity is within the
velocities of the narrow lines.

\subsection{3C 227  \label{ch2-sec8}}

The mean profile of 3C 227 (Figure \ref{227:fig:meanrms}a) has a
strong central blue peak with  $V$(B) = $-$1500 \kms, and a broad base
with a red and blue shelf.  The rms profile reveals a strong and
narrow region of variability on the blue side of the central peak,
with a FWHM $\simeq$ 2200 \kms, and strong variability redward of the
central peak, as well as some variability in the shape of the blue and
red shelf of the profile.  Figure \ref{227:fig:tp_227_3} shows the a)
flux calibrated relative to narrow H$\alpha$, b) FWQM, c) QMS, d)
$F$(R)/$F$(B), e) QMC, and f) $V$(B) of the broad H$\alpha$ line over
the entire duration of the monitoring program.  The $F$(R)/$F$(B)
ratio is measured from the integrated flux in the regions of strong
variability measured by the rms spectrum redward of the central peak
(+1700 to +3400 \kms) and blueward of the central peak ($-$1200 to
$-$4400 \kms) of the mean spectrum.

{\it Excess feature in the blue peak.} -- The $F$(R)/$F$(B) ratio
increases monotonically from 1990 to 2000, revealing an excess feature
in the central blue peak that fades away.  The historical velocity
curve for 3C 227 (Figure \ref{227:fig:vels}), which includes data
points in 1974 from Osterbrock, Koski \& Phillips (1976), and in 1982
from Gaskell (1983), indicates that the excess in the central blue
peak persisted from 1974 to 1998.  The QMS and QMC increase by +800
\kms between 1995 and 2000, also a result of the fading away of the
blue feature.

\subsection{3C 382  \label{ch2-sec9}}

The mean profile of 3C 382 (Figure \ref{382:fig:meanrms}a) has a red
peak and a blue shoulder that are superimposed by the H$\alpha$+[N II]
narrow-lines.  The mean and rms profiles are measured without the
narrow emission lines subtracted from the spectrum, because of the
large errors in fitting the weak [O I] $\lambda$ 6300 narrow line and
the crowded H$\alpha$+[N II] narrow-line complex. The rms profile
(Figure \ref{382:fig:meanrms}b) shows variability in the shape of the
red wing, as well as a narrow region of profile variability at the
velocity of the red peak of the mean profile, and in the far blue
wing.  Figure \ref{382:fig:tp_382_3} shows the integrated flux in the
regions of maximum variability measured by the rms spectrum, relative
to the total integrated flux of the line, avoiding regions
superimposed by narrow lines, near the red peak (+2100 to +3800 \kms),
the red wing (+4400 to +6200, +8000 to +15,000 \kms), and the blue
wing ($-$7500 to $-$11,300 \kms) of the mean spectrum, over the entire
duration of the monitoring program.  $F$(R)/$F$(tot) drops by 15\% in
1994 to 1995, and then returns back to its mean value in 1997.  During
the drop in $F$(R)/$F$(tot), the normalized flux in the red wing
increases by 25\%, and then steadily decreases.  After this jump in
$F$(RW)/$F$(tot), there is a sudden increase in $F$(BW)/$F$(tot) in
1998 by 40\%,  after which the normalized flux in the blue wing
steadily decreases.  After 2002, while the normalized flux in the red
and blue wing is decreasing, the normalized flux in the red peak
increases.

{\it Fluctuation of $F$(RW)/$F$(BW).}-- The one cycle of fluctuation
in the  $F$(RW)/$F$(BW) ratio between 1994 and 2000,  is a result of
the sharp increase in the relative strength of the red wing in 1995
followed by an increase in the blue wing in 1998.  This fluctuation of
$F$(RW)/$F$(BW) does not repeat during the 16 yr of our spectroscopic
monitoring program, and thus if this trend is periodic, its timescale
is $>$ 20 yr.

\section{Comparison with Models \label{ch2-sec3-ssec2}}

\subsection{Arp 102B}
\subsubsection{Circular Disk \label{ch2-sec3-cd}}

The characteristic timescale of the $F$(R)/$F$(B) oscillation in Arp
102B ($\sim$ 2 yr), is on a much shorter timescale than the slow,
step-like variations in the FWQM (see Figure \ref{arp:fig:tp_arp_3}).
In order to decouple the oscillation of the peaks of the profile from
the slower, systematic changes in the wings, the step-like variation
in the FWQM of the profiles was successfully modeled by changing the
inner and outer radii of the best-fit circular disk model from
1990--1994 ($i=31\degr$, $\xi_{i}=300$, $\xi_{o}=825$, $q=3$,
$\sigma=1050$ km s$^{-1}$).  Figure \ref{arp:fig:modeldisks} shows the
three model disk profiles that fit the mean FWQMs of the profiles
plotted with solid lines in Figure \ref{arp:fig:tp_arp_3}b.  The
step-like variations of the inner and outer radius (from $\xi_{i}$ =
400 and $\xi_{o}$ = 700, to $\xi_{i}$ = 325 and $\xi_{o}$ = 825, to
$\xi_{i}$ = 300 and $\xi_{o}$ = 825) of the disk are either a result
of changes in the emissivity of the disk, perhaps due to changes in
the illumination, or changes in the structure of the accretion flow
itself.  The radial structure of the accretion disk evolves on the
viscous timescale, $\tau_{visc} \sim r^{2}/(\tau_{dyn} \alpha
c_{s}^{2})$, where $\tau_{dyn}$ is the dynamical time, $\alpha$
($\sim$ 0.1) is the Shakura -Sunyaev viscosity parameter (Shakura \&
Sunyaev 1973), and $c_{s}$ is the sound speed.  This timescale is
several orders of magnitude larger than the time over which the
apparent radii of the disk change ($\sim$ 10 yr); thus the structure
of the disk cannot be varying fast enough to cause the changes in the
radii of the line-emitting region of the disk.  If illumination is the
driver for the changes in radii, then it follows that since the
broad-line flux does not correlate well with the changes in radii, it
may be that the geometry, not the flux, of the continuum source is
changing the emissivity profile of the disk.

The profiles were fitted with one of the three model disk profiles
with the closest FWQM, by scaling the model to minimize the sum of the
squares of the errors from the model (least-squares fit).  Regions of
the profile contaminated by the narrow lines were not used in the fit.
The scaled model disk profile for each spectrum is shown in blue in
the spectra in the Appendix.  During times of the oscillation of
$F$(R)/$F$(B) the profiles were fitted with a model disk profile,
allowing for a Gaussian excess above the disk profile.   Our method is
similar to that of Newman et al. (1997) for modeling the profiles with
a disk plus a Gaussian, except that we keep our disk profile fixed at
any given time, and allow the Gaussian to vary, while Newman \etal
(1997) allowed both components to vary, by allowing the inner and
outer radii of the disk model to change.  Figure \ref{arp:fig:hotvel}
plots the peak velocity of the Gaussian excesses during the
oscillations of $F$(R)/$F$(B) as a function of time.  The velocity of
the excesses are well constrained when they have a velocity near the
peaks of the profile.  However, there are large uncertainties when the
excesses are located within the velocities of the narrow
H$\alpha$+[N~II] complex, since the shape of the broad line underneath
the subtracted narrow lines is not well constrained due to errors in
the narrow-line subtraction.  The FWHM of the Gaussians fits from 1991
to 1994 are $\simeq$ 4000 km s$^{-1}$, while the Gaussians are
narrower from 1999 to 2003 with a FWHM of $\simeq$ 2000 km s$^{-1}$,
and narrow even further to 1000 km s$^{-1}$ in 2004.  The most natural
explanation for an excess that periodically oscillates between the red
and blue sides of the profile is rotation.

\subsubsection{Orbiting Bright Spot?\label{ch2-sec2-ssec3-sssec1}}

Newman \etal (1997) fitted the first episode of the $F$(R)/$F$(B)
oscillation from 1991 to 1995 with a bright spot at a radius of 455
r$_{g}$,  with an angular width of 8$\degr$ (FWHM = 2000 km s$^{-1}$),
a radial width of 5 r$_{g}$ and an orbital period of 2.2 yr.  Newman
\etal (1997) determined the radius  of the bright spot from the peak
velocity of the Gaussian fit to the excess in the blue peak in the
1991 June 17--20 spectra.  If the bright spot is assumed to be at a
phase angle ($\phi$) of 270$\degr$ in the disk (when the bright spot
is at its maximum blue radial velocity), the gravitational radius of
the bright spot can be calculated from the relativistic Doppler
factor, $D = \nu/\nu_{e}$, which in the weak-field approximation is

\begin{equation}
D = (1-3/\xi)^{1/2}(1 + \xi^{-1/2}sin i ~sin \phi)^{-1},
\label{ch2:eq:doppler}
\end{equation}

\noindent where $i$ is the inclination determined from the disk fit,
and the observed velocity is then

\begin{equation}
v = c\Big{(}\frac{1}{D}- 1\Big{)}.
\label{ch2:eq:veldoppler}
\end{equation}
  
\noindent The radius, period, and phase, $\phi$ = $(2\pi/P)(t -
t_{0}$), of the bright spot for each time of oscillation of
$F$(R)/$F$(B) is determined from a least-squares fit of the observed
velocity curve of the Gaussian excesses to the velocity curve of a
bright spot in the disk (eq.~[3]) , with
weights set to 1/$\sigma^2$.  The resulting radii for the first and
second episode of the oscillation are 427 $\pm$ 5 and 757 $\pm$ 4
\rg~respectively, and periods of 2.247 $\pm$ 0.003 and 2.031 $\pm$
0.001 yr, with quoted errors equal to 1~$\sigma$.  The quoted errors
are the standard deviation of the best-fit parameters for the model
velocity curve and do not reflect systematic errors from the
assumption that the data behave like the model.  Combining the orbital
periods of the bright spots with their radius in units of \rg, a
dynamical measurement of the central black hole can be made.  Using
the complete form of Kepler's third law for a circular orbit,

\begin{equation}
P = 2\pi\Big{(}\frac{r^{3}}{GM_{bh}}\Big{)}^{1/2} = \frac{2\pi G M
\xi^{3/2}}{c^{3}},
\label{ch2:eq:kepler3}
\end{equation}

\noindent this yields central black hole masses of 2.60 $\pm$ 0.05
$\times 10^{8}$ \msun and 1.00 $\pm$ 0.01 $\times$ 10$^{8}$ \msun.
Our measurement of the black hole mass from the first oscillating
bright spot is consistent with Newman \etal's (1997) measurement;
however, it is discouraging that modeling the second episode of the
$F$(R)/$F$(B) oscillation with an orbiting bright spot yields a
discrepant black hole mass.   An independent estimate of Arp 102B's
central black hole mass was made by Lewis \& Eracleous (2006) from the
measured stellar velocity dispersion (Barth \etal 2002), and the
correlation between the velocity dispersion and the mass of the
central black hole (Tremaine \etal 2002).  Including the uncertainty
in the slope of the \mbh-$\sigma$ relation, they determined a central
black hole mass of ($1.1 \pm 0.2$)$ \times 10^{8}$ \msun.  It may be
that a single orbiting bright spot is not a  correct interpretation of
the Gaussian excesses, and the  excesses may be the consequence of a
more complex configuration of emissivity enhancements orbiting in the
disk.  The transient nature of the oscillations suggests a
time-dependent configuration.  One mechanism to produce transient
bright spots in the disk is spiral shocks induced by tidal
perturbations.  Chakrabarti \& Wiita (1993) conducted two-dimensional
hydrodynamical simulations to investigate the tidal effects on an AGN
accretion disk by passing massive perturbers ranging from
M$_{pert}$/\mbh = 10$^{-3}$ -- 10$^{-1}$.  Stellar-mass perturbers,
with M$_{pert}$/\mbh $\le$ 10$^{-5}$, were not included in the
simulation due to limitations in their grid resolution.  The
simulations revealed density waves in the disk that steepened into
spiral arms and knotty structures of increased density, temperature,
and emissivity, that formed, fragmented, and reformed.  The lifetimes
of these features was longer than the orbital period in the inner
regions of the disk, and during their lifetime, they propagated
outward in the disk.  This scenario may be a better physical model for
the ``bright spot'' in Arp 102B, which appears for two orbital periods
with a radius of 427 \rg and then reappears several orbital periods
later, further out in the disk at a radius of 757 \rg.

\subsubsection{Collection of Orbiting Clouds in a Disk?\label{ch2-sec2-ssec3-sssec2}}

An alternative model that predicts an oscillation of
$F$(R)/$F$(B) is not a single orbiting bright spot, but rather a disk-like distribution
of clouds in nearly Keplerian orbits, with randomly distributed clouds
having enhanced emissivity (Sergeev \etal 2000, hereafter S00).  Monte Carlo
simulations by S00 of clouds with a distribution of emissivities
produce times of sinusoidal variation of the red and blue peaks, as well as a
sinusoidal variation of the central portion of the profile with half
the period of the $F$(R)/$F$(B) variation (since an orbiting cloud will
pass through the center twice per orbit).   Contrary to this, S00's measurement of the
central-to-total flux ratio in their spectra of H$\alpha$ from 1992 to 1995 had a sinusoidal
variation with the same period within the errors to that of the
$F$(R)/$F$(B) oscillation.  In order to explain this discrepancy with the
predictions of the model, they postulated that the line emission originates 
in the face of the cloud illuminated by the central continuum source.  Thus, if the disk is edge-on, 
clouds traversing in front of 
the continuum source will have
their emitting face hidden from the line of sight, and their contribution to the line flux is seen
to traverse the central portion of the profile only once per orbit.

Figure \ref{arp:fig:s00} shows the red, blue, and central flux ratios defined by
S00, measured for  the spectra in our monitoring program.  Plotted
with solid lines are the best-fit sine curves for their data set from
1992 to 1995.  These curves are extended with dotted lines over the
baseline of our observing program.  It should be noted that these flux
ratios include regions  superimposed by narrow lines, and thus are
subject to systematic errors  introduced by the narrow-line
subtraction, and typical error bars are shown.  Our data fit their model for all the ratios from 1991
to 1994 [during the first oscillation of $F$(R)/$F$(B)], and follow
reasonably well for $F$(R)/$F$(tot), $F$(B)/$F$(tot), and $F$(R)/$F$(B) during the
later oscillation of $F$(R)/$F$(B) between 1999 and 2004.  Contrary to their model, the
$F$(C)/$F$(tot) ratio appears to stop oscillating after 2001, although the error
bars in the $F$(C)/$F$(tot) ratio are too large to make a convincing
argument either way.

S00 simulated the residuals from the mean normalized profile with a
collection of  1500 clouds with an internal velocity dispersion
resulting in a FWHM of 2000 km s$^{-1}$.  Their observed residuals from
the scaled mean profile 
show enhancements traversing from the red to the
blue side of the profile only, with no tracks from the blue to the red
side.  They again attribute this to the phase effects that explain the
equal periods of $F$(C)/$F$(tot) and $F$(R)/$F$(B).  Due to the uncertainty in
the narrow-line subtraction in the center of the profile, as well as
the unknown nature and variability of the central broad component of
the profile not accounted for by the disk-fit (see \S \ref{intro}), it is not clear whether a
reliable detection of tracks traversing through the central part of
the profile is possible.  Our measurement of $F$(C)/$F$(tot) and the residuals
from the disk profile do not show definitive evidence for phase effects,
and thus it seems that the edge-on
configuration does not seem necessary, in either the cloud model or
the bright spot model (in which the bright spot is also expected to traverse
the center twice per orbit).  In general the cloud model is
problematic, because the clouds must have large internal velocity
dispersions and a large covering factor, which would lead to
dissipation of the clouds or destruction by collisions on the
timescale of decades (S00), unless there was a mechanism
to produce new clouds on this timescale.  Thus, without the
necessity for phase effects, the orbiting cloud model seems less
compelling than emissivity enhancements orbiting in an accretion disk.

\subsubsection{Precessing Spiral Arm, Eccentric Disk, or Warp?\label{ch2-sec2-ssec3-sssec3}}

The bright spot model is not the only variation of the disk model that can produce an
oscillation in $F$(R)/$F$(B) and an excess above the disk profile that
shifts from one peak to the other periodically.  A precessing eccentric disk,
and a disk with a precessing spiral arm or warp, will have periodic changes of $F$(R)/$F$(B).   In the
eccentric disk model, the precession of the disk causes a sinusoidal
variation of $F$(R)/$F$(B) as well as a sinusoidal variation of the peak
velocities with the same period, with a minimum blue peak velocity
during the minimum of $F$(R)/$F$(B) and a minimum red peak velocity during
the maximum of $F$(R)/$F$(B) (for emissivity power-law index $q \le$ 3; see Lewis \etal 2004).  
The lack of a shift in
the velocity of the peaks on the same timescale as the $F$(R)/$F$(B) oscillation in our
data does not support this model.  Most importantly, the timescale of
the precession of the disk is determined by either the general
relativistic advance of pericenter or the tidal effects of a binary
companion, and is on the timescale of a few centuries for a 10$^{8}$
M$_{\sun}$ black hole (Eracleous \etal 1995).  The 2 yr period
of the $F$(R)/$F$(B) oscillation is much too short to be associated with a
precessing disk or a precessing warp (see Storchi-Bergmann \etal 1997, eq.~[4]), and its 
transience could not be associated with a persistent phenomenon.

The timescale of the $F$(R)/$F$(B) oscillation due to a precessing spiral
arm is determined by the pattern precession period, which is typically
an order of magnitude longer than the dynamical time, and less than
the sound-crossing time (see discussion in Storchi Bergmann \etal 2003, and references
therein).  For a
10$^{8}$ M$_{\sun}$ black hole and an inner radius of 300 r$_{g}$,
that translates to a period between 1 and 20 yr (using eqs.~[5] and [6] in
Storchi-Bergmann \etal 2003).  The 2 yr period of the
$F$(R)/$F$(B) oscillation is within the range of plausible
periods for a spiral arm; however, another consequence of a spiral arm
are changes in the velocity and width of the peaks of the profile as the arm
precesses.  For a one-armed spiral that spans the entire radial extent of the line-emitting
disk, during the minima of $F$(R)/$F$(B) the
blue peak narrows and is at a maximum velocity, while during the
maxima of $F$(R)/$F$(B) the red peak narrows and is at a maximum velocity
(see Gilbert \etal 1999, Storchi-Bergmann \etal 2003, and Lewis \etal 2004 for examples).  In our data,
the width of the blue peak does not appear to narrow during minima of
the $F$(R)/$F$(B) oscillation, nor does the velocity of the blue peak
increase, except during the minimum of $F$(R)/$F$(B) in 1991.4.  Thus,
with the only spiral-like behavior being the periodic variation of
$F$(R)/$F$(B), there are no compelling reasons to favor a spiral arm over a bright spot or other configuration of emissivity enhancements orbiting in the disk.

\subsubsection{Reverberation of Flares in the Continuum?\label{ch2-sec2-ssec3-sssec4}}

The change in shape of the line profile during the small amplitude,
rapid flare in  1989 July (Figure 2.13), is on a short-enough
timescale to be the result of light travel time delays of the
reverberation of the variable ionizing continuum.  Using the black
hole mass estimated from two independent methods: the bright spot model
and the stellar velocity dispersion (Lewis \& Eracleous 2006) which both yield a mass 
of $\sim$ 1 x 10$^{8}$
M$_{\sun}$, and the outer radius of the best-fit disk profile in 1989 of 825
r$_{g}$, the estimated light-travel time across the broad-line region
is 5 days.  A flare in the continuum with a duration on the order of
the light-crossing time and much greater than the recombination time 
[$\tau_{flare} \sim \tau_{LT} >> \tau_{rec} = (n_{e}\alpha_{B})^{-1}$, 
where $n_{e}$ is the electron density and $\alpha_{B}$ is the case B recombination coefficient], 
can cause a detectable light echo to
propogate across the profile (Peterson 1988).  A light echo from such a
flare is expected to produce spikes in the profile
that drift from higher to lower velocities as it propagates outward in the disk (Stella 1990).  During the
flare in 1989, the line profile has a dramatically weaker red wing 
(Figure \ref{arp:fig:flare1}), 
and a narrower and stronger blue peak.  The flare in 1998 occurs over a much longer
timescale than the light-crossing time, and appears to have the same changes in the
profile shape, namely a narrower red wing and stronger blue peak.  Thus light-travel time
effects are unlikely to be causing the dramatic changes in the profile shape during both flares.

During a flare, the increase in central continuum
emission can temporarily illuminate the regions further out in the disk,
increasing the contribution of emission from lower velocities.  Increasing the outer
radius of the circular disk model mainly broadens the peaks of the profile and shifts
them to lower velocities but does not significantly affect the wings
of the profile.  Alternatively, the rapid flare could temporarily
fully ionize the inner regions of the disk, causing an increase in the effective
inner radius of the line-emitting portion of the disk or a flatter emissivity law at small
radii.  The result is a
decrease in the strength of the wings, a shift of the peaks to
lower velocities, and an increase in $F$(R)/$F$(B) (shown in Figure \ref{arp:fig:modeldisks}b).
The decrease of $F$(R)/$F$(B) during the flares is not compatible with this scenario.
In general, the decrease in $F$(R)/$F$(B) and in the FWQM cannot be 
modeled by increasing the central continuum illumination of a circular disk.

\subsubsection{Gravitational Lensing?\label{ch2-sec2-ssec3-sssec5}}

Gravitational lensing can amplify a broad emission-line if the
Einstein radius of the microlens is of a comparable size to the
accretion disk producing the line emission.  An interesting
consequence of this, is that the profile can be amplified
asymmetrically if the lens is off-center.  Popovic \etal (2001)
calculated two scenarios in which a microlens could produce sinusoidal
variations in $F$(R)/$F$(B) similar to those observed in Arp 102B: (1) a
supermassive binary companion at a distance of 0.3--1pc, which takes a
few years to transit in front of the disk, but has an orbital period
of $\sim$ 1000 yr; (2) a 1 \msun~star in an intervening galaxy
(1/20th the distance of Arp 102B) with a relative velocity of 200
\kms.  The important difference between these scenarios is the
frequency with which the lensing events would occur.  In the
supermassive binary case, the lens should only transit in front of the
disk once every 1000 yr, and thus only one period of the $F$(R)/$F$(B)
sinusoidal variation should be observed during the entirety of our
monitoring program.  For the stellar-mass microlens case, the lensing
transit events would randomly cause one period of an oscillation of $F$(R)/$F$(B)
during the transit, with a frequency depending on the
density of microlenses in the intervening galaxy.  The multiple
consecutive periods of the $F$(R)/$F$(B) oscillation observed after 1999 rule out a
transiting scenario for the source of the oscillation.  However,
microlensing could be a mechanism to amplify the line profile
asymmetrically and cause a flare of broad-line flux.  This could be a possible
explanation of the stronger blue peak of the profiles during the
flares, but the decrease in flux of the red wing during the flares is not
consistent with this picture, nor is the coincidence of both flares
having a stronger blue peak, which in the gravitational lens model,
should be a random consequence of the alignment of the microlens with
respect to the disk.

\subsubsection{Drift of Peaks due to Changing Outer Radius?\label{ch2-sec2-ssec3-sssec6}}

During the systematic drift of the blue
peak velocity from $-$5300 to $-$4800 km s$^{-1}$ and back to $-$5300 km
s$^{-1}$ again from 1995 to 1996, it is difficult to determine whether the velocity of
the red peak is also shifting, because of the [S II] narrow lines superimposed on the red peak.   In the circular disk model, the peak velocities
will shift when the outer radius of the disk increases.  If both of the peaks were
drifting to lower velocities, then this could be modeled as an increase in the outer radius of
the line emitting portion of the disk, which causes the peak separation
to decrease.  Figure \ref{arp:fig:veldrift2} shows the spectra during the drift of the blue peak from 1995 to 1996 in comparison to the change in the profile shape caused by an increase
in outer radius from 825 to 1300 \rg.  Although the shift in the blue peak velocity in the model
matches the data, the shift of the red peak and broadening of both peaks in the model
do not.  This drift in the blue peak velocity occurs during a time of low $F$(R)/$F$(B).  In general,
there appears to be a trend towards a lower blue peak velocity when the blue peak is stronger, as
is the case during flares in the broad-line flux.

\subsection{3C 390.3 \label{ch2-sec4-ssec2}}

\subsubsection{Binary Broad-Line Region? \label{ch2-sec4-ssec2-sssec1}}

Eracleous \etal (1997) created yearly averaged velocity curves of the
red and blue peaks of 3C 390.3 through 1996 and found that the red peak
velocities were not consistent with the motion of a spectroscopic
binary, and the best-fit model yielded a lower limit on the binary
mass of $>$ 2.6 $\times 10^{11}$ \msun, larger than any known central black
hole mass, and over 2 orders of magnitude larger than the black hole mass inferred from the stellar
velocity dispersion, \mbh = 5 $\pm~1 \times 10^{8}$ \msun~(Lewis \& Eracleous 2006).    Eracleous \etal (1997) noted that the velocity of the
red and blue peaks are highly sensitive to changes in the shape of the
peaks, which is evident in our spectra of the transient sharp features
in the blue peak in 1989 and
1998 and in the flattening of the shape of the red peak from 1997 onwards, and occur on much shorter
timescales than that predicted by an orbiting binary model.
The reverberation mapping results for 3C 390.3 (see \S \ref{intro}) also contradict the binary
broad-line region scenario for the source of the displaced red and
blue peaks of the profile, since no lag is detected in the response of
the red and blue peaks, which would be expected if they were produced
by a binary separated by 1.5 pc, as is determined from the best-fit
binary model from Eracleous \etal (1997).  It is also noteworthy that the widths of the two peaks
of the H$\alpha$ line are narrower than their separation, which is opposite from what one expects for
a binary broad-line region governed by Kepler's laws (M.~V. Penston 1998, private communication;
see footnote 3 in Chen \etal 1989).

\subsubsection{Bipolar Outflow or Orbiting Bright Spot? \label{ch2-sec4-ssec2-sssec2}}

In general, it is difficult to fit the low $F$(R)/$F$(B), narrow FWQM, and
strong central component of the broad H$\alpha$ profile of 3C 390.3 with a simple circular
disk.  The best-fitting disk model to the broad H$\alpha$ line by Eracleous \& Halpern (1994) 
in a spectrum obtained on 1988 July 9 successfully fits the wings of the profile but underestimates the strength of the blue peak.  Even more difficult, is reconciling the circular disk model
with the 1976 and 1980 H$\alpha$ profiles reported in Zheng \etal (1991), 
in which $F$(R)/$F$(B) changed from 0.8 to 1.5, indicating a
stronger red peak than the blue peak in 1980.  Zheng \etal (1991)
modeled the two epochs of the H$\alpha$ profile with an edge-on
(non-relativistic) circular disk plus a bright spot with an orbital
period of 10.4 yr, and also with a double-stream outflow. In the
bipolar outflow model, the changes in $F$(R)/$F$(B) are due to variations
of the relative intensity of the ionizing continuum illuminating the
two cones, or due to changes in orientation of the cones, perhaps resulting
from a precession of the rotation axis caused by its misalignment with the axis
of the accretion disk.  However, the theoretical timescale for the alignment of the disk (and hence the outflow axis) is much too long, 10$^{5}-10^{6}$ yr, to account for the timescales observed (Natarajan \& Pringle 1998; Natarajan \& Armitage 1999).   Regular periodicity of
$F$(R)/$F$(B) on the timescale of 10 yr is not evident in our
monitoring data, ruling out the possibility of regularly
precessing jets.   The bipolar jet model is disfavored
by reverberation mapping campaigns of 3C 390.3 starting in 1992, which show no lag between the
response of the red side of the profile compared to the blue side in
the H$\alpha$ and H$\beta$ lines, which would be observed in the case
of radial motion (Dietrich \etal 1998; Shapovalova \etal 2001; Sergeev
\etal 2002).  Thus, a bipolar outflow, if present from 1976 to 1980, must have disappeared
by 1992 and would not be
able to explain the fluctuation of $F$(R)/$F$(B) from 1994 to 1998.  The mass
of the central black hole can be estimated by combining the light crossing time of the
broad-line region of 20 days (Dietrich \etal 1998) with the best-fit outer
radius of the accretion disk model fit of 1300
\rg~from Eracleous \& Halpern (1994), yielding a black hole
mass of 3 x 10$^{8}$ \msun, which is in relatively good agreement with the black hole
mass derived from the stellar velocity dispersion.
For a central black hole mass of 3 x 10$^{8}$ \msun,
and an
inner radius of 380 \rg, this translates to orbital periods ranging
from 2 to 14 yr.  The two episodes of an oscillation of $F$(R)/$F$(B) in
1976 to 1980, and 1994 to 1998, have periods that are both within the range
of orbital periods in the disk and could be attributed to transient
emissivity enhancements orbiting in the disk.

\subsubsection{Precessing Eccentric Disk, Spiral Arm, or Warp? \label{ch2-sec4-ssec2-sssec3_01}}

The transient and irregular nature of the $F$(R)/$F$(B) oscillations
disfavors a precessing eccentric disk, for which the period of the
oscillation does not change, and the oscillation persists for longer
than one cycle.  Similary to Arp 102B, the estimated mass of the
central black hole also implies a disk precession timescale of at
least a few centuries, which is not compatible with the periods of the
$F$(R)/$F$(B) transient oscillations.  Gilbert \etal (1999) found the
best-fit pattern period for a spiral arm to produce the oscillation in
$F$(R)/$F$(B) between 1974 and 1998 to be greater than 25 yr.  Although
the spiral-arm model reproduces the sparsely sampled trend of
$F$(R)/$F$(B) to oscillate from 0.8 in 1976 to 1.5 in 1980 and then back
to $\sim$ 0.7 after 1990, it does not match the profile shapes in
detail.  A spiral arm precessing with this period would predict a
subsequent rise to a maximum in $F$(R)/$F$(B) of 1.5 in 2005, which is not
consistent with the relatively constant $F$(R)/$F$(B) ratio of $\sim$ 0.6 after
2000.   Thus, any regular, periodic phenomena such as precessing
spiral arms, eccentric disks, and warps are not able to reproduce the
irregular changes in $F$(R)/$F$(B), as well as the extreme changes in the
shape of the wings of the profile and the central red component.
For any of the above scenarios to be viable, the perturbation must have a 
life time comparable to the pattern precession.

\subsubsection{Binary Companion with Shared Accretion Disk? \label{ch2-sec4-ssec2-sssec3_02}}

The smooth changes in the red and blue wings and the central component
have a large amplitude, and occur on a timescale of $\sim$ 5 yr.
The normalized flux in the blue wing increases monotonically by 50\%
from 1995 to 2001, while the normalized flux in the red wing decreases
by 65\% from 1998 to 2002, and the flux in the normalized central red
component decreases by 25\% from 1993 to 1998.  Since a complete cycle
of variation is not observed in the normalized flux of any of these
regions, it is not yet possible to determine if these changes are in
fact periodic or even if they repeat.  If they do repeat, the recurrence time would be
$\gtrsim$ 10 yr.  A two-armed spiral does produce a peak in the
central part of the profile, which could then drift toward the blue
peak (Chakrabarti \& Wiita 1993, 1994).  However, such a spiral
pattern would also produce dramatic changes in the velocities and
intensities of the red and blue peaks as the central peak drifts,
which are not observed.  The physical mechanism producing the observed
dramatic variability must change the shape of the high-velocity wings
of the profile, and the central red component, without similarly
affecting the lower velocity gas that produces the peaks of the
profile.  One possible scenario in which changes would occur only in
the high-velocity wings of the profile, is an orbiting binary
companion with a common circumbinary accretion disk with the central black hole.
The orbital motion of the companion perturbs the disk, and creates a non-axisymmetric instability 
that causes enhanced accretion in the inner regions of the disk close to the
companion.  In addition, the binary will excite spiral waves in the inner parts
of the circumbinary disk (Artymowicz \& Lubow 1996; Armitage \& Natarajan 2002).  Both
phenomena will affect the wings of the profile.
For a binary separation less than the inner radius of the
line-emitting portion of the disk, D $<$ 380 \rg, and a total binary
mass of 3 x 10$^{8}$ \msun, this would imply a binary separation of
$<$ 0.005 pc, and an orbital period of $<$ 2.2 yr.  The orbital period
of the secondary is too short to produce the $\gtrsim$ 10 yr
characteristic periods of the variability observed in the wings of the
profile.  Regardless of this,  the stability of such a binary
configuration is unlikely, since  at such a small separation, the
expected time for coalescence of the binary due to gravitational
radiation is short, $<$ 5 $\times$ 10$^{4}$ yr (Lightman \etal 1979).

\subsubsection{Orbiting Features in the Disk? \label{ch2-sec4-ssec2-sssec4}}

The lifetimes of the relatively stationary narrow features in the blue
peak constrain whether they can be orbiting within the disk.  The
sharp spike that appears in 1989.50, and persists for $>$ 2 yr at
the same velocity, has a lifetime of at least 15\% of an orbital
period in the disk.  If in orbital motion in the disk, the velocity of
the feature would have been expected to drift by at least 2000 km
s$^{-1}$ during its lifetime.  The feature that appears in 1998.27
drifts by 400 km s$^{-1}$ over 1.5 months and has a lifetime that is
short enough to be only 1\% of an orbital period in the disk, and thus
its drift of 400 km s$^{-1}$ could be consistent with orbital motion.
Another possibility is that the sharp features in the line profiles are shocks
that propagate through the disk slowly (at the speed of sound).  Thus they
can cool and dissipate quickly before their apparent radial velocity changes significantly.

\subsection{3C 332 \label{ch2-sec5-ssec2}}

\subsubsection{Changes in the Emissivity of a Circular Disk? \label{ch2-sec5-ssec2-sssec1}}

The decrease in FWHM and peak velocity separation in the profiles
of 3C 332 during the increase in broad-line flux is easily accounted for
by the disk model.  When a disk is illuminated by a brighter central
continuum source, regions farther out in the disk are photoionized,
increasing the outer radius of the line-emitting portion of the disk.
The increase in continuum flux can also fully ionize the inner regions
of the disk, effectively causing an increase in the inner radius of
the disk.  The increase in inner and outer radius both cause lower
velocity material to contribute more to the emission line, thus
weakening the wings, and shifting the velocities of the peaks of the
profile to lower velocities.  A disk interpretation was also offered for
the increase in peak velocity separation with decreasing broad-line
flux in the double-peaked H$\alpha$ profile of the LINER galaxy NGC
1097 (Storchi-Bergmann \etal 2003).  By decreasing the inner radius of
their disk model from 1300 to 450 \rg, they could reproduce the drift
in peaks of the H$\alpha$ profile with the dimming broad-line flux.
The decrease in inner radius was interpreted as a change in emissivity
of the line-emitting gas, such that when the central continuum flux
decreased, the region of maximum line emission shifted towards the
inner regions of the disk.

Figure \ref{332:fig:disks} shows examples of the broad H$\alpha$
profiles with the model disk fit for a low-state ($\xi_{i}$ = 190,
$\xi_{o}$ = 540) and high-state ($\xi_{i}$ = 260, $\xi_{o}$ = 620).
Both disks have $i$ = 36$\degr$, $\sigma$ = 1600 \kms, and $q$ = 3.
The spectra in the Appendix show the
least-squares scaled circular disk models in comparison with the
profiles.  Spectra before 1998 were fitted with the low-state model disk,
and after 1998 with the high-state model disk.  The disk profiles
match well the decrease in FWHM and peak velocity separation, but the
high-state profile appears to underestimate the flux in the very far
blue wing and overestimate the flux in the red wing from 1998 to 2001.
Beginning in 2004 there also appears to be an excess in the blue side
of the profile, near $-$4000 \kms, that is not accounted for by the disk
model profile.

\subsubsection{Precessing Spiral Arm? \label{ch2-sec5-ssec2-sssec2}}
  
The rise and fall of $F$(R)/$F$(B) from 1990 to 1998 was modeled by Gilbert
\etal (1999) as a one-armed spiral emissivity enhancement with a
pattern period of 14 yr.  Although this model reproduces the
behavior of $F$(R)/$F$(B) from 1990 to 1998, it does not match the historical
$F$(R)/$F$(B) measurements in 1976 and 1978, and predicts a minimum in
$F$(R)/$F$(B) in 2002, which is clearly not consistent with our continued
monitoring of $F$(R)/$F$(B), which shows it to be nearly constant [$F$(R)/$F$(B) $\sim$
0.9] for over 8 yr after 1997 (Figure \ref{332:fig:tp_332_3}).
Thus the oscillation in $F$(R)/$F$(B) does not repeat, nor does it appear
to be sinusoidal in shape.

The maximum in
$F$(R)/$F$(B) in 1995 can be attributed to a transient excess in the red
wing at +12,250 \kms with a FWHM $\simeq$ 3500 \kms, and a maximum
strength of 40\% of the red peak.  The excess in the red wing appears to
drift to +10,800 \kms in 20 months.    This excess may be related to
the excess in the blue wing in 1999.46 with a width of 5000 \kms and a
peak velocity of -11,900 \kms; this could indicate a feature rotating
with the disk that only completes one orbit before fading away in
2002.  The maximum red velocity of the excess of +12,250 \kms
at an assumed phase angle of 90$\degr$ and an inclination of the disk
of 36$\degr$, corresponds to a bright spot at a radius of 280 \rg.
The excess in
the blue wing appears to persist for longer than expected for an
orbiting feature, and the excess near -4000 \kms does not occur when
the bright spot would be expected to be traversing the central part of
the profile.

\subsection{PKS 0235+023 \label{ch2-sec6-ssec2}}

\subsubsection{Circular Disk? \label{ch2-sec6-ssec2-sssec1}}

The broad H$\alpha$ profile of PKS 0235+023 in 1991 is well fitted with a circular accretion disk model
with $i = 70\degr, \xi_{i} = 1020, \xi_{o} = 7000, \sigma = 800$ \kms, and $q
= 1.7$ (Eracleous \& Halpern 1994).  The scaled disk model is shown in
comparison with the profiles in the spectra in the Appendix.  For all
the spectra after 1991, the blue peak is shifted to a higher
velocity, and the strength of the peak has increased in comparison to
the disk model.  In contrast, the red peak appears to have decreased
in strength (and in some cases completely disappeared), and the red wing does
not extend to very high velocities.  Even if the centroid of the entire disk profile is shifted
blueward, the circular disk model is not a good fit to the profiles
after 1991.

\subsection{Mkn 668  \label{ch2-sec7-ssec2}}

\subsubsection{Binary Broad-Line Region? \label{ch2-sec7-ssec2-sssec1_01}}

The systematic drift of the central red peak of Mkn 668 could be interpreted as
the result of the orbital motion of binary supermassive black holes
with separate broad-line regions, in which the central peak is
emission from the broad-line region of one of the orbiting black
holes.  Although the central red peak repeatedly drifts by $-$900 \kms
from 1982 to 1985, and then by +900 \kms from 1985 to 1989, and then
again by $\gtrsim$ $-$1000 \kms from 1989 to 1995, the velocity of the
peak remains $<$ +1700 \kms from 1995 to 2004.  The velocity curve does
not have a regular periodic shape, and the peak velocity never passes
through zero velocity, and thus in the case of a binary, the velocity
of its center of mass would not be coincident with the velocity of
the host galaxy.  Thus the shift of the red peak is not consistent
with orbital motion of a binary broad-line region.

\subsubsection{Radiatively Accelerated Outflow? \label{ch2-sec7-ssec2-sssec1_02}}

Marziani \etal (1993) reported a positive correlation between the peak
velocity and the total broad-line flux for their spectra from 1985 to 1991.  
They then modeled the profiles as emission from a
radiatively accelerated biconical outflow of clouds, in which the
positive correlation between the peak velocity and the continuum flux
is a consequence of radiative acceleration, and the asymmetry of the
profile is attributed to self-absorption in the Balmer line or dust
absorption from the backsides of the clouds in the outflow.  Although from 1989 to 1996
the peak velocity and total broad-line flux decrease monotonically, we find that after 2000
this correlation breaks down.
During a high state of broad-line flux in
2001--2002, during which the flux increases by a factor of 6, the
central peak remains at a minimum peak velocity of $<$ +1700 \kms.  
Thus the large-amplitude changes of the
velocity of the peak are irregular, and are unrelated to the changes
in the broad-line flux.   The radiatively accelerated outflow model is
thus no longer observationally motivated.

\subsubsection{Elliptical Accretion Disk with an Added Perturbation? \label{ch2-sec7-ssec2-sssec2}}

Figure \ref{668:fig:spikes} shows that the shift in the red central
peak can be modeled as a Gaussian excess at a peak velocity of +3300
\kms and a FWHM $\simeq$ 2000 \kms in 1990--1995 that drifts to a peak
velocity of $-$1800 \kms in 1996--2000, and then back to a peak
velocity of +4400 \kms in 2003.   We were able to fit the non-varying
portion of the profile with an eccentric disk, with $i = 35\degr,
\xi_{i} = 200, \xi_{o} = 3000, q = 1.7, \sigma = 1500$ \kms, $e =
0.6$, and $\phi = 60\degr$ (shown in blue in Figure
\ref{668:fig:spikes}).   The maximum radial velocity of the excess and
the inclination of the disk fit can be used to determine the radius of
the excess in gravitational radii.  The maximum redshifted radial
velocity for a disk with an inclination of $35 \degr$ and an
eccentricity of 0.6 occurs at a phase angle of the excess of $\phi =
25\degr$ (determined from the Doppler factor of an eccentric disk from
equation 15 in Eracleous \etal 1995).  Thus, for a maximum red radial
velocity of +3300 \kms, this corresponds to a radius of 2125 \rg.  The
maximum blueshifted radial velocity for an excess at this radius would
then occur at $\phi = 255\degr$ and is -4300 \kms.  The observed time
for the peak to shift from the maximum red velocity to a maximum blue
velocity corresponds to an orbital period $\approx$ 12 yr.  This model
yields a central black hole mass from equation \ref{ch2:eq:kepler3} of
1.2 x 10$^{8}$ \msun.  The velocity curve of a feature orbiting in the
eccentric disk at a radius of 2125 \rg~is shown in comparison with the
peak velocities of the Gaussian fits to the excess above the model
disk profile in Figure \ref{668:fig:velspot}.  The period of the
velocity curve has been fixed to 12 yr.  It is clear that the observed
blue velocity of the feature does not match the amplitude of the model
velocity curve,  and when the excess returns to the red side of the
disk, the radial velocity is too large.  In general, the velocity of
the excesses appears step-like, and not smooth, as would be expected
for projected orbital motion.  The historical velocity curve suggests
that there was an excess in the red side of the central peak from 1974
to 1982.  This would be consistent with an excess that oscillates,
non-sinusoidally, from one side of the peak to another with a period
of $\sim$ 12--16 yr.

\subsection{3C 227 \label{ch2-sec8-ssec1}}

\subsubsection{Orbiting Bright Spot?}

The complex profile of 3C 227 cannot be modeled with a single disk
component (circular or elliptical), since the profile has a
low-velocity blue peak as well as a high-velocity blue shelf.  The
changes in the shape of the profile from 1990--1996 to 1999--2004 can
be modeled as a sum of two Gaussians, one with a peak velocity of
$-$2400 \kms and a FWHM $\simeq$ 1500 \kms plus a Gaussian at a
velocity of $-$5000 \kms and a FWHM $\simeq$ 2500 \kms, The two
Gaussians drift to the red side of the profile to peak velocities of
+1900 and +4800 \kms, respectively, from 1990 to 2000 (Figure
\ref{227:fig:spikes}).  Since the feature is present in the blue peak
starting in 1974, this would imply a lifetime of the feature in the
blue peak of $>$ 14 yr.  If the feature were orbiting from the blue
side of the profile to the red side, then the time it takes for it to
complete half an orbit, $\approx$ 5 yr,  is inconsistent with its
persistence in the blue peak from 1974 to 1990.  Thus, although the
feature appears to drift across the profile, this drift is not
consistent with rotation.

\subsection{3C 382: Elliptical Accretion Disk? \label{ch2-sec9-ssec2}}

The broad H$\alpha$ profile of 3C 382 was successfully modeled with an
elliptical disk by Eracleous \etal (1995), with $i = 45\degr, \xi_{i}
= 470, \xi_{o} = 3000, q=1.7, \sigma = 1200$ \kms, $e = 0.33$, and
$\phi_{o} = 110 \degr$.  This disk profile scaled to the total broad
line flux is shown in comparison with the observed profiles in the
spectra in the Appendix.  Although the disk model fits very well for
most of the observations, the drop in the red peak of the profile in
1994--1995, with no corresponding change in the blue peak, is not
behavior easily produced with a disk.  An orbiting emissivity
enhancement in the disk that increases the strength of the red peak
would be expected to traverse the profile and subsequently increase
the strength of the blue peak.  It is possible that the perturbation
moves slowly in the azimuthal direction, and dissipates before it can
affect the other peak.  This may also be the case for the transient
bumps that only appear in the far blue wing.  However, the large
amplitude increase in the red wing that subsequently occurs in the
blue wing may be the consequence of an orbiting feature in the inner
region of the disk, such as a bright spot or spiral arm.  However, the
excesses in the wings do not appear to traverse the lower velocity
portions of the profile, which is not consistent with rotation.  The
relative constancy of the central part of the profile well fitted by
the elliptical disk profile can be used to place a lower limit to the
black hole mass, since the lack of precession of the elliptical disk
profile over 15 yr of the monitoring program implies \mbh $>1 \times
10^{6} $\msun.

\section{Summary and Discussion \label{ch2-sec10}}

The $F$(R)/$F$(B) oscillation in Arp 102B that occurred from 1991 to
1995, recurs in 1999 to 2004 with the same period.  The period of the
oscillations is on the dynamical timescale, and is much shorter than
the precession period of a spiral arm, eccentric disk, or
warp. Modeling the recurring oscillation as a single transient bright
spot yields discrepant central black hole mass estimates.  Thus, the
excess in the profile that drifts from one side to the other may be
the result of a more complex, and non-stable configuration of
emissivity enhancements such as transient shocks induced by tidal
perturbations.  We do not find strong evidence for phase effects in
the residuals of the line profiles during the oscillation of the
emissivity enhancements from one side of the profile to the other, and
therefore we disfavor the configuration of a collection of clouds
orbiting in a disk proposed by S00.   The relatively slow changes in
the wings of the profile can be reproduced by changing the inner and
outer radii of the line-emitting portion of the disk.  In the
self-consistent photoionized accretion disk model of Chen \& Halpern
(1989), the inner radius of the line-emitting portion of the accretion
disk is determined by the boundary between the thin disk and the
elevated structure illuminating the outer portions of the disk in the
form of an ion torus or a radiatively inefficient accretion flow
(RIAF; see Narayan \etal 1998).  The outer radius is determined by the
radius at which the disk is no longer stable to self-gravity, and the
disk fragments.  It is interesting that both of these radii are
sensitive to the accretion rate. The timescale for changes in the
accretion rate is on the viscous timescale.  This timescale is orders
of magnitude longer than the timescale of the observed variations in
the radii of the disk.  Thus changes in the accretion rate cannot be
physically driving the changes in radii.
  
The successful fit of an eccentric disk to the non-variable portion of
the profile of Mkn 668 makes it tempting to attribute its oscillation
of $F$(R)/$F$(B) to an orbiting emissivity enhancement.  The velocity
curve of the residuals from the disk profile model are not consistent
with the velocity curve predicted for a feature orbiting in the
outskirts of the disk.  The historical velocity curve indicates that
the oscillation of $F$(R)/$F$(B) and the central peak velocity is
erratic, and not a regular phenomenon.  We find that the correlation
between total broad-line flux and peak velocity reported by Marziani
\etal (1993) is not compatible with the high state profiles observed
in later spectra, during which the peak velocity remained at a
minimum.  Thus, the radiatively accelerated outflow model for the
source of the double-peaked emission line is no longer observationally
motivated.   An orbiting binary is also ruled out because of the
irregularity of the peak velocity curve, and the fact that the radial
velocity of the central peak never crosses zero velocity, requiring
the velocity of the center of mass of the binary to not be coincident
with that of the host nucleus.  The sharp decrease of the velocity of
the blue peak of 3C 227 appears to be caused by an excess feature in
the blue peak that traverses from one side of the profile to the
other. The inferred orbital period of the feature of $\approx$ 10 yr,
however, is inconsistent with the persistence of the feature in the
blue peak for over 15 yr before drifting to the red side of the
profile.

The decrease in FWHM and peak velocity separation of the profile shape
of 3C 332 during its high-state of broad-line flux is well accounted
for by the circular disk model, and is similar to the behavior
reported in NGC 1097, for which the peak velocity separation increased
during a low-state in broad-line flux.  Arp 102B, on the other hand,
demonstrates a peculiar, asymmetric response to flares in the
broad-line flux, that are not compatible with a simple response of the
disk to a changing illumination.  The repeated tendency of the blue
H$\alpha$ peak to be stronger and narrower during high-states of flux
also makes a random asymmetric gravitational lensing event an unlikely
explanation.

Sharp features that appear in the blue peak of the broad H$\alpha$
profiles of Arp 102B and 3C 390.3 are narrow ($\lesssim$ 1000 \kms ),
and persist for months to years.  The relatively stationary velocities
of the features are inconsistent with orbital motion within the disk,
and require a well-organized velocity field.  Stellar collisions with
the accretion disk are expected to be frequent (several per day) and
should produce a shock in the disk that can emit X-ray emission
(Nayakshin \etal 2004), as well as remove a narrow plume of material.
The expelled gas could produce the transient narrow features observed
in the profiles, although the rate of such events is not consistent
with the persistence of features over months and years.  The
appearance of the features preferentially in the blue peak of the
profiles suggest that their blueshift may be a signature of an
outflow.  This would be expected from stellar collisions, since we
would preferentially see the plumes raised over the surface of the
disk.

3C 390.3 displays a large amplitude of profile variability, with five
coherent regions of the profile that appear to vary on the timescale
of decades, and with no clear periodic correlation between each other.
Even more difficult to model is the sharpening of the blue peak, and
the dramatic flattening of the red peak and decrease in the red wing
to produce a sharp shelf after 1995.  Although a circular disk model
provides a reasonable fit to the line profile in 1988, the subsequent
changes in the profile shape make an acceptable fit to a circular disk
model rather difficult.  The excellent circular disk fit to PKS
0235+023 in 1991 also breaks down for subsequent spectra, in which the
shape of the profile changes drastically.  Similarly, although an
elliptical disk model provides an excellent fit for many of the
profiles of 3C 382 during the monitoring program, the variability of
the profile reveals inconsistencies with this simple disk model.  The
extreme variability seen in the red and blue wings is never detected
in the lower velocity peaks of the profile, and although the red peak
does show variability, the relative flux of the blue peak remains
unchanged.

\section{Conclusions \label{sec-conclusion}}

Of all the models discussed in the literature, the photoionized
accretion disk is the most robust model for the origin of the
double-peaked Balmer emission lines.  The disk model provides an
excellent fit to the line profiles with relatively few free
parameters, and successfully predicts the observed properties of the
UV emission lines.  The accretion disk model is also compatible with
the spatial scales of the broad line region measured from
reverberation mapping experiments.

Variability adds an extra dimension to discerning the structure and
kinematics of the broad-line gas, and probes the structure of the
accretion disk, the most likely source of the line emission.  Several
models for physical processes in the accretion disk have been proposed
to explain the ubiquitous profile variability of double-peaked
emitters, including precessing eccentric disks and spiral arms, and
orbiting bright spots.  Individual spectra of the H$\alpha$ profiles
of six of the seven objects in this study are successfully fitted by
line emission from a circular or elliptical accretion disk, however
the variability of their line profiles reveals behavior that is not
always consistent with the model predictions.  There is no persistent,
regular variability observed in the objects that would require a
precessing elliptical disk, one-armed spiral, or warp.  Although the
timescale of the oscillation in the relative flux of the peaks of the
profiles in many objects appears to occur on the orbital timescales
estimated from the disk model fits to the profiles, the velocity
curves of the residuals from the model profiles are often not
consistent with the projected orbital motion of a bright spot.   In
many of the objects, longer-timescale systematic variability is seen
in the wings of the profile that appears to be uncorrelated with the
variability of the lower velocity regions of the profile may reflect
changes in the emissivity of the inner accretion flow.   The negative
correlation between peak velocity separation and total broad-line flux
observed in 3C 332 is well accounted for by the disk model, although
the asymmetric response of the profile during high-states of flux in
Arp 102B, and possibly 3C 390.3, is problematic.  We find that,
although the accretion disk remains the most favorable model for the
origin of the double-peaked emission lines, most of the models for
physical processes in the accretion disk, previously used in the
literature to explain the variability of the profiles, break down when
the baseline of spectroscopic monitoring is extended.

Our results suggest that future efforts to model the variability of
double-peaked emission lines should explore more sophisticated models.
More specifically, models in which the perturbations have short life
times, as well as finite onset and decay times (on the order of a few
dynamical times, $\tau_{dyn} = 6\; M_8\; \xi_3^{3/2}$~months) are
highly desirable. Another family of models that warrant careful
consideration are ``stochastic models.'' In such models the observed
variability is due to the collective effect of an ensemble of
perturbations, such as shocks in the disk or spiral arms that are not
continuous but rather are comprised of discrete segments.

A promising method to probe the structure and kinematics of the
broad-line region is two-dimensional reverberation mapping.  By
measuring the time delay of the response of the broad emission line
region to the ionizing continuum as a function of velocity, high
fidelity velocity delay maps can be constructed to discriminate
between different geometries and velocity fields for the broad-line
gas (Horne \etal 2004).  This method requires high temporal
resolution, high throughput, and long and homogenous monitoring, which
would be possible with a future dedicated satellite mission with a
UV/optical spectrophotometer.  In the meantime, a less demanding but
important experiment would be to monitor a few double-peaked emitters
on timescales shorter than the dynamical time (e.g.~sampling every
1--2 weeks).  Such observations could be carried out with 2--4m
ground-based telescopes would probe phenomena such as the development
and propagation of shocks in the disk.  We have, in fact, seen
evidence of such phenomena in our observations (see, e.g., \S
\ref{ch2-sec4} and Figure \ref{390:fig:spike1}) but we were not able
to follow them closely and study them in detail.  Continued
spectroscopic monitoring of double-peaked emitters will continue to
provide challenges to our current understanding of the nature of the
broad-line region in double-peaked emitters and AGNs in general.

\acknowledgments
We would like to thank Alexei V. Filippenko for providing spectra from the Lick 3m telescope.

The Hobby-Eberly Telescope (HET) is a joint project of the University
of Texas at Austin, the Pennsylvania State University, Stanford
University, Ludwig-Maximillians-Universit\"at M\"unchen, and
Georg-August-Universit\"at G\"ottingen. The HET is named in honor of
its principal benefactors, William P. Hobby and Robert E. Eberly.

The Marcario Low-Resolution Spectrograph is named for Mike Marcario of
High Lonesome Optics, who fabricated several optics for the instrument
but died before its completion; it is a joint project of the
Hobby-Eberly Telescope partnership and the Instituto de
Astronom\'{\i}a de la Universidad Nacional Aut\'onoma de M\'exico.

Many of the spectra from the 3m Shane reflector at Lick ovservatory
were taken with the Kast double spectrograph, which was made possible
by a generous gift from William and Marina Kast.

M.E. acknowledges support from NASA through Hubble Fellowship grant
HF-01068.01-94A awarded by the space telescope science institute,
which is operated by the Association for Universities for Research in
Astronomy, Inc.  for NASA, under contract NAS 5-26555. This award
covered a significant fraction of the observing-related costs from
1995 September to 1998 August.

\clearpage


\begin{deluxetable}{llll}
\tablecaption{BLRGs in our Monitoring Program\label{table:dp}}
\tablewidth{0pt}
\tablehead{
\colhead{Object} &
\colhead{{\it m$_{V}$}} &
\colhead{{\it z}}  &
\colhead{E(B-V)\tablenotemark{a}}\\
}
\startdata
Arp 102B & 14.8 & 0.024 & 0.024 \\
3C 390.3 & 15.4 & 0.056 & 0.071 \\
3C 332 & 17.3 & 0.151 & 0.024 \\
PKS 0235+023 & 17.1 & 0.209 & 0.033 \\
Mkn 668 &15.4 &  0.077 & 0.018 \\
3C 227 & 16.3 & 0.086 & 0.026 \\
3C 382 & 15.4 & 0.059 & 0.070 \\
\enddata
\tablenotetext{a}{Galactic reddening in magnitudes from Schlegel et al. (1998).}
\end{deluxetable}

\begin{deluxetable}{lcl}
\tablecaption{Arp 102B: Log of Spectroscopic Observations \label{table:arp}}
\tablewidth{0pt}
\tablehead{
\colhead{UT Date} &
\colhead{Exp Time} &
\colhead{Telescope} \\
\colhead{} &
\colhead{(sec)} &
\colhead{}
}
\startdata
1983 Jun 06 &     ...  & Palomar 5m\\
1983 Jun 21 &     ...  & Palomar 5m\\
1985 Jun 28 &     ...  & Palomar 5m\\
1986 Jul 13 &     3000 & Lick 3m \\
1987 May 04 &     1000 & Lick 3m \\
1987 Aug 08 &     1300 & Lick 3m \\
1988 Jun 30 &     ...  & Lick 3m\\
1988 Jul 17 &     ...  & Lick 3m\\
1988 Sep 15 &     ...  & Lick 3m\\
1989 Apr 27 &      900 & Lick 3m \\
1989 May 30 &     ... & Lick 3m\\
1989 Jul 01 &     3600 & KPNO 2.1m\\
1989 Jul 05 &      835 & KPNO 2.1m\\
1989 Jul 10 &      900 & Lick 3m \\
1990 Feb  23 &     5400 & KPNO 2.1m\\
1990 May 01 &     ...  & Lick 3m\\
1990 Jun 14 &     ...  & Lick 3m\\
1990 Jul 17 &     1800 & Lick 3m \\
1990 Jul 30 &     ...  & Lick 3m\\
1990 Aug 30 &     1800 & Lick 3m \\
1990 Nov 11 &     1800 & Lick 3m \\
1991 Jun 17 &     9790 & KPNO 2.1m\\
1991 Jun 18 &     1897 & KPNO 2.1m\\
1991 Jun 19 &     7200 & KPNO 2.1m\\
1991 Jun 20 &     7200 & KPNO 2.1m\\
1991 Jul 03 &     7200 & KPNO 2.1m\\
1991 Jul 04 &     4950 & KPNO 2.1m\\
1991 Aug 05 &     1500 & Lick 3m \\
1991 Oct 31 &     1800 & Lick 3m \\
1992 May 06 &     7200 & KPNO 2.1m\\
1992 Oct 03 &     1800 & Lick 3m \\
1992 Nov 19 &     1200 & Lick 3m \\
1993 Apr 14 &      900 & Lick 3m \\
1993 May 16 &     3600 & KPNO 2.1m\\
1993 Jun 28 &      900 & Lick 3m \\
1993 Jul 28 &     2100 & Lick 3m \\
1993 Sep 10 &     3000 & Lick 3m \\
1993 Sep 25 &     1800 & Lick 3m \\
1993 Oct 22 &     1800 & Lick 3m \\
1993 Nov 08 &     3000 & Lick 3m \\
1994 Apr 18 &     1200 & Lick 3m \\
1994 Jun 16 &     2700 & Lick 3m \\
1994 Jul 04 &     3600 & KPNO 2.1m\\
1994 Jul 05 &     3600 & KPNO 2.1m\\
1994 Jul 15 &     3000 & Lick 3m \\
1994 Aug 04 &     3600 & Lick 3m \\
1994 Sep 03 &     3600 & Lick 3m \\
1994 Oct 01 &     4800 & Lick 3m \\
1994 Nov 12 &     3600 & Lick 3m \\
1995 Jan 23 &     1800 & KPNO 2.1m\\
1995 Mar  25 &     1500 & Lick 3m \\
1996 Jun 02 &     3600 & KPNO 2.1m\\
1995 Jun 03 &     3600 & KPNO 2.1m\\
1995 Jun 04 &     3600 & KPNO 2.1m\\
1995 Sep 26 &     1800 & Lick 3m \\
1996 Feb  13 &     3200 & KPNO 2.1m\\
1996 Apr 22 &     2400 & Lick 3m \\
1996 Jun 13 &     1711 & KPNO 2.1m\\
1996 Aug 10 &     ...  & Lick 3m\\
1996 Sep 10 &     ...  & Lick 3m\\
1996 Oct 10 &     2400 & Lick 3m \\
1997 Feb  06 &     2700 & KPNO 2.1m\\
1997 Jun 08 &     3600 & KPNO 2.1m\\
1997 Aug 04 &     ...  & Lick 3m\\
1997 Sep 06 &     ...  & Lick 3m\\
1997 Sep 27 &     3600 & KPNO 2.1m\\
1998 Jan 28 &     3000 & KPNO 2.1m\\
1998 Apr 08 &     2700 & MDM 2.4m\\
1998 May 19 &     3000 & KPNO 2.1m\\
1998 Jun 27 &     3000 & MDM 2.4m\\
1998 Jul 23 &     ...  & Lick 3m\\
1998 Sep 21 &     ...  & Lick 3m\\
1998 Oct 15 &     3000 & KPNO 2.1m\\
1999 Jun 18 &     2000 & MDM 2.4m\\
1999 Sep 12 &     2400 & MDM 2.4m\\
2000 Jun 03 &     3000 & KPNO 2.1m\\
2000 Sep 25 &     3600 & KPNO 2.1m\\
2001 Jan 24 &     1500 & MDM 2.4m\\
2001 Jul 20 &     3600 & KPNO 2.1m\\
2001 Sep 20 &     ...  & Lick 3m\\
2001 Oct 23 &     2700 & KPNO 2.1m\\
2002 Jan 08 &     1690 & MDM 2.4m\\
2002 Mar  07 &     ...  & Lick 3m\\
2002 Apr 13 &     2700 & MDM 2.4m\\
2002 Jun 03 &      300 & HET 9.2m\\
2002 Jun 12 &     3600 & KPNO 2.1m\\
2002 Oct 09 &     3600 & KPNO 2.1m\\
2003 Jun 08 &     3590 & MDM 2.4m\\
2004 Mar  29 &     2700 & MDM 2.4m\\
2004 Jun 10 &     2700 & MDM 2.4m\\
2004 Jun 22 &      300 & HET 9.2m\\
2004 Aug 05 &      300 & HET 9.2m\\
2004 Sep 21 &     2700 & MDM 2.4m\\
\enddata
\end{deluxetable}

\begin{deluxetable}{lcl}
\tablecaption{3C 390.3: Log of Spectroscopic Observations\label{table:390}}
\tablewidth{0pt}
\tablehead{
\colhead{UT Date} &
\colhead{Exp Time} &
\colhead{Telescope} \\
\colhead{} &
\colhead{(sec)} &
\colhead{}
}
\startdata
1988 Jul 9 &     1200 & KPNO 4m\\
1989 Jul 01 &     1800 & KPNO 2.1m\\
1989 Jul 02 &     1500 & KPNO 2.1m\\
1990 Feb  22 &     1800 & KPNO 2.1m\\
1990 May 30 &     2400 & KPNO 2.1m\\
1991 Jun 18 &     3600 & KPNO 2.1m\\
1991 Jun 20 &     3600 & KPNO 2.1m\\
1991 Jun 21 &     3600 & KPNO 2.1m\\
1991 Jul 03 &     3600 & KPNO 2.1m\\
1991 Jul 04 &     3600 & KPNO 2.1m\\
1991 Jul 05 &     5000 & KPNO 2.1m\\
1992 May 06 &     2700 & KPNO 2.1m\\
1994 Jul 03 &     2700 & KPNO 2.1m\\
1994 Jul 05 &     3600 & KPNO 2.1m\\
1995 Mar  25 &     3000 & KPNO 2.1m\\
1995 Jun 04 &     3000 & KPNO 2.1m\\
1995 Jun 05 &     3600 & KPNO 2.1m\\
1996 Feb  15 &     3000 & KPNO 2.1m\\
1996 Apr 21 &     1200 & Lick 3m \\
1996 Jun 14 &     1831 & KPNO 2.1m\\
1996 Oct 10 &     1800 & Lick 3m \\
1997 Feb  06 &     1800 & KPNO 2.1m\\
1997 Jun 08 &     2700 & KPNO 2.1m\\
1997 Sep 28 &     2700 & KPNO 2.1m\\
1998 Apr 08 &     1200 & MDM 2.4m \\
1998 May 20 &     2540 & KPNO 2.1m\\
1998 Jun 27 &     1800 & MDM 2.4m \\
1998 Oct 14 &     2700 & KPNO 2.1m\\
1999 Jun 18 &     1800 & MDM 2.4m\\
1999 Sep 11 &     1200 & MDM 2.4m\\
1999 Dec 02 &     1800 & KPNO 2.1m\\
2000 Mar  16 &     3000 & MDM 1.3m\\
2000 Jun 03 &     2400 & KPNO 2.1m\\
2000 Sep 24 &     3000 & KPNO 2.1m\\
2001 Jul 23 &     3600 & KPNO 2.1m\\
2001 Oct 24 &     2700 & KPNO 2.1m\\
2002 Jun 12 &     3000 & KPNO 2.1m\\
2002 Oct 10 &     3000 & KPNO 2.1m\\
2003 Jun 08 &     1800 & MDM 2.4m\\
2004 Mar  28 &     1800 & MDM 2.4m\\
2004 Jun 10 &     1800 & MDM 2.4m\\
2004 Sep 21 &     1800 & MDM 2.4m\\
\enddata
\end{deluxetable}

\begin{deluxetable}{lcl}
\tablecaption{3C 332: Log of Spectroscopic Observations\label{table:332}}
\tablewidth{0pt}
\tablehead{
\colhead{UT Date} &
\colhead{Exp Time} &
\colhead{Telescope} \\
\colhead{} &
\colhead{(sec)} &
\colhead{}
}
\startdata
1990 May 28 &     2700 & KPNO 2.1m\\
1990 May 30 &     2400 & KPNO 2.1m\\
1990 Jun 12 &     2700 & Lick 3m \\
1991 Feb  04 &     3000 & KPNO 2.1m\\
1991 Feb  05 &     3000 & KPNO 2.1m\\
1992 Jul 07 &     3200 & KPNO 2.1m\\
1994 Jul 05 &     3600 & KPNO 2.1m\\
1995 Jun 04 &     3600 & KPNO 2.1m\\
1996 Feb  16 &     3600 & KPNO 2.1m\\
1996 Apr 21 &     1800 & Lick 3m \\
1996 Jun 15 &     3600 & KPNO 2.1m\\
1997 Feb  07 &     3600 & KPNO 2.1m\\
1997 Jun 09 &     3600 & KPNO 2.1m\\
1998 Apr 09 &     3600 & MDM 2.4m\\
1998 Jun 27 &     3600 & MDM 2.4m\\
1998 Jun 28 &     3600 & MDM 2.4m\\
1999 Jun 15 &     6000 & KPNO 2.1m\\
2000 Jun 05 &     3600 & KPNO 2.1m\\
2000 Sep 26 &     3600 & KPNO 2.1m\\
2001 Jul 21 &     3600 & KPNO 2.1m\\
2001 Jul 22 &     1008 & KPNO 2.1m\\
2002 Apr 12 &     3600 & MDM 2.4m\\
2002 Jun 14 &     3600 & KPNO 2.1m\\
2002 Oct 11 &     3600 & KPNO 2.1m\\
2003 Jun 07 &     4800 & MDM 2.4m\\
2003 Jul 01 &      600 & HET 9.2m\\
2004 Mar  28 &     6423 & MDM 2.4m\\
2004 Jun 09 &     7200 & MDM 2.4m\\
2004 Jun 22 &      600 & HET 9.2m\\
2004 Aug 04 &      600 & HET 9.2m\\
\enddata
\end{deluxetable}

\begin{deluxetable}{lcl}
\tablecaption{PKS 0235+023: Log of Spectroscopic Observations\label{table:pks}}
\tablewidth{0pt}
\tablehead{
\colhead{UT Date} &
\colhead{Exp Time} &
\colhead{Telescope} \\
\colhead{} &
\colhead{(sec)} &
\colhead{}
}
\startdata
1991 Feb  05 &     1800 & KPNO 2.1m\\
1996 Oct 11 &     2400 & Lick 3m \\
1997 Sep 29 &     3600 & KPNO 2.1m\\
1998 Oct 13 &     3600 & KPNO 2.1m\\
1999 Sep 13 &     3600 & MDM 2.4m\\
1999 Dec 04 &     3600 & KPNO 2.1m\\
2001 Jan 24 &     6000 & MDM 2.4m\\
2001 Oct 25 &     7200 & KPNO 2.1m\\
2002 Jan 08 &     3000 & MDM 2.4m\\
2002 Jan 09 &     3600 & MDM 2.4m\\
2002 Oct 11 &     3600 & KPNO 2.1m\\
2002 Dec 28 &      900 & HET 9.2m\\
2003 Oct 23 &     1800 & HET 9.2m\\
2003 Dec 20 &      900 & HET 9.2m\\
2004 Sep 09 &     1800 & HET 9.2m\\
2004 Dec 13 &      900 & HET 9.2m\\
\enddata
\end{deluxetable}

\begin{deluxetable}{lcl}
\tablecaption{Mkn 668: Log of Spectroscopic Observations\label{table:668}}
\tablewidth{0pt}
\tablehead{
\colhead{UT Date} &
\colhead{Exp Time} &
\colhead{Telescope} \\
\colhead{} &
\colhead{(sec)} &
\colhead{}
}
\startdata
1989 Jul 03 &     1800 & KPNO 2.1m\\
1990 Feb  22 &     1800 & KPNO 2.1m\\
1990 May 30 &     2400 & KPNO 2.1m\\
1995 Jan 21 &     1691 & KPNO 2.1m\\
1995 Mar  25 &     3600 & KPNO 2.1m\\
1995 Jun 04 &     3600 & KPNO 2.1m\\
1996 Feb  14 &     3200 & KPNO 2.1m\\
1997 Feb  07 &     2700 & KPNO 2.1m\\
1997 Jun 09 &     2700 & KPNO 2.1m\\
1998 Jan 29 &     3600 & KPNO 2.1m\\
1998 Apr 08 &     1800 & MDM 2.4m\\
1998 May 19 &     2700 & KPNO 2.1m\\
1998 Jun 27 &     2000 & MDM 2.4m\\
1999 Jun 15 &     2700 & KPNO 2.1m\\
2000 Jun 03 &     2000 & KPNO 2.1m\\
2001 Jan 24 &     1800 & MDM 2.4m\\
2001 Jul 21 &     3000 & KPNO 2.1m\\
2002 Jan 08 &     1800 & MDM 2.4m\\
2003 Jan 26 &     4400 & MDM 1.3m\\
2004 Mar  29 &     1800 & MDM 2.4m\\
2004 Jun 09 &     1800 & MDM 2.4m\\
\enddata
\end{deluxetable}

\begin{deluxetable}{lcl}
\tablecaption{3C 227: Log of Spectroscopic Observations\label{table:227}}
\tablewidth{0pt}
\tablehead{
\colhead{UT Date} &
\colhead{Exp Time} &
\colhead{Telescope} \\
\colhead{} &
\colhead{(sec)} &
\colhead{}
}
\startdata
1990 Feb  22 &     7200 & KPNO 2.1m\\
1993 Dec 13 &     1500 & KPNO 2.1m\\
1995 Jan 23 &     3600 & KPNO 2.1m\\
1995 Mar  24 &     3600 & KPNO 2.1m\\
1996 Feb  15 &     3200 & KPNO 2.1m\\
1997 Feb  06 &     3600 & KPNO 2.1m\\
1998 Jan 29 &     3600 & KPNO 2.1m\\
1998 May 20 &     2700 & KPNO 2.1m\\
1998 Dec 20 &     2400 & MDM 2.4m\\
1999 Dec 04 &     3600 & KPNO 2.1m\\
2001 Jan 24 &     4800 & MDM 2.4m\\
2002 Jan 08 &     2700 & MDM 2.4m\\
2002 Jan 11 &     2700 & MDM 2.4m\\
2002 Apr 12 &     2400 & MDM 2.4m\\
2002 Dec 27 &      300 & HET 9.2m\\
2003 Jan 26 &     4800 & MDM 1.3m\\
2003 Apr 05 &      300 & HET 9.2m\\
2003 Oct 31 &      300 & HET 9.2m\\
2003 Dec 22 &      300 & HET 9.2m\\
2004 Mar  28 &     2700 & MDM 2.4m\\
\enddata
\end{deluxetable}

\begin{deluxetable}{lcl}
\tablecaption{3C 382: Log of Spectroscopic Observations\label{table:382}}
\tablewidth{0pt}
\tablehead{
\colhead{UT Date} &
\colhead{Exp Time} &
\colhead{Telescope} \\
\colhead{} &
\colhead{(sec)} &
\colhead{}
}
\startdata
1989 Jul 02 &     1500 & KPNO 2.1m\\
1989 Jul 04 &     1800 & KPNO 2.1m\\
1989 Jul 05 &     1200 & KPNO 2.1m\\
1990 May 30 &     2400 & KPNO 2.1m\\
1991 Jun 18 &     3600 & KPNO 2.1m\\
1991 Jun 20 &     3600 & KPNO 2.1m\\
1991 Jun 21 &     3600 & KPNO 2.1m\\
1994 Jul 05 &     2100 & KPNO 2.1m\\
1995 Jun 05 &     1800 & KPNO 2.1m\\
1996 Apr 21 &     1200 & Lick 3m \\
1996 Apr 21 &     1200 & Lick 3m \\
1996 Jun 14 &      442 & KPNO 2.1m\\
1996 Oct 10 &      900 & Lick 3m \\
1997 Mar  21 &      600 & MDM 2.4m\\
1997 Jun 08 &     1200 & KPNO 2.1m\\
1997 Sep 29 &     1800 & KPNO 2.1m\\
1998 Apr 08 &      900 & MDM 2.4m\\
1998 Jun 27 &     1200 & MDM 2.4m\\
1998 Oct 15 &     1800 & KPNO 2.1m\\
1999 Jun 15 &     1200 & KPNO 2.1m\\
1999 Sep 11 &      900 & MDM 2.4m\\
1999 Dec 03 &     1800 & KPNO 2.1m\\
2000 Jun 04 &     1500 & KPNO 2.1m\\
2000 Sep 25 &     1800 & KPNO 2.1m\\
2001 Jul 21 &     1800 & KPNO 2.1m\\
2001 Oct 23 &     1200 & KPNO 2.1m\\
2002 Apr 12 &     1200 & MDM 2.4m\\
2002 Jun 03 &      300 & HET 9.2m\\
2002 Jun 13 &     2100 & KPNO 2.1m\\
2002 Oct 10 &     1800 & KPNO 2.1m\\
2003 Apr 01 &      300 & HET 9.2m\\
2003 Jun 07 &     3600 & MDM 2.4m\\
2004 Mar  29 &     1200 & MDM 2.4m\\
2004 Jun 09 &     1200 & MDM 2.4m\\
2004 Jun 25 &      300 & HET 9.2m\\
2004 Sep 09 &      150 & HET 9.2m\\
\enddata
\end{deluxetable}


\begin{deluxetable}{lcccccccc}
\tablecaption{Arp 102B: Broad H$\alpha$ Properties \label{tbl:arp:ha_props}}
\tablewidth{0pt}
\tablehead{
\colhead{UT Date} & \colhead{Flux} & \colhead{V(B)} & \colhead{P(B)} & \colhead{F(R)/F(B)} & \colhead{FWQM} & \colhead{QMS} & \colhead{QMC}\\
\colhead{(1)} & \colhead{(2)} & \colhead{(3)} & \colhead{(4)} & \colhead{(5)} & \colhead{(6)} & \colhead{(7)} & \colhead{(8)}
}
\startdata
1983 Jun 06 & 21.0$\pm$0.6 & $-$5.0 & 0.069$\pm$0.002 & 0.98 & 16.9 &  0.9 &  0.9\\
1983 Jun 21 & 21.0$\pm$0.6 & $-$5.0 & 0.069$\pm$0.002 & 0.98 & 16.9 &  0.9 &  0.9\\
1985 Jun 28 & 16.6$\pm$0.6 & $-$4.8 & 0.059$\pm$0.002 & 0.80 & 17.2 &  0.7 &  0.4\\
1986 Jul 13 & 19.1$\pm$0.8 & $-$5.2 & 0.079$\pm$0.003 & 0.72 & 16.4 &  0.9 &  0.4\\
1987 May 04 & 23$\pm$1 & $-$4.8 & 0.087$\pm$0.005 & 0.76 & 16.5 &  0.9 &  0.4\\
1987 Aug 08 & 20.3$\pm$0.8 & $-$4.9 & 0.077$\pm$0.003 & 0.72 & 16.8 &  1.0 &  0.4\\
1988 Jun 30 & 13.8$\pm$0.7 & $-$5.4 & 0.060$\pm$0.003 & 0.70 & 16.3 &  0.8 &  0.1\\
1988 Jul 17 & 13.9$\pm$0.9 & $-$5.2 & 0.056$\pm$0.004 & 0.70 & 16.6 &  0.7 &  0.1\\
1988 Sep 15 & 16.0$\pm$0.7 & $-$5.3 & 0.065$\pm$0.003 & 0.71 & 16.6 &  0.7 &  0.1\\
1989 Apr 27 & 17.1$\pm$0.9 & $-$5.0 & 0.068$\pm$0.004 & 0.74 & 17.2 &  1.2 &  0.7\\
1989 May 30 & 14.7$\pm$0.7 & $-$5.3 & 0.057$\pm$0.003 & 0.74 & 18.0 &  1.2 &  0.7\\
1989 Jul 01 & 14.4$\pm$0.7 & $-$5.2 & 0.056$\pm$0.003 & 0.74 & 17.8 &  1.3 &  0.7\\
1989 Jul 05 & 19.8$\pm$0.6 & $-$5.2 & 0.085$\pm$0.003 & 0.68 & 16.3 &  0.8 &  0.2\\
1989 Jul 10 & 18$\pm$2 & $-$5.2 & 0.068$\pm$0.007 & 0.73 & 18.3 &  1.1 &  0.6\\
1990 Feb 23 & 14.8$\pm$0.6 & $-$5.3 & 0.052$\pm$0.002 & 0.70 & 18.3 &  1.0 &  0.5\\
1990 May 01 & 13.7$\pm$0.6 & $-$5.0 & 0.045$\pm$0.002 & 0.72 & 19.0 &  0.8 &  0.3\\
1990 May 30 & 15.1$\pm$0.9 & $-$5.0 & 0.051$\pm$0.003 & 0.70 & 19.0 &  0.8 &  0.3\\
1990 Jun 14 & 13.8$\pm$0.8 & $-$5.3 & 0.049$\pm$0.003 & 0.64 & 18.8 &  0.7 &  0.1\\
1990 Jul 17 & 17$\pm$1 & $-$5.1 & 0.054$\pm$0.003 & 0.74 & 18.7 &  1.2 &  0.8\\
1990 Jul 30 & 12.4$\pm$0.7 & $-$5.1 & 0.044$\pm$0.003 & 0.65 & 18.5 &  1.1 &  0.4\\
1990 Aug 30 & 14.7$\pm$0.7 & $-$5.0 & 0.050$\pm$0.002 & 0.75 & 18.3 &  1.0 &  0.6\\
1990 Nov 11 & 16$\pm$1 & $-$4.7 & 0.053$\pm$0.003 & 0.84 & 17.6 &  1.0 &  0.8\\
1991 Jun 17 & 12.5$\pm$0.5 & $-$5.4 & 0.050$\pm$0.002 & 0.62 & 17.9 &  0.7 &  0.0\\
1991 Jun 18 & 13.6$\pm$0.5 & $-$5.5 & 0.053$\pm$0.002 & 0.64 & 18.2 &  0.9 &  0.1\\
1991 Jun 19 & 13.8$\pm$0.5 & $-$5.5 & 0.053$\pm$0.002 & 0.64 & 18.2 &  0.9 &  0.1\\
1991 Jun 20 & 14.6$\pm$0.6 & $-$5.5 & 0.057$\pm$0.002 & 0.60 & 18.3 &  0.7 & $-$0.1\\
1991 Jul 03 & 12.9$\pm$0.5 & $-$5.5 & 0.051$\pm$0.002 & 0.64 & 18.0 &  0.7 &  0.0\\
1991 Jul 04 & 12.4$\pm$0.5 & $-$5.5 & 0.050$\pm$0.002 & 0.61 & 18.2 &  0.8 & 0.0\\
1991 Aug 05 & 16.9$\pm$0.5 & $-$5.2 & 0.062$\pm$0.002 & 0.66 & 18.6 &  1.0 &  0.3\\
1991 Oct 31 & 12.9$\pm$0.7 & $-$5.4 & 0.047$\pm$0.002 & 0.63 & 17.7 &  0.4 & $-$0.2\\
1992 May 06 & 17.0$\pm$0.8 & $-$5.2 & 0.056$\pm$0.003 & 0.85 & 19.0 &  1.4 &  0.9\\
1992 Oct 03 & 13.0$\pm$0.6 & $-$5.2 & 0.048$\pm$0.002 & 0.83 & 17.9 &  1.0 &  0.7\\
1992 Nov 19 & 15$\pm$1 & $-$5.1 & 0.048$\pm$0.003 & 0.84 & 18.6 &  1.0 &  0.7\\
1993 Apr 14 & 14.0$\pm$0.7 & $-$5.1 & 0.054$\pm$0.003 & 0.63 & 17.2 &  0.9 &  0.1\\
1993 May 16 & 13.9$\pm$0.8 & $-$5.0 & 0.051$\pm$0.003 & 0.69 & 17.3 &  1.1 &  0.4\\
1993 Jun 28 & 16.4$\pm$0.6 & $-$5.0 & 0.059$\pm$0.002 & 0.74 & 17.4 &  0.9 &  0.4\\
1993 Jul 28 & 14.9$\pm$0.7 & $-$5.1 & 0.059$\pm$0.003 & 0.65 & 17.4 &  0.9 &  0.2\\
1993 Sep 10 & 15.7$\pm$0.6 & $-$5.2 & 0.062$\pm$0.002 & 0.64 & 17.4 &  0.8 &  0.1\\
1993 Sep 25 & 18.2$\pm$0.8 & $-$5.1 & 0.073$\pm$0.003 & 0.66 & 17.4 &  0.8 &  0.1\\
1993 Oct 22 & 15.1$\pm$0.6 & $-$5.2 & 0.062$\pm$0.002 & 0.67 & 17.5 &  0.7 &  0.3\\
1993 Nov 08 & 15.4$\pm$0.6 & $-$5.3 & 0.060$\pm$0.002 & 0.65 & 17.7 &  1.0 &  0.3\\
1994 Apr 18 & 14.1$\pm$0.6 & $-$5.1 & 0.056$\pm$0.002 & 0.76 & 17.4 &  1.2 &  0.9\\
1994 Jun 16 & 17.3$\pm$0.7 & $-$5.3 & 0.068$\pm$0.003 & 0.79 & 17.6 &  1.3 &  0.9\\
1994 Jul 04 & 15.1$\pm$0.6 & $-$5.4 & 0.059$\pm$0.002 & 0.80 & 17.4 &  1.2 &  0.9\\
1994 Jul 05 & 15.8$\pm$0.6 & $-$5.4 & 0.059$\pm$0.002 & 0.85 & 17.4 &  1.1 &  0.9\\
1994 Jul 15 & 16.2$\pm$0.8 & $-$5.3 & 0.061$\pm$0.003 & 0.83 & 17.5 &  1.2 &  0.9\\
1994 Aug 04 & 18.7$\pm$0.7 & $-$5.3 & 0.068$\pm$0.003 & 0.86 & 17.3 &  1.0 &  0.9\\
1994 Sep 03 & 15.9$\pm$0.8 & $-$5.3 & 0.062$\pm$0.003 & 0.78 & 17.1 &  1.0 &  0.7\\
1994 Sep 08 & 15.9$\pm$0.8 & $-$5.3 & 0.062$\pm$0.003 & 0.78 & 17.1 &  1.0 &  0.7\\
1994 Oct 01 & 15.3$\pm$0.7 & $-$5.3 & 0.058$\pm$0.002 & 0.82 & 17.4 &  1.1 &  0.8\\
1994 Nov 12 & 14.5$\pm$0.9 & $-$5.3 & 0.058$\pm$0.004 & 0.71 & 17.3 &  0.9 &  0.5\\
1995 Jan 23 & 14.8$\pm$0.6 & $-$5.2 & 0.056$\pm$0.002 & 0.72 & 16.9 &  0.8 &  0.5\\
1995 Mar 25 & 15.3$\pm$0.7 & $-$5.2 & 0.056$\pm$0.003 & 0.82 & 17.4 &  1.2 &  0.9\\
1995 Jun 02 & 15.2$\pm$0.6 & $-$5.1 & 0.055$\pm$0.002 & 0.71 & 17.6 &  1.0 &  0.5\\
1995 Jun 03 & 16.5$\pm$0.5 & $-$5.1 & 0.057$\pm$0.002 & 0.75 & 17.7 &  1.1 &  0.7\\
1995 Jun 04 & 16.5$\pm$0.5 & $-$5.1 & 0.057$\pm$0.002 & 0.75 & 17.7 &  1.1 &  0.7\\
1995 Jul 06 & 13.4$\pm$0.8 & $-$5.1 & 0.048$\pm$0.003 & 0.72 & 18.0 &  1.1 &  0.4\\
1995 Sep 26 & 12.7$\pm$0.7 & $-$5.0 & 0.051$\pm$0.003 & 0.64 & 17.6 &  1.2 &  0.3\\
1996 Feb 13 & 16.3$\pm$0.6 & $-$4.8 & 0.060$\pm$0.002 & 0.70 & 17.4 &  0.7 &  0.3\\
1996 Apr 22 & 12.8$\pm$0.3 & $-$4.8 & 0.051$\pm$0.001 & 0.64 & 17.7 &  1.2 &  0.4\\
1996 Jun 13 & 14.2$\pm$0.8 & $-$5.1 & 0.053$\pm$0.003 & 0.74 & 17.8 &  1.3 &  0.7\\
1996 Aug 10 & 17$\pm$1 & $-$5.2 & 0.066$\pm$0.004 & 0.69 & 17.4 &  1.1 &  0.4\\
1996 Sep 10 & 13.4$\pm$0.6 & $-$5.3 & 0.059$\pm$0.003 & 0.62 & 17.4 &  1.0 &  0.2\\
1996 Oct 10 & 14.3$\pm$0.7 & $-$5.3 & 0.060$\pm$0.003 & 0.64 & 16.9 &  0.9 &  0.3\\
1997 Feb 06 & 12.7$\pm$0.6 & $-$5.3 & 0.056$\pm$0.003 & 0.65 & 17.1 &  0.5 & $-$0.1\\
1997 Jun 08 & 14.1$\pm$0.6 & $-$5.2 & 0.055$\pm$0.002 & 0.69 & 17.9 &  0.8 &  0.3\\
1997 Aug 04 & 15.2$\pm$0.8 & $-$5.3 & 0.062$\pm$0.003 & 0.65 & 17.2 &  0.5 & 0.0\\
1997 Sep 06 & 16$\pm$1 & $-$5.3 & 0.066$\pm$0.005 & 0.64 & 16.6 &  0.6 & $-$0.1\\
1997 Sep 27 & 18.5$\pm$0.7 & $-$5.3 & 0.070$\pm$0.003 & 0.78 & 16.7 &  0.7 &  0.3\\
1998 Jan 28 & 21$\pm$1 & $-$5.0 & 0.079$\pm$0.005 & 0.85 & 16.7 &  0.9 &  0.6\\
1998 Apr 08 & 21.0$\pm$0.9 & $-$5.1 & 0.083$\pm$0.004 & 0.73 & 16.4 &  0.7 &  0.2\\
1998 May 19 & 19.8$\pm$0.6 & $-$5.2 & 0.085$\pm$0.003 & 0.68 & 16.3 &  0.8 &  0.2\\
1998 Jun 27 & 31$\pm$2 & $-$5.2 & 0.141$\pm$0.007 & 0.70 & 15.9 &  0.7 &  0.2\\
1998 Jul 23 & 23$\pm$2 & $-$4.8 & 0.098$\pm$0.006 & 0.67 & 16.0 &  0.7 &  0.1\\
1998 Sep 21 & 17.7$\pm$0.8 & $-$4.9 & 0.078$\pm$0.004 & 0.58 & 15.8 &  0.5 & $-$0.4\\
1998 Oct 15 & 23.3$\pm$0.8 & $-$4.9 & 0.092$\pm$0.003 & 0.68 & 16.4 &  0.7 &  0.1\\
1999 Jun 18 & 17.0$\pm$0.7 & $-$5.1 & 0.075$\pm$0.003 & 0.67 & 15.7 &  0.4 & $-$0.1\\
1999 Sep 12 & 17.0$\pm$0.6 & $-$5.0 & 0.077$\pm$0.003 & 0.62 & 15.9 &  0.4 & $-$0.3\\
2000 Jun 03 & 16.8$\pm$0.7 & $-$5.0 & 0.067$\pm$0.003 & 0.77 & 16.8 &  1.0 &  0.5\\
2000 Sep 25 & 15.5$\pm$0.8 & $-$5.1 & 0.061$\pm$0.003 & 0.83 & 16.9 &  0.8 &  0.5\\
2001 Jan 24 & 14.6$\pm$0.8 & $-$5.0 & 0.060$\pm$0.003 & 0.75 & 16.3 &  0.6 &  0.2\\
2001 Jul 20 & 13.3$\pm$0.7 & $-$5.0 & 0.056$\pm$0.003 & 0.72 & 16.0 &  0.5 & 0.0\\
2001 Sep 20 & 14.1$\pm$0.7 & $-$5.1 & 0.058$\pm$0.003 & 0.61 & 16.5 &  0.4 & $-$0.3\\
2001 Oct 23 & 13.5$\pm$0.5 & $-$5.0 & 0.052$\pm$0.002 & 0.68 & 16.7 &  0.4 & $-$0.1\\
2002 Jan 08 & 13.8$\pm$0.7 & $-$5.0 & 0.058$\pm$0.003 & 0.68 & 16.4 &  0.5 &  0.0\\
2002 Mar 07 & 13.9$\pm$0.8 & $-$4.9 & 0.056$\pm$0.003 & 0.64 & 16.6 &  0.6 & $-$0.1\\
2002 Apr 13 & 16.1$\pm$0.6 & $-$5.0 & 0.060$\pm$0.002 & 0.75 & 17.1 &  0.5 &  0.1\\
2002 Jun 03 & 13$\pm$1 & $-$5.1 & 0.050$\pm$0.004 & 0.78 & 17.1 &  0.7 &  0.3\\
2002 Jun 12 & 15.8$\pm$0.7 & $-$5.0 & 0.058$\pm$0.003 & 0.86 & 16.5 &  0.7 &  0.5\\
2002 Oct 09 & 13.1$\pm$0.6 & $-$5.0 & 0.051$\pm$0.002 & 0.78 & 16.4 &  0.6 &  0.1\\
2003 Jun 08 & 15.5$\pm$0.7 & $-$5.2 & 0.063$\pm$0.003 & 0.70 & 16.4 &  0.8 &  0.3\\
2004 Mar 29 & 12$\pm$1 & $-$5.1 & 0.049$\pm$0.004 & 0.71 & 16.4 &  0.6 &  0.1\\
2004 Jun 10 & 14.0$\pm$0.7 & $-$5.1 & 0.058$\pm$0.003 & 0.72 & 16.1 &  0.6 &  0.1\\
2004 Jun 22 & 11.5$\pm$0.7 & $-$5.1 & 0.047$\pm$0.003 & 0.76 & 16.2 &  0.5 &  0.0\\
2004 Aug 05 & 11.8$\pm$0.7 & $-$5.2 & 0.048$\pm$0.003 & 0.82 & 15.8 &  0.2 & $-$0.1\\
2004 Sep 21 & 13.8$\pm$0.6 & $-$5.0 & 0.051$\pm$0.002 & 0.82 & 16.9 &  0.4 &  0.1\\
\enddata
\tablecomments{Col.~(1): Date of observation. Col.~(2): Flux of broad H$\alpha$ relative to [O I]$\lambda 6300$ with 1$\sigma$ errors. Col.~(3): Flux-weighted velocity centroid of top 10\% of blue peak in units of 10$^{3}$ km s$^{-1}$. Col.~(4): Average flux in top 10\% of blue peak relative to [O I]$\lambda 6300$ with 1$\sigma$ errors. Col.~(5): Ratio of average flux in fixed velocity intervals corresponding to the top 10\% of the red peak and blue peak in the mean profile. Col.~(6): Full width at quarter-maximum in units of $10^{3}$ km s$^{-1}$. Col.~(7): Quarter-maximum shift measured from central velocity at quarter maximum in units of $10^{3}$ km s$^{-1}$. Col.~(8): Flux-weighted centroid at quarter-maximum in units of $10^{3}$ km s$^{-1}$}
\end{deluxetable}

\begin{deluxetable}{lcccccccccc}
\tablecaption{3C 390.3: Broad H$\alpha$ Properties\label{tbl:390:ha_props}}
\tablewidth{0pt}
\tablehead{
\colhead{UT Date} & \colhead{Flux} & \colhead{V(R)} & \colhead{V(B)} & \colhead{P(R)} & \colhead{P(B)} & \colhead{F(R)/F(B)} & \colhead{FWQM} & \colhead{QMS} & \colhead{QMC}\\
\colhead{(1)} & \colhead{(2)} & \colhead{(3)} & \colhead{(4)} & \colhead{(5)} & \colhead{(6)} & \colhead{(7)} & \colhead{(8)} & \colhead{(9)} & \colhead{(10)}
}
\startdata
1988 Jul 9 & 29$\pm$3 &  4.3 & $-$2.8 & 0.09$\pm$0.01 & 0.12$\pm$0.01 & 0.77 & 13.9 &  0.8 &  0.5\\
1989 Jul 01 & 45$\pm$1 &  4.1 & $-$3.6 & 0.136$\pm$0.003 & 0.207$\pm$0.005 & 0.63 & 13.5 &  0.4 & $-$0.1\\
1989 Jul 02 & 50$\pm$1 &  4.1 & $-$3.5 & 0.150$\pm$0.003 & 0.224$\pm$0.005 & 0.63 & 13.3 &  0.3 & $-$0.1\\
1990 Feb 22 & 53$\pm$4 &  4.2 & $-$3.4 & 0.16$\pm$0.01 & 0.23$\pm$0.02 & 0.63 & 13.3 &  0.1 & $-$0.2\\
1990 May 30 & 45$\pm$1 &  4.1 & $-$3.2 & 0.134$\pm$0.003 & 0.211$\pm$0.005 & 0.59 & 13.0 &  0.2 & $-$0.2\\
1991 Jun 18 & 48$\pm$2 &  4.3 & $-$3.1 & 0.147$\pm$0.007 & 0.21$\pm$0.01 & 0.68 & 13.1 &  0.1 & $-$0.1\\
1991 Jun 20 & 47$\pm$1 &  4.4 & $-$3.1 & 0.143$\pm$0.003 & 0.208$\pm$0.005 & 0.68 & 13.4 &  0.3 & 0.0\\
1991 Jun 21 & 51$\pm$2 &  4.3 & $-$3.2 & 0.152$\pm$0.007 & 0.22$\pm$0.01 & 0.67 & 13.5 &  0.4 & 0.0\\
1991 Jul 03 & 50$\pm$2 &  4.4 & $-$3.2 & 0.156$\pm$0.007 & 0.22$\pm$0.01 & 0.70 & 13.4 &  0.5 &  0.1\\
1991 Jul 04 & 50$\pm$3 &  4.3 & $-$3.2 & 0.148$\pm$0.008 & 0.22$\pm$0.01 & 0.67 & 13.3 &  0.3 & 0.0\\
1991 Jul 05 & 52$\pm$3 &  4.4 & $-$3.2 & 0.156$\pm$0.008 & 0.23$\pm$0.01 & 0.68 & 13.5 &  0.3 & 0.0\\
1992 May 06 & 58$\pm$5 &  4.2 & $-$3.4 & 0.18$\pm$0.01 & 0.26$\pm$0.02 & 0.68 & 13.0 &  0.1 & $-$0.2\\
1994 Jul 03 & 35$\pm$2 &  4.2 & $-$3.3 & 0.115$\pm$0.007 & 0.17$\pm$0.01 & 0.68 & 13.4 &  0.5 &  0.0\\
1994 Jul 05 & 34$\pm$2 &  4.2 & $-$3.4 & 0.113$\pm$0.005 & 0.167$\pm$0.007 & 0.69 & 13.4 &  0.5 &  0.1\\
1995 Mar 25 & 42$\pm$3 &  4.4 & $-$3.3 & 0.141$\pm$0.009 & 0.19$\pm$0.01 & 0.74 & 14.5 &  1.1 &  0.5\\
1995 Jun 04 & 38$\pm$2 &  4.3 & $-$3.4 & 0.134$\pm$0.005 & 0.180$\pm$0.007 & 0.74 & 14.1 &  0.9 &  0.3\\
1995 Jun 05 & 42$\pm$2 &  4.2 & $-$3.3 & 0.145$\pm$0.007 & 0.190$\pm$0.009 & 0.76 & 14.3 &  1.0 &  0.4\\
1996 Feb 15 & 56$\pm$2 &  4.0 & $-$2.7 & 0.193$\pm$0.008 & 0.218$\pm$0.009 & 0.82 & 14.2 &  0.6 &  0.3\\
1996 Apr 21 & 47$\pm$1 &  4.0 & $-$2.9 & 0.161$\pm$0.004 & 0.192$\pm$0.005 & 0.78 & 13.6 &  0.3 &  0.0\\
1996 Oct 10 & 49$\pm$3 &  3.8 & $-$2.9 & 0.16$\pm$0.01 & 0.23$\pm$0.01 & 0.69 & 13.6 &  0.4 &  0.0\\
1997 Feb 06 & 47$\pm$4 &  3.3 & $-$3.0 & 0.14$\pm$0.01 & 0.22$\pm$0.02 & 0.61 & 13.3 &  0.3 & $-$0.2\\
1997 Jun 08 & 40$\pm$2 &  3.3 & $-$3.2 & 0.117$\pm$0.006 & 0.19$\pm$0.01 & 0.60 & 14.3 &  0.7 & 0.0\\
1997 Sep 28 & 41$\pm$1 &  3.2 & $-$3.2 & 0.121$\pm$0.003 & 0.193$\pm$0.005 & 0.60 & 14.4 &  0.8 &  0.0\\
1998 Apr 08 & 43$\pm$1 &  3.6 & $-$3.2 & 0.124$\pm$0.003 & 0.198$\pm$0.005 & 0.64 & 14.4 &  0.6 & $-$0.1\\
1998 May 20 & 49$\pm$6 &  3.6 & $-$3.1 & 0.15$\pm$0.02 & 0.22$\pm$0.03 & 0.66 & 14.7 &  0.8 & 0.0\\
1998 Jun 27 & 58$\pm$5 &  3.8 & $-$3.2 & 0.18$\pm$0.01 & 0.25$\pm$0.02 & 0.71 & 14.9 &  0.8 &  0.2\\
1998 Oct 14 & 51$\pm$4 &  3.2 & $-$3.4 & 0.14$\pm$0.01 & 0.24$\pm$0.02 & 0.58 & 14.7 &  0.6 & $-$0.2\\
1999 Jun 18 & 40$\pm$3 &  2.8 & $-$3.3 & 0.118$\pm$0.009 & 0.20$\pm$0.02 & 0.53 & 14.9 &  0.7 & $-$0.2\\
1999 Sep 11 & 40$\pm$2 &  3.0 & $-$3.4 & 0.110$\pm$0.006 & 0.19$\pm$0.01 & 0.54 & 14.9 &  0.5 & $-$0.3\\
1999 Dec 02 & 39$\pm$4 &  3.0 & $-$3.4 & 0.12$\pm$0.01 & 0.19$\pm$0.02 & 0.59 & 14.9 &  0.4 & $-$0.3\\
2000 Mar 16 & 46$\pm$4 &  3.7 & $-$3.4 & 0.14$\pm$0.01 & 0.22$\pm$0.02 & 0.65 & 15.1 &  0.4 & $-$0.2\\
2000 Jun 03 & 49$\pm$4 &  3.6 & $-$3.5 & 0.14$\pm$0.01 & 0.23$\pm$0.02 & 0.62 & 14.8 &  0.3 & $-$0.3\\
2000 Sep 24 & 49$\pm$2 &  3.7 & $-$3.6 & 0.139$\pm$0.007 & 0.24$\pm$0.01 & 0.61 & 14.0 &  0.0 & $-$0.4\\
2001 Jul 23 & 42$\pm$3 &  3.7 & $-$3.5 & 0.127$\pm$0.008 & 0.21$\pm$0.01 & 0.64 & 14.8 &  0.4 & $-$0.1\\
2001 Oct 24 & 34$\pm$2 &  3.0 & $-$3.5 & 0.113$\pm$0.006 & 0.19$\pm$0.01 & 0.61 & 13.9 & $-$0.3 & $-$0.7\\
2002 Jun 12 & 32$\pm$2 &  3.4 & $-$3.1 & 0.108$\pm$0.006 & 0.157$\pm$0.009 & 0.69 & 14.3 & $-$0.1 & $-$0.3\\
2002 Oct 10 & 41$\pm$2 &  3.5 & $-$3.3 & 0.142$\pm$0.008 & 0.21$\pm$0.01 & 0.70 & 13.5 & $-$0.1 & $-$0.3\\
2003 Jun 08 & 44$\pm$4 &  3.5 & $-$3.2 & 0.14$\pm$0.01 & 0.20$\pm$0.02 & 0.66 & 13.8 & $-$0.1 & $-$0.5\\
2004 Mar 28 & 66$\pm$8 &  3.4 & $-$3.7 & 0.19$\pm$0.02 & 0.36$\pm$0.04 & 0.56 & 13.3 & $-$0.1 & $-$0.7\\
2004 Jun 10 & 57$\pm$4 &  3.7 & $-$3.3 & 0.18$\pm$0.01 & 0.30$\pm$0.02 & 0.61 & 13.7 &  0.2 & $-$0.4\\
2004 Sep 21 & 63$\pm$5 &  3.6 & $-$3.3 & 0.19$\pm$0.01 & 0.33$\pm$0.03 & 0.61 & 13.8 &  0.1 & $-$0.5\\
\enddata
\tablecomments{Col.~(1): Date of observation. Col.~(2): Flux of broad H$\alpha$ relative to [O I]$\lambda 6300$ with 1$\sigma$ errors. Col.~(3-4): Flux-weighted velocity centroid of top 10\% of the red peak and blue peak in units of 10$^{3}$ km s$^{-1}$. Col.~(5-6): Average flux in top 10\% of the red peak and blue peak relative to [O I]$\lambda 6300$ with 1$\sigma$ errors. Col.~(7): Ratio of average flux in fixed velocity intervals corresponding to the top 10\% of the red peak and blue peak in the mean profile. Col.~(8): Full width at quarter-maximum in units of $10^{3}$ km s$^{-1}$. Col.~(9): Quarter-maximum shift measured from central velocity at quarter maximum in units of $10^{3}$ km s$^{-1}$. Col.~(10): Flux-weighted centroid at quarter-maximum in units of $10^{3}$ km s$^{-1}$}
\end{deluxetable}

\begin{deluxetable}{lcccccccccc}
\tablecaption{3C 332: Broad H$\alpha$ Properties \label{tbl:332:ha_props}}
\tablewidth{0pt}
\tablehead{
\colhead{UT Date} & \colhead{Flux} & \colhead{V(R)} & \colhead{V(B)} & \colhead{P(R)} & \colhead{P(B)} & \colhead{F(R)/F(B)} & \colhead{FWHM} & \colhead{HMS} & \colhead{HMC}\\
\colhead{(1)} & \colhead{(2)} & \colhead{(3)} & \colhead{(4)} & \colhead{(5)} & \colhead{(6)} & \colhead{(7)} & \colhead{(8)} & \colhead{(9)} & \colhead{(10)}
}
\startdata
1990 May 28 &  6.8$\pm$0.4 &  8.9 & $-$7.4 & 0.020$\pm$0.001 & 0.024$\pm$0.001 & 0.64 & 21.8 &  1.3 &  0.6\\
1990 May 30 &  9.7$\pm$0.6 & 11.4 & $-$7.5 & 0.027$\pm$0.002 & 0.033$\pm$0.002 & 0.66 & 22.4 &  1.3 &  0.4\\
1990 Jun 12 & 12.1$\pm$0.4 &  9.4 & $-$8.3 & 0.035$\pm$0.001 & 0.048$\pm$0.002 & 0.68 & 20.2 &  0.9 &  0.2\\
1991 Feb 04 &  5.7$\pm$0.3 &  8.6 & $-$7.4 & 0.012$\pm$0.001 & 0.017$\pm$0.001 & 0.80 & 23.3 &  1.9 &  1.1\\
1991 Feb 05 &  5.9$\pm$0.3 &  9.6 & $-$7.6 & 0.012$\pm$0.001 & 0.018$\pm$0.001 & 0.72 & 22.6 &  1.5 &  0.8\\
1992 Jul 07 &  6.6$\pm$0.4 & 10.2 & $-$7.5 & 0.014$\pm$0.001 & 0.017$\pm$0.001 & 0.88 & 23.9 &  1.8 &  1.6\\
1994 Jul 05 &  6.4$\pm$0.4 &  9.9 & $-$7.0 & 0.018$\pm$0.001 & 0.013$\pm$0.001 & 1.20 & 23.7 &  1.8 &  2.6\\
1995 Jun 04 & 10.8$\pm$0.4 & 11.2 & $-$6.0 & 0.028$\pm$0.001 & 0.027$\pm$0.001 & 1.01 & 24.2 &  2.1 &  2.5\\
1996 Feb 16 & 11.1$\pm$0.4 & 10.1 & $-$6.9 & 0.027$\pm$0.001 & 0.028$\pm$0.001 & 0.90 & 24.0 &  1.6 &  1.8\\
1996 Apr 21 &  9.7$\pm$0.4 & 10.2 & $-$6.9 & 0.026$\pm$0.001 & 0.025$\pm$0.001 & 0.86 & 24.0 &  1.5 &  1.5\\
1996 Jun 15 & 10.2$\pm$0.4 &  9.8 & $-$7.1 & 0.026$\pm$0.001 & 0.024$\pm$0.001 & 1.01 & 24.5 &  1.6 &  2.0\\
1997 Feb 07 & 10.2$\pm$0.4 & 10.0 & $-$7.4 & 0.022$\pm$0.001 & 0.027$\pm$0.001 & 0.84 & 23.3 &  1.5 &  1.4\\
1997 Jun 09 & 10.4$\pm$0.4 &  9.6 & $-$7.1 & 0.024$\pm$0.001 & 0.028$\pm$0.001 & 0.80 & 22.8 &  1.2 &  1.0\\
1998 Apr 09 & 17.0$\pm$0.6 &  8.0 & $-$7.1 & 0.036$\pm$0.001 & 0.046$\pm$0.002 & 0.81 & 21.6 &  0.9 &  0.5\\
1998 Jun 27 & 18.5$\pm$0.7 &  7.9 & $-$7.1 & 0.041$\pm$0.002 & 0.049$\pm$0.002 & 0.83 & 21.7 &  0.8 &  0.4\\
1998 Jun 28 & 16.1$\pm$0.6 &  8.2 & $-$7.0 & 0.036$\pm$0.001 & 0.041$\pm$0.002 & 0.87 & 21.9 &  0.8 &  0.5\\
1999 Jun 15 & 36$\pm$2 &  8.1 & $-$6.0 & 0.085$\pm$0.005 & 0.083$\pm$0.005 & 0.90 & 20.6 &  0.5 &  0.3\\
2000 Jun 05 & 32$\pm$1 &  7.4 & $-$5.5 & 0.072$\pm$0.003 & 0.080$\pm$0.003 & 0.88 & 20.8 &  0.1 &  0.4\\
2000 Sep 26 & 33.4$\pm$0.6 &  7.5 & $-$5.6 & 0.074$\pm$0.001 & 0.086$\pm$0.002 & 0.85 & 20.0 &  0.6 &  0.3\\
2001 Jul 21 & 18.3$\pm$0.6 &  8.8 & $-$5.6 & 0.041$\pm$0.001 & 0.049$\pm$0.002 & 0.77 & 20.2 &  0.4 &  0.0\\
2001 Jul 22 & 26.4$\pm$0.8 &  8.0 & $-$5.3 & 0.061$\pm$0.002 & 0.077$\pm$0.002 & 0.76 & 19.7 &  0.6 &  0.1\\
2002 Apr 12 & 33$\pm$3 &  7.5 & $-$5.4 & 0.076$\pm$0.006 & 0.076$\pm$0.006 & 0.93 & 20.0 &  0.6 &  0.2\\
2002 Jun 14 & 18.9$\pm$0.6 &  7.9 & $-$5.8 & 0.041$\pm$0.001 & 0.045$\pm$0.002 & 0.88 & 20.7 &  0.6 &  0.3\\
2002 Oct 11 & 26$\pm$1 &  7.9 & $-$6.2 & 0.059$\pm$0.002 & 0.063$\pm$0.002 & 0.89 & 20.8 &  0.7 &  0.4\\
2003 Jun 07 & 43$\pm$4 &  7.3 & $-$5.7 & 0.11$\pm$0.01 & 0.12$\pm$0.01 & 0.81 & 21.1 &  0.2 &  0.1\\
2003 Jul 01 & 49$\pm$2 &  7.3 & $-$5.3 & 0.120$\pm$0.005 & 0.115$\pm$0.004 & 0.95 & 19.8 &  0.8 &  0.5\\
2004 Mar 28 & 35$\pm$3 &  7.1 & $-$5.0 & 0.072$\pm$0.005 & 0.095$\pm$0.007 & 0.79 & 20.0 &  1.2 &  0.6\\
2004 Jun 09 & 59$\pm$6 &  7.6 & $-$5.2 & 0.14$\pm$0.01 & 0.16$\pm$0.02 & 0.81 & 19.6 &  1.0 &  0.3\\
2004 Jun 22 & 46$\pm$2 &  7.6 & $-$5.0 & 0.106$\pm$0.004 & 0.119$\pm$0.005 & 0.91 & 20.0 &  1.1 &  0.7\\
2004 Aug 04 & 37$\pm$1 &  7.9 & $-$5.1 & 0.083$\pm$0.003 & 0.096$\pm$0.003 & 0.87 & 19.9 &  1.2 &  0.8\\
\enddata
\tablecomments{Col.~(1): Date of observation. Col.~(2): Flux of broad H$\alpha$ relative to narrow H$\alpha$ with 1$\sigma$ errors. Col.~(3-4): Flux-weighted velocity centroid of top 10\% of the red peak and blue peak in units of 10$^{3}$ km s$^{-1}$. Col.~(5-6): Average flux in top 10\% of the red peak and blue peak relative to narrow H$\alpha$ with 1$\sigma$ errors. Col.~(7): Ratio of average flux in fixed velocity intervals corresponding to the top 10\% of the red peak and blue peak in the mean profile. Col.~(8): Full width at half-maximum in units of $10^{3}$ km s$^{-1}$. Col.~(9): Half-maximum shift measured from central velocity at half maximum in units of $10^{3}$ km s$^{-1}$. Col.~(10): Flux-weighted centroid at half-maximum in units of $10^{3}$ km s$^{-1}$}
\end{deluxetable}

\begin{deluxetable}{lcccccccc}
\tablecaption{PKS 0235+023: Broad H$\alpha$ Properties \label{tbl:0235:ha_props}}
\tablewidth{0pt}
\tablehead{
\colhead{UT Date} & \colhead{Flux} & \colhead{V(B)} & \colhead{P(B)} & \colhead{F(R)/F(B)} & \colhead{FWQM} & \colhead{QMS} & \colhead{QMC}\\
\colhead{(1)} & \colhead{(2)} & \colhead{(3)} & \colhead{(4)} & \colhead{(5)} & \colhead{(6)} & \colhead{(7)} & \colhead{(8)}
}
\startdata
1991 Feb 05 &  9.0$\pm$0.9 & $-$3.6 & 0.034$\pm$0.003 & 1.37 & 14.7 &  0.3 &  0.2\\
1996 Oct 11 &  7.9$\pm$0.8 & $-$4.3 & 0.030$\pm$0.003 & 0.70 & 14.6 & $-$1.3 & $-$1.3\\
1997 Sep 29 &  3.8$\pm$0.4 & $-$4.4 & 0.020$\pm$0.002 & 0.52 & 13.5 & $-$1.5 & $-$2.1\\
1998 Oct 13 &  7.1$\pm$0.7 & $-$4.7 & 0.042$\pm$0.004 & 0.52 & 12.0 & $-$1.0 & $-$1.7\\
1999 Sep 13 & 12$\pm$1 & $-$4.7 & 0.059$\pm$0.006 & 0.56 & 12.9 & $-$1.0 & $-$1.5\\
1999 Dec 04 &  8.4$\pm$0.8 & $-$4.5 & 0.046$\pm$0.004 & 0.67 & 13.5 &  0.1 & $-$0.9\\
2001 Jan 24 &  7.3$\pm$0.7 & $-$4.1 & 0.036$\pm$0.003 & 0.67 & 13.4 & $-$0.8 & $-$1.4\\
2001 Oct 25 & 15$\pm$1 & $-$4.7 & 0.066$\pm$0.006 & 0.71 & 13.8 & $-$0.7 & $-$1.2\\
2002 Jan 09 & 11$\pm$1 & $-$4.6 & 0.052$\pm$0.005 & 0.70 & 14.5 & $-$0.4 & $-$1.1\\
2002 Oct 11 & 17$\pm$2 & $-$4.5 & 0.066$\pm$0.006 & 0.77 & 14.7 & $-$0.7 & $-$1.0\\
2002 Dec 28 &  8.2$\pm$0.8 & $-$4.2 & 0.036$\pm$0.003 & 0.75 & 14.4 & $-$1.0 & $-$1.4\\
2003 Oct 23 & 13$\pm$1 & $-$4.6 & 0.051$\pm$0.005 & 0.76 & 15.7 & $-$1.5 & $-$1.5\\
2003 Dec 20 & 12$\pm$1 & $-$4.9 & 0.046$\pm$0.004 & 0.74 & 15.2 & $-$1.6 & $-$1.5\\
2004 Sep 09 & 12$\pm$1 & $-$4.9 & 0.054$\pm$0.005 & 0.72 & 14.3 & $-$1.1 & $-$1.4\\
2004 Dec 13 &  5.4$\pm$0.5 & $-$5.0 & 0.028$\pm$0.003 & 0.68 & 13.8 & $-$1.3 & $-$1.7\\
\enddata
\tablecomments{Col.~(1): Date of observation. Col.~(2): Flux of broad H$\alpha$ relative to narrow H$\alpha$ with 1$\sigma$ errors. Col.~(3): Flux-weighted velocity centroid of top 10\% of blue peak in units of 10$^{3}$ km s$^{-1}$. Col.~(4): Average flux in top 10\% of blue peak relative to narrow H$\alpha$ with 1$\sigma$ errors. Col.~(5): Ratio of integrated flux in fixed velocity intervals corresponding to the regions of variability in the red and blue sides of the rms profile. Col.~(6): Full width at quarter-maximum in units of $10^{3}$ km s$^{-1}$. Col.~(7): Quarter-maximum shift measured from central velocity at quarter maximum in units of $10^{3}$ km s$^{-1}$. Col.~(8): Flux-weighted centroid at quarter-maximum in units of $10^{3}$ km s$^{-1}$}
\end{deluxetable}

\begin{deluxetable}{lcccccccc}
\tablecaption{Mkn 668: Broad H$\alpha$ Properties\label{tbl:668:ha_props}}
\tablewidth{0pt}
\tablehead{
\colhead{UT Date} & \colhead{Flux} & \colhead{V(R)} & \colhead{P(R)} & \colhead{F(R)/F(B)} & \colhead{FWQM} & \colhead{QMS} & \colhead{QMC}\\
\colhead{(1)} & \colhead{(2)} & \colhead{(3)} & \colhead{(4)} & \colhead{(5)} & \colhead{(6)} & \colhead{(7)} & \colhead{(8)}
}
\startdata
1989 Jul 03 & 22$\pm$1 &  2.7 & 0.121$\pm$0.006 & 2.16 & 11.6 & $-$0.5 &  0.4\\
1990 Feb 22 & 14.2$\pm$0.8 &  2.7 & 0.090$\pm$0.005 & 2.48 & 11.0 & $-$0.3 &  0.6\\
1990 May 30 & 13.7$\pm$0.7 &  2.6 & 0.089$\pm$0.005 & 2.50 & 10.8 & $-$0.3 &  0.6\\
1995 Jan 21 & 10.4$\pm$0.6 &  2.4 & 0.062$\pm$0.003 & 1.59 & 11.0 & $-$0.5 &  0.2\\
1995 Mar 24 & 12.8$\pm$0.7 &  2.0 & 0.070$\pm$0.004 & 1.65 & 11.0 & $-$0.1 &  0.4\\
1995 Jun 04 & 11.2$\pm$0.6 &  2.0 & 0.060$\pm$0.003 & 1.53 & 11.2 & $-$0.4 &  0.2\\
1996 Feb 14 & 14.1$\pm$0.8 &  1.4 & 0.077$\pm$0.004 & 1.48 & 11.2 & $-$0.6 & 0.0\\
1997 Feb 07 &  7.3$\pm$0.4 &  1.3 & 0.041$\pm$0.002 & 1.48 & 11.7 & $-$0.4 &  0.0\\
1997 Jun 09 &  8.8$\pm$0.5 &  1.1 & 0.052$\pm$0.003 & 1.48 & 10.8 & $-$0.5 &  0.0\\
1998 Jan 29 &  6.0$\pm$0.3 &  1.8 & 0.034$\pm$0.002 & 1.46 & 11.0 & $-$0.3 &  0.1\\
1998 Apr 08 &  7.6$\pm$0.4 &  1.3 & 0.040$\pm$0.002 & 1.40 & 11.6 & $-$0.4 & $-$0.1\\
1998 May 19 &  4.7$\pm$0.3 &  0.8 & 0.024$\pm$0.001 & 1.46 & 11.8 & 0.0 &  0.2\\
1998 Jun 27 &  8.6$\pm$0.5 &  0.5 & 0.048$\pm$0.003 & 1.41 & 10.9 & $-$0.3 &  0.1\\
1999 Jun 15 &  6.0$\pm$0.3 &  1.2 & 0.034$\pm$0.002 & 1.44 & 11.2 & $-$0.3 &  0.1\\
2000 Jun 03 &  6.3$\pm$0.4 &  1.8 & 0.031$\pm$0.002 & 1.48 & 12.2 & $-$0.2 &  0.1\\
2001 Jul 21 & 24$\pm$1 &  1.1 & 0.130$\pm$0.007 & 1.81 & 11.4 & $-$0.3 &  0.2\\
2002 Jan 08 & 29$\pm$2 &  1.3 & 0.154$\pm$0.008 & 2.17 & 11.9 &  0.0 &  0.5\\
2003 Jan 26 & 15.9$\pm$0.8 &  1.6 & 0.088$\pm$0.005 & 2.30 & 12.6 &  0.3 &  0.8\\
2004 Mar 29 & 21$\pm$1 &  1.5 & 0.111$\pm$0.006 & 2.25 & 12.1 & $-$0.1 &  0.5\\
2004 Jun 09 & 17.8$\pm$0.9 &  2.0 & 0.093$\pm$0.005 & 2.34 & 12.4 & $-$0.1 &  0.5\\
\enddata
\tablecomments{Col.~(1): Date of observation. Col.~(2): Flux of broad H$\alpha$ relative to narrow H$\alpha$ with 1$\sigma$ errors. Col.~(3): Flux-weighted velocity centroid of top 10\% of red peak in units of 10$^{3}$ km s$^{-1}$. Col.~(4): Average flux in top 10\% of red peak relative to narrow H$\alpha$ with 1$\sigma$ errors. Col.~(5): Ratio of integrated flux in fixed velocity intervals corresponding to regions of strong variability on the red and blue side determined from the rms proifle.  Col.~(6): Full width at quarter-maximum in units of $10^{3}$ km s$^{-1}$. Col.~(7): Quarter-maximum shift measured from central velocity at quarter maximum in units of $10^{3}$ km s$^{-1}$. Col.~(8): Flux-weighted centroid at quarter-maximum in units of $10^{3}$ km s$^{-1}$}
\end{deluxetable}

\begin{deluxetable}{lcccccccc}
\tablecaption{3C 227: Broad H$\alpha$ Properties\label{tbl:227:ha_props}}
\tablewidth{0pt}
\tablehead{
\colhead{UT Date} & \colhead{Flux} & \colhead{V(B)} & \colhead{P(B)} & \colhead{F(R)/F(B)} & \colhead{FWQM} & \colhead{QMS} & \colhead{QMC}\\
\colhead{(1)} & \colhead{(2)} & \colhead{(3)} & \colhead{(4)} & \colhead{(5)} & \colhead{(6)} & \colhead{(7)} & \colhead{(8)}
}
\startdata
1990 Feb 22 & 17.8$\pm$0.8 & $-$2.1 & 0.120$\pm$0.005 & 0.25 & 13.1 & $-$0.3 & $-$0.8\\
1993 Dec 13 & 22$\pm$1 & $-$2.0 & 0.145$\pm$0.006 & 0.35 & 12.3 & $-$1.2 & $-$1.1\\
1995 Jan 23 &  9.8$\pm$0.5 & $-$2.1 & 0.067$\pm$0.003 & 0.33 & 11.6 & $-$1.6 & $-$1.3\\
1995 Mar 24 & 14.3$\pm$0.6 & $-$2.1 & 0.097$\pm$0.004 & 0.33 & 11.1 & $-$1.3 & $-$1.1\\
1996 Feb 15 & 11.4$\pm$0.5 & $-$2.1 & 0.074$\pm$0.003 & 0.36 & 12.4 & $-$1.5 & $-$1.3\\
1997 Feb 06 & 10.6$\pm$0.5 & $-$2.1 & 0.072$\pm$0.003 & 0.36 & 12.8 & $-$1.0 & $-$1.1\\
1998 Jan 29 & 15.5$\pm$0.7 & $-$1.8 & 0.096$\pm$0.004 & 0.42 & 12.8 & $-$1.0 & $-$0.8\\
1998 May 20 &  7.9$\pm$0.4 & $-$2.0 & 0.056$\pm$0.003 & 0.37 & 13.4 & $-$0.9 & $-$1.0\\
1998 Dec 20 & 14.0$\pm$0.6 & $-$1.6 & 0.078$\pm$0.003 & 0.44 & 14.4 & $-$0.7 & $-$0.6\\
1999 Dec 04 & 19.1$\pm$0.8 & $-$1.3 & 0.115$\pm$0.005 & 0.51 & 14.4 & $-$0.5 & $-$0.4\\
2001 Oct 25 & 16.6$\pm$0.7 & $-$1.1 & 0.103$\pm$0.005 & 0.45 & 14.2 & $-$0.9 & $-$0.7\\
2002 Jan 08 & 15.8$\pm$0.7 & $-$1.1 & 0.095$\pm$0.004 & 0.50 & 14.9 & $-$0.3 & $-$0.3\\
2002 Jan 11 & 16.3$\pm$0.7 & $-$1.0 & 0.098$\pm$0.004 & 0.50 & 14.0 & $-$0.8 & $-$0.6\\
2002 Apr 12 & 11.6$\pm$0.5 & $-$1.1 & 0.072$\pm$0.003 & 0.50 & 13.9 & $-$0.6 & $-$0.5\\
2002 Dec 27 & 17.4$\pm$0.8 & $-$0.9 & 0.107$\pm$0.005 & 0.52 & 13.2 & $-$0.8 & $-$0.5\\
2003 Jan 26 & 17.2$\pm$0.7 & $-$1.1 & 0.104$\pm$0.005 & 0.52 & 13.8 & $-$1.0 & $-$0.7\\
2003 Apr 05 & 15.2$\pm$0.7 & $-$0.9 & 0.090$\pm$0.004 & 0.53 & 13.4 & $-$0.8 & $-$0.5\\
2003 Oct 31 & 10.9$\pm$0.5 & $-$0.6 & 0.058$\pm$0.003 & 0.52 & 13.4 & $-$0.7 & $-$0.4\\
2003 Dec 22 & 11.5$\pm$0.5 & $-$1.2 & 0.075$\pm$0.003 & 0.43 & 12.7 & $-$0.8 & $-$0.6\\
2004 Mar 28 & 11.1$\pm$0.5 & $-$0.9 & 0.067$\pm$0.003 & 0.51 & 12.6 & $-$0.8 & $-$0.5\\
\enddata
\tablecomments{Col.~(1): Date of observation. Col.~(2): Flux of broad H$\alpha$ relative to narrow H$\alpha$ with 1$\sigma$ errors. Col.~(3): Flux-weighted velocity centroid of top 10\% of blue peak in units of 10$^{3}$ km s$^{-1}$. Col.~(4): Average flux in top 10\% of blue peak relative to [O I]$\lambda 6300$ with 1$\sigma$ errors. Col.~(5): Ratio of integrated flux in fixed velocity intervals corresponding to regions on the red and blue side of the profile corresponding to strong variations in the rms spectrum. Col.~(6): Full width at quarter-maximum in units of $10^{3}$ km s$^{-1}$. Col.~(7): Quarter-maximum shift measured from central velocity at quarter maximum in units of $10^{3}$ km s$^{-1}$. Col.~(8): Flux-weighted centroid at quarter-maximum in units of $10^{3}$ km s$^{-1}$}
\end{deluxetable}


\clearpage

\begin{figure} [tbp]\centering
\includegraphics[width=0.9\linewidth]{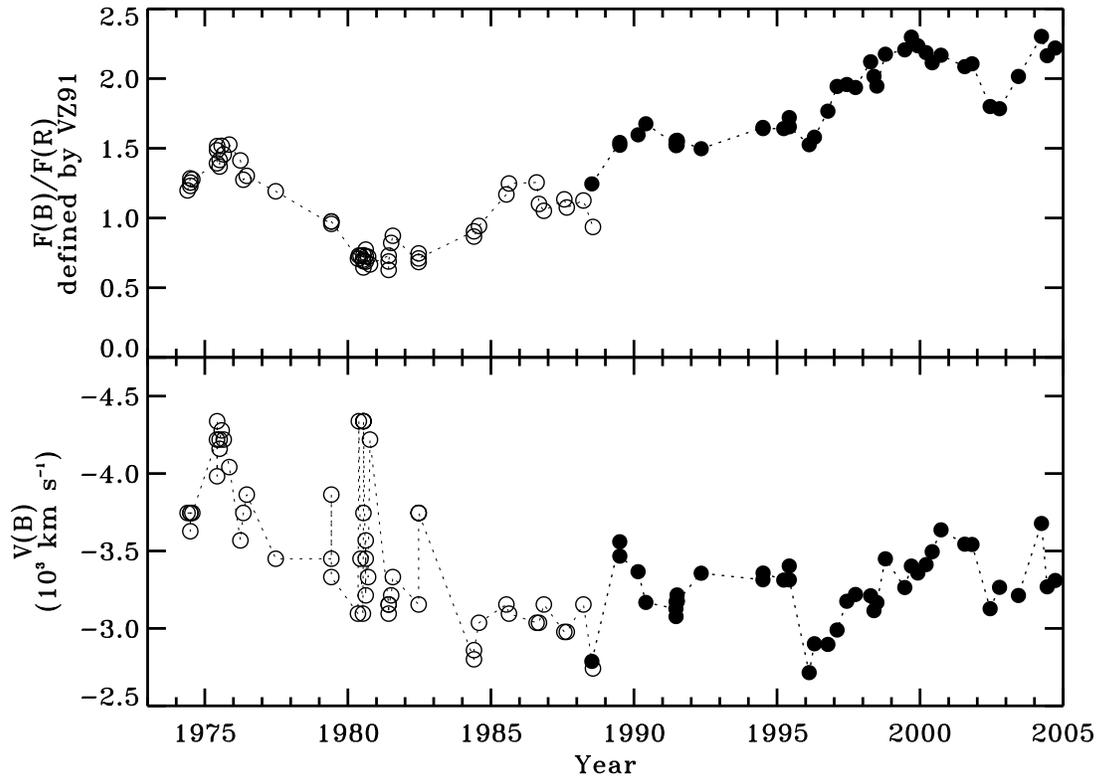}
\caption[3C 390.3: Historical velocity curve and $F$(B/$F$(R) ratio.]{3C 390.3: $F$(B)/$F$(R) as defined 
by Veilleux \& Zheng (1991) (VZ91), and the velocity of the blue peak
for over 30 yr.  VZ91 data measured from the broad H$\beta$ profile
are plotted with open circles, and our data measured from the
H$\alpha$ profiles in  our monitoring program are plotted with filled
circles.
\label{390:fig:vz91}}
\end{figure}

\begin{figure} [tbp]\centering
\includegraphics[width=0.9\linewidth]{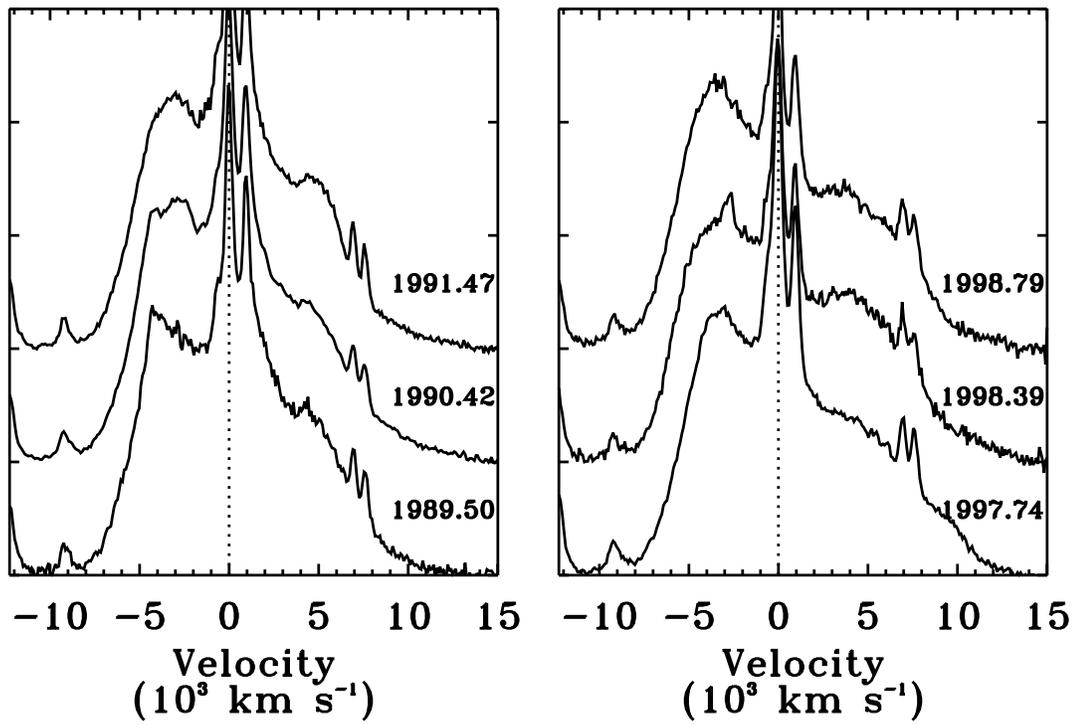}
\caption[3C 390.3: Transient sharp features in the blue peak.] {3C 390.3: A sequence of broad 
H$\alpha$ line spectra, without the narrow emission lines removed, which reveal transient sharp features in the blue peak.
\label{390:fig:spike1}}
\end{figure}

\begin{figure}[tbp]
\begin{center}
\includegraphics[width=0.9\linewidth]{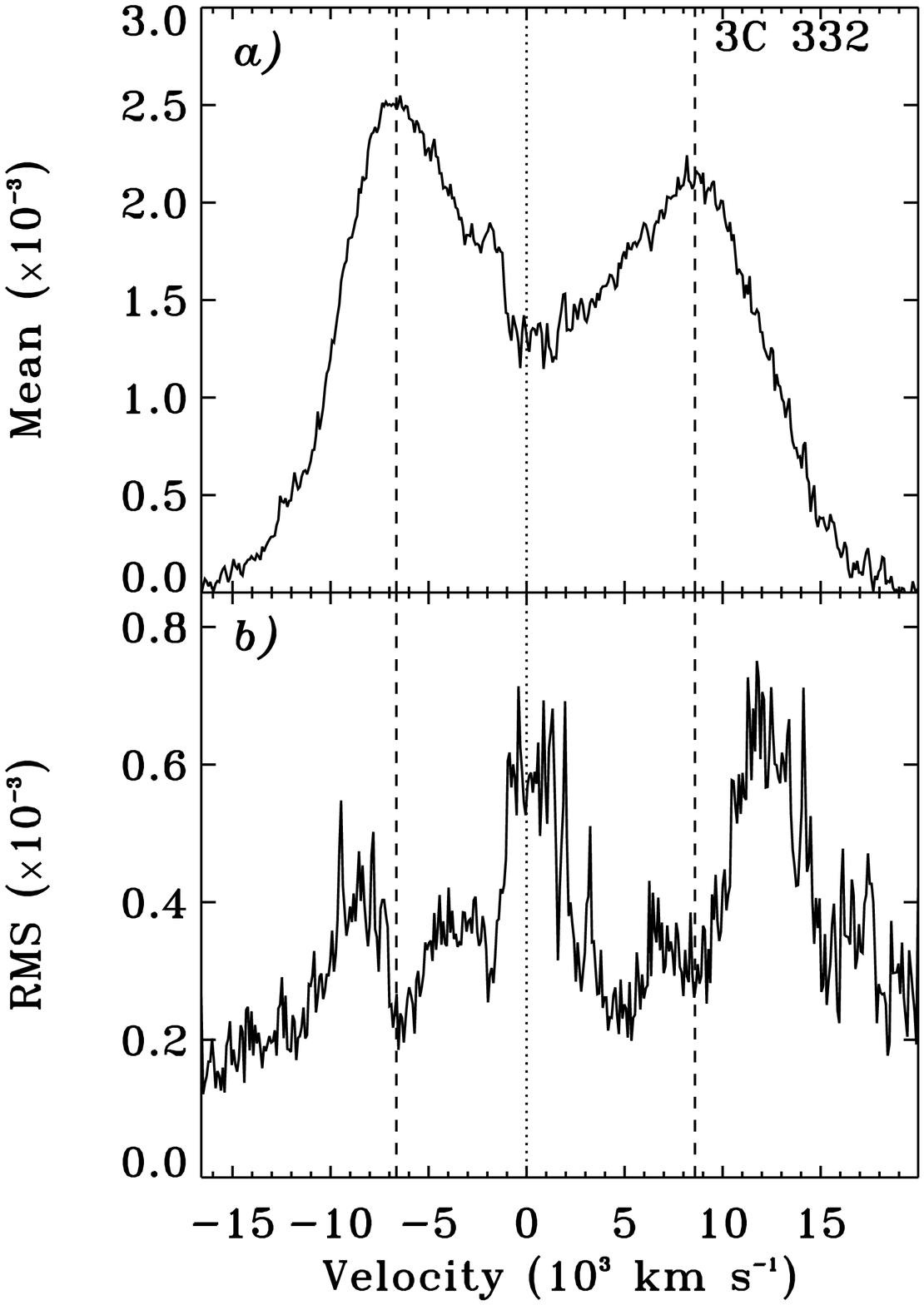}
\caption[3C 332: Mean and rms profile.]{3C 332: (a) Mean and (b) rms profile of the broad H$\alpha$ line in units of $f_{\nu}$ normalized by the total broad-line flux, with dashed lines indicating 
the velocities of the red and blue peaks of the mean profile, measured from the flux-weighted velocity
centroid of the top 10\% of the peaks. \label{332:fig:meanrms}}
\end{center}
\end{figure}

\begin{figure} [tbp]\centering
\includegraphics[width=0.8\linewidth]{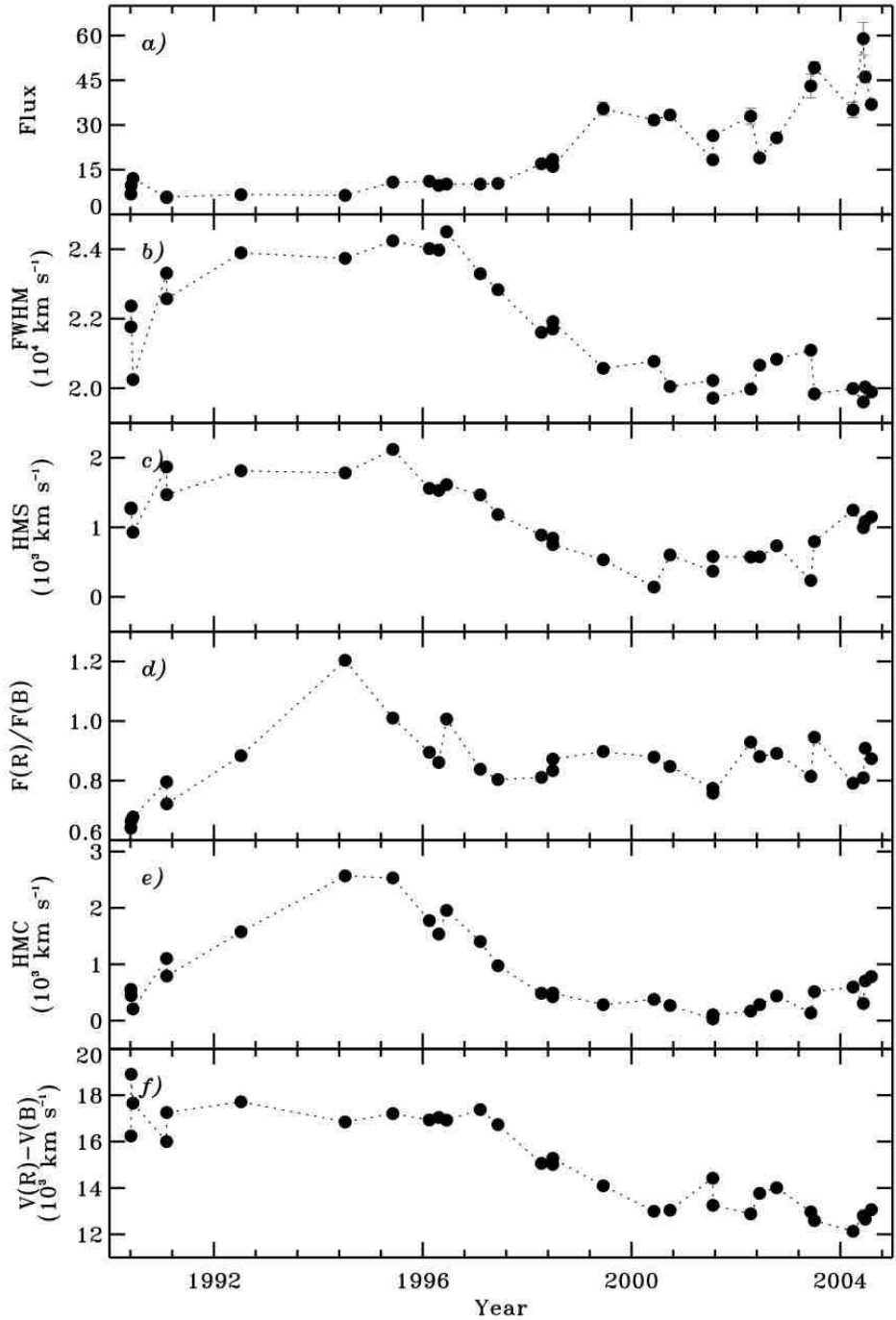}
\caption[3C 332: Characterization of the profiles.]{3C 332: 
(a) Total broad-line flux calibrated relative to the narrow H$\alpha$ line, (b) full width at half-maximum, (c) shift at half-maximum, 
(d) ratio of the average flux in fixed intervals in the red (+7200 to +10,000 \kms) and blue($-$5300 to $-$8000 \kms) peaks, (e) flux-weighted centroid at quarter-maximum,
and (f) peak velocity separation of the broad H$\alpha$ profiles over the entire duration of our monitoring program.
\label{332:fig:tp_332_3}}
\end{figure}

\begin{figure} [tbp]\centering
\includegraphics[width=0.9\linewidth]{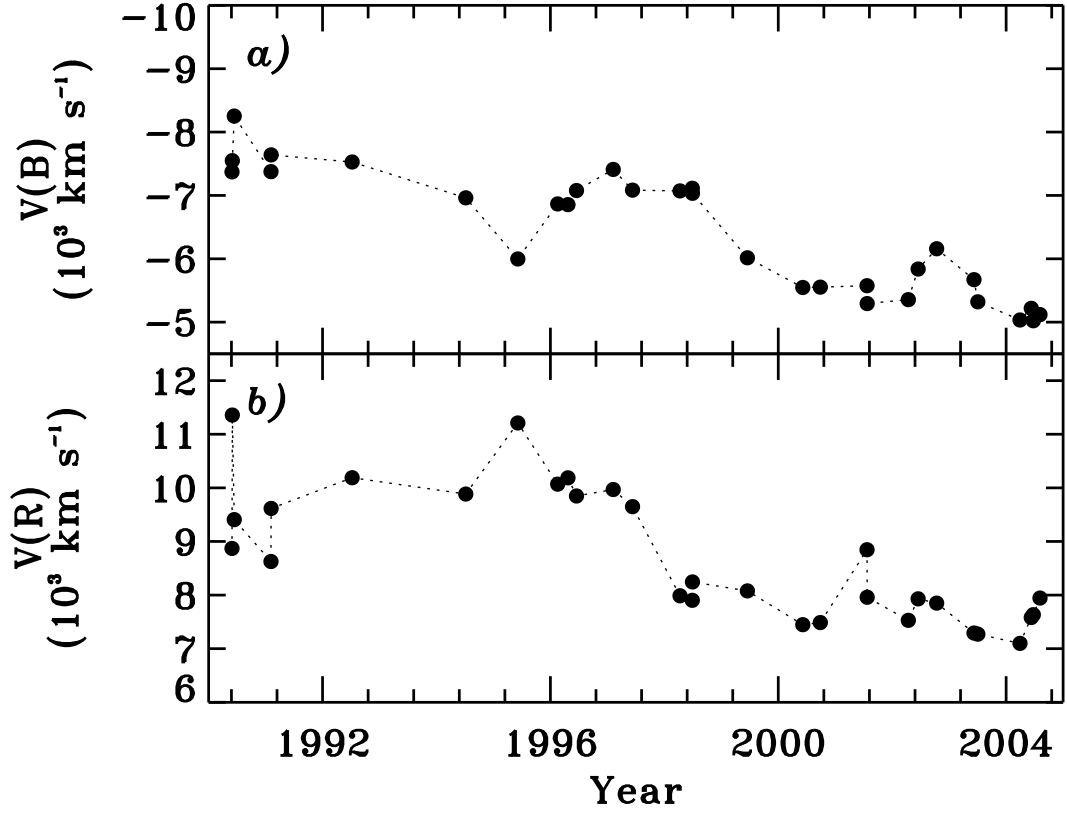}
\caption[3C 332: Characterization of the red and blue peaks.]{3C 332: 
Velocity of the (a) red and (b) blue peaks measured from the flux-weighted velocity centroid of
the top 10\% of the peaks of the broad H$\alpha$ profiles over the entire duration of our monitoring program.
\label{332:fig:velsig}}
\end{figure}

\begin{figure}[tbp]\centering
\includegraphics[width=0.9\linewidth]{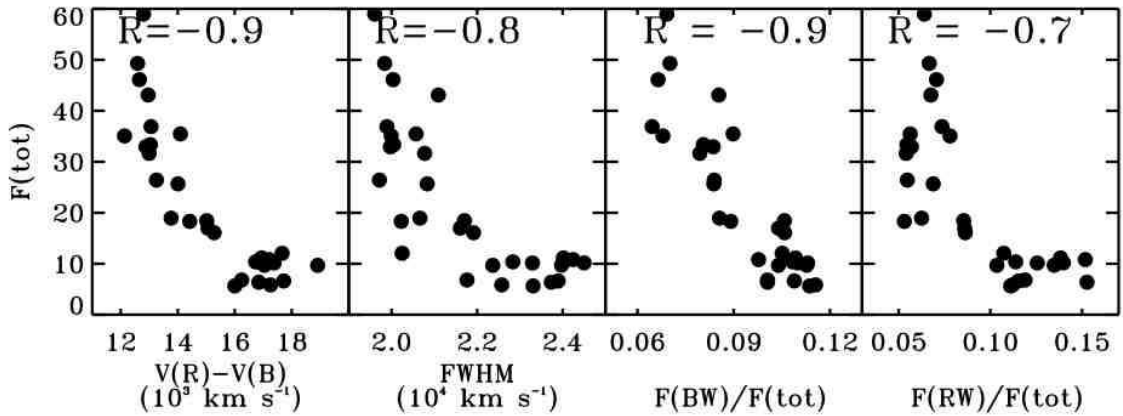}
\caption[3C 332: Grid of correlations between measured properties of the 
profiles.]{3C 332: Strong negative correlations, measured from the linear correlation coefficient $R$, 
between the total broad-line flux [$F$(tot)] and the peak velocity separation [$V$(R)-$V$(B)], 
the full width at half-maximum (FWHM), and the integrated flux in
fixed intervals in the blue wing ($-$8000 to $-$10,800 \kms) and the red wing (+10,000 to +12,800 \kms) relative to $F$(tot).  \label{332:fig:corrmatrix}}
\end{figure}

\begin{figure}[tbp]
\begin{center}
\includegraphics[width=0.9\linewidth]{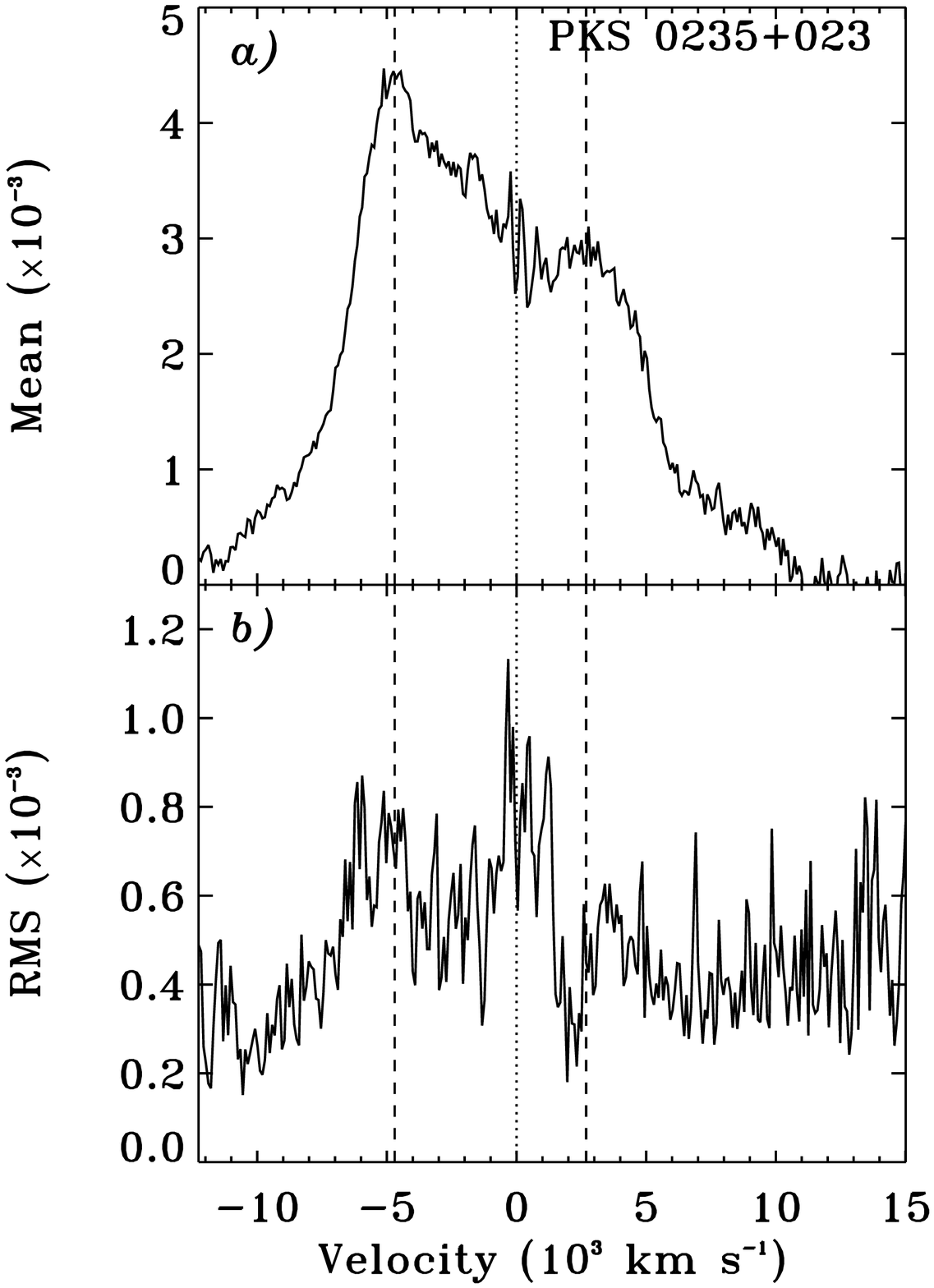}
\caption[PKS 0235+023: Mean and rms profile.]{PKS 0235+023: (a) Mean and (b) rms profile of the broad H$\alpha$ line in units of $f_{\nu}$ normalized by the total broad-line flux, with dashed lines indicating 
the velocities of the red and blue peaks of the mean profile, measured from the flux-weighted velocity
centroid of the top 10\% of the peaks. \label{0235:fig:meanrms}}
\end{center}
\end{figure}

\clearpage
\begin{figure} [tbp]
\begin{center}
\includegraphics[width=0.8\linewidth]{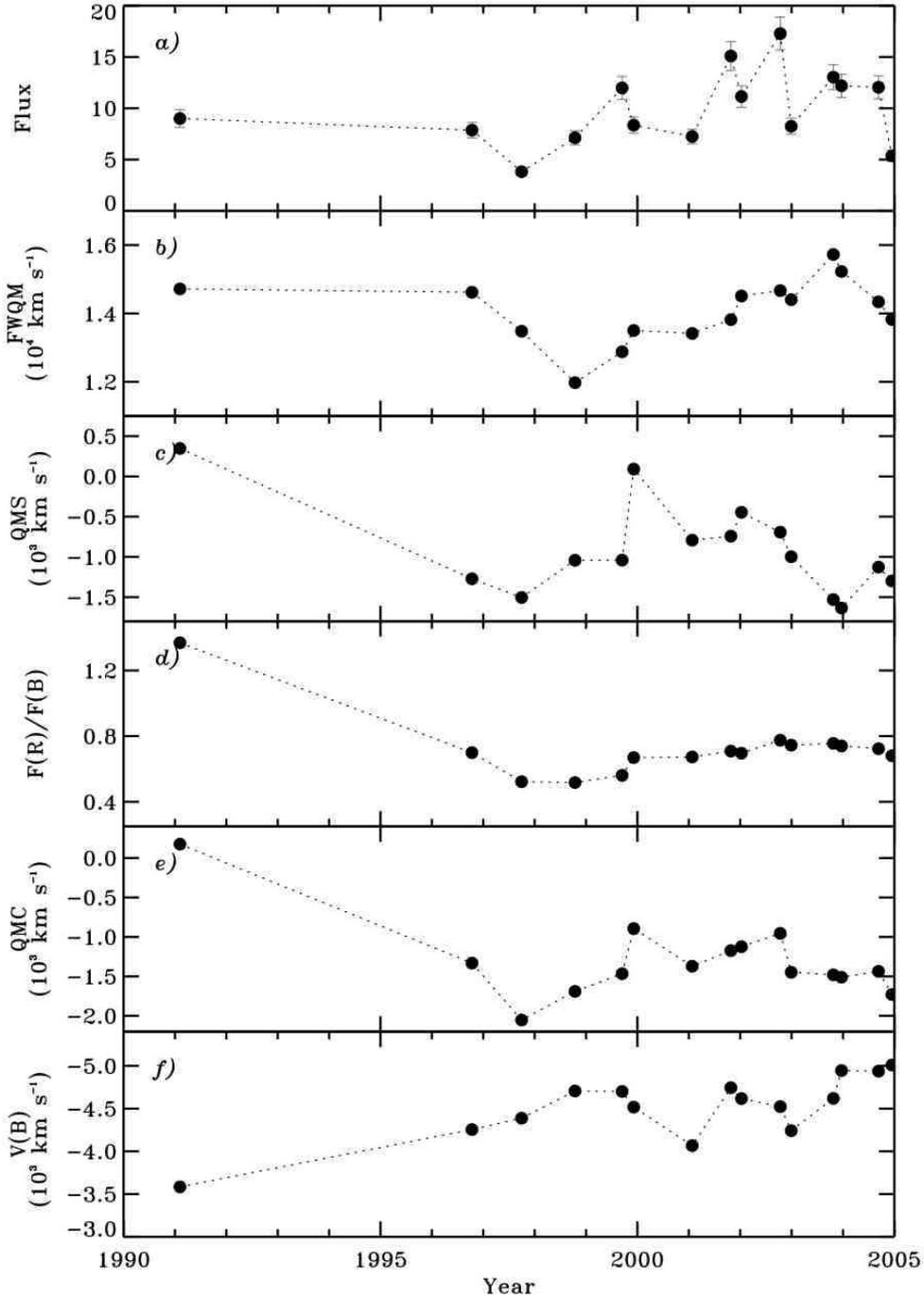}
\caption[PKS 0235+023: Characterization of the profiles.]{PKS 0235+023: 
(a) Total broad-line flux calibrated relative to the narrow H$\alpha$ line, (b) full width at quarter-maximum, (c) shift at quarter-maximum, 
(d) ratio of the integrated flux in fixed intervals in the red (+2300 to +5200 \kms) and blue($-$3900 to $-$ 6900 \kms) peaks, (e) flux-weighted centroid at quarter-maximum,
(f) flux-weighted velocity centroid and (g) dispersion of top 10\% of the blue peak of the broad H$\alpha$ profiles over the entire duration of our monitoring program.
\label{0235:fig:tp_0235_3}}
\end{center}
\end{figure}

\begin{figure}[tbp]
\begin{center}
\includegraphics[width=0.9\linewidth]{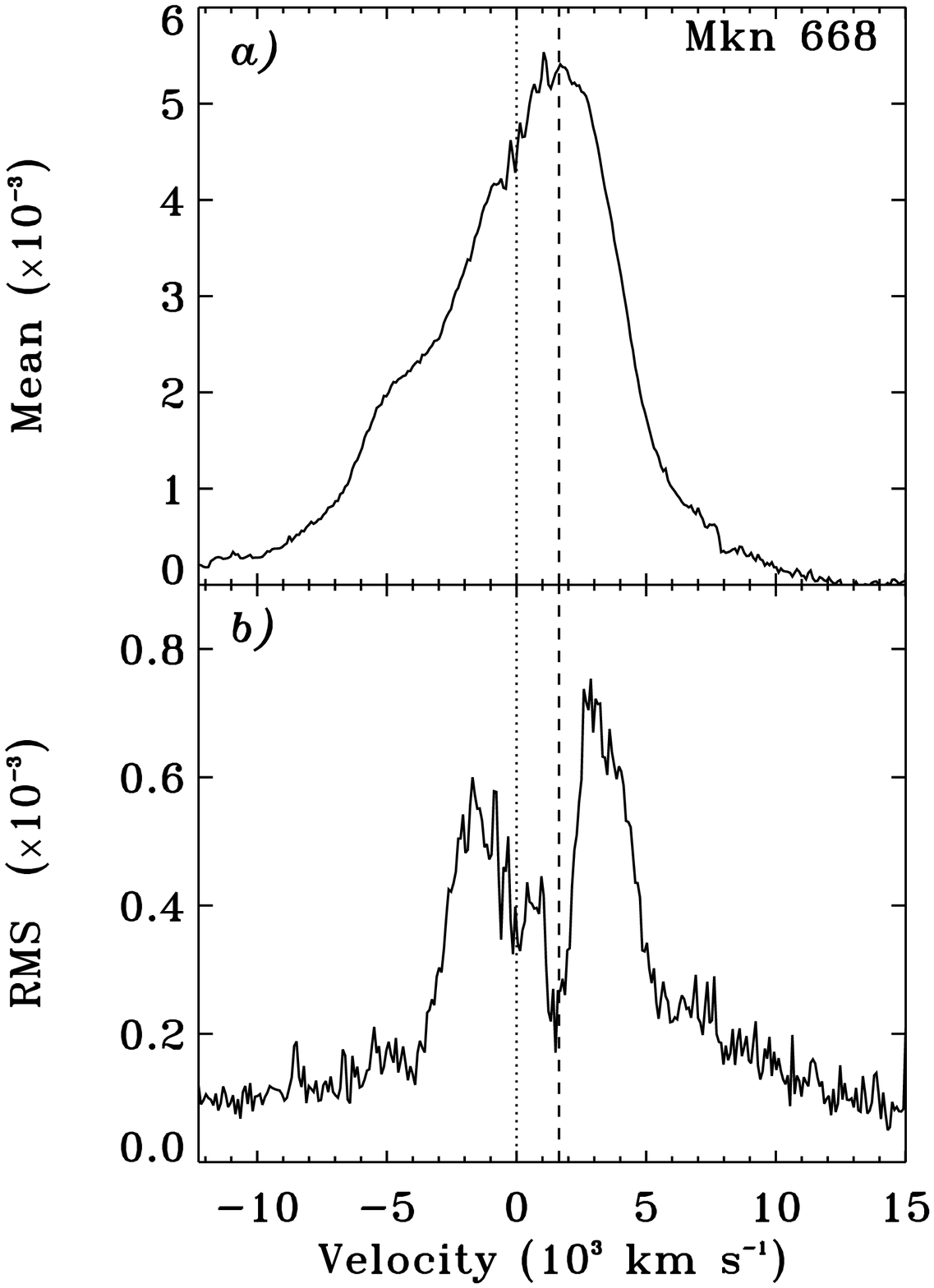}
\caption[Mkn 668: Mean and rms profile.]{Mkn 668: (a) Mean and (b) rms profile of the broad H$\alpha$ line in units of $f_{\nu}$ normalized by the total broad-line flux, with a dashed line indicating 
the velocity of the central red peak of the mean profile, measured from the flux-weighted velocity
centroid of the top 10\% of the peak. 
\label{668:fig:meanrms}}
\end{center}
\end{figure}
  
\clearpage
\thispagestyle{empty}
\begin{figure} [tbp]\centering
\vspace*{-20mm}
\includegraphics[width=0.9\linewidth]{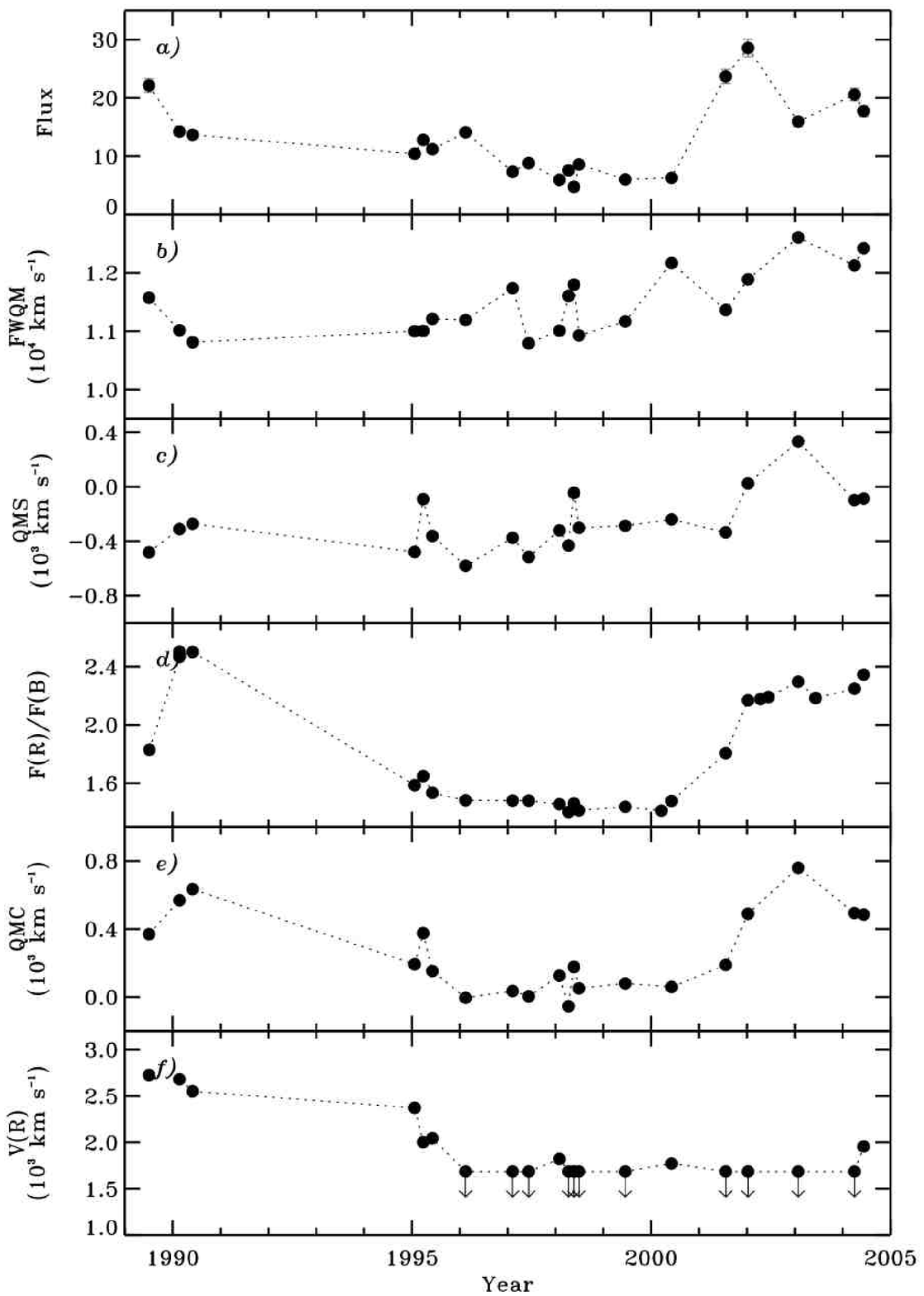}
\caption[Mkn 668: Characterization of the profiles.]{Mkn 668:  (a) Total broad-line flux calibrated relative to the narrow H$\alpha$ line, 
(b) full width at quarter-maximum, (c) shift at quarter-maximum, 
(d) ratio of the integrated flux in fixed intervals in the red side (+3200 to +5600 \kms) and blue side 
($-$800 to $-$3200 \kms) of the profile, (e) flux-weighted centroid at quarter-maximum,
(f) flux-weighted velocity centroid of top 10\% of the central red peak of the broad H$\alpha$ profiles over the entire duration of our monitoring program.\label{668:fig:tp_668_3}}
\end{figure}
  
\begin{figure} [tbp]\centering
\includegraphics[width=0.9\linewidth]{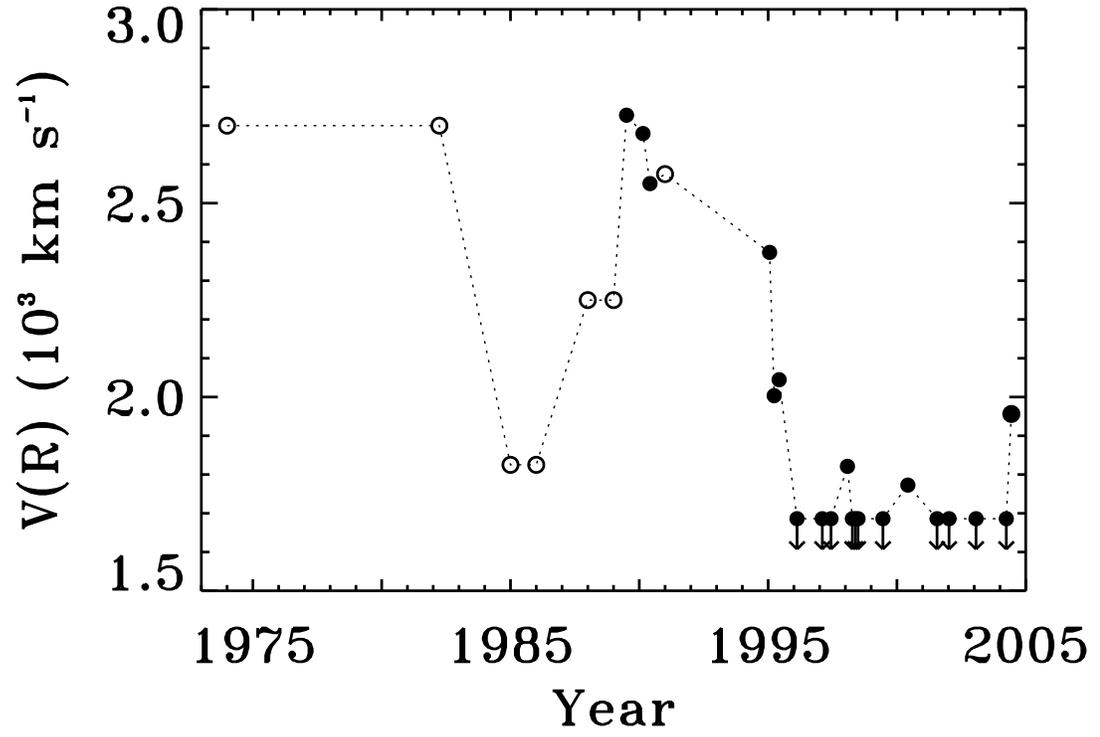}
\caption[Mkn 668: Historical velocity curve of red peak.]{Mkn 668: Historical velocity curve of the red peak.  Data measured for the H$\beta$ profile in 1974 from
Osterbrock \& Cohen (1979), in 1982 from Gaskell (1983), and from 1985 to 1991 from Marziani \etal (1993) are plotted with
open circles.  Our data measured from the H$\alpha$ profiles in our monitoring program plotted with filled circles.
\label{668:fig:vels}}
\end{figure}

\clearpage
\begin{figure}[tbp]
\begin{center}
\includegraphics[width=0.9\linewidth]{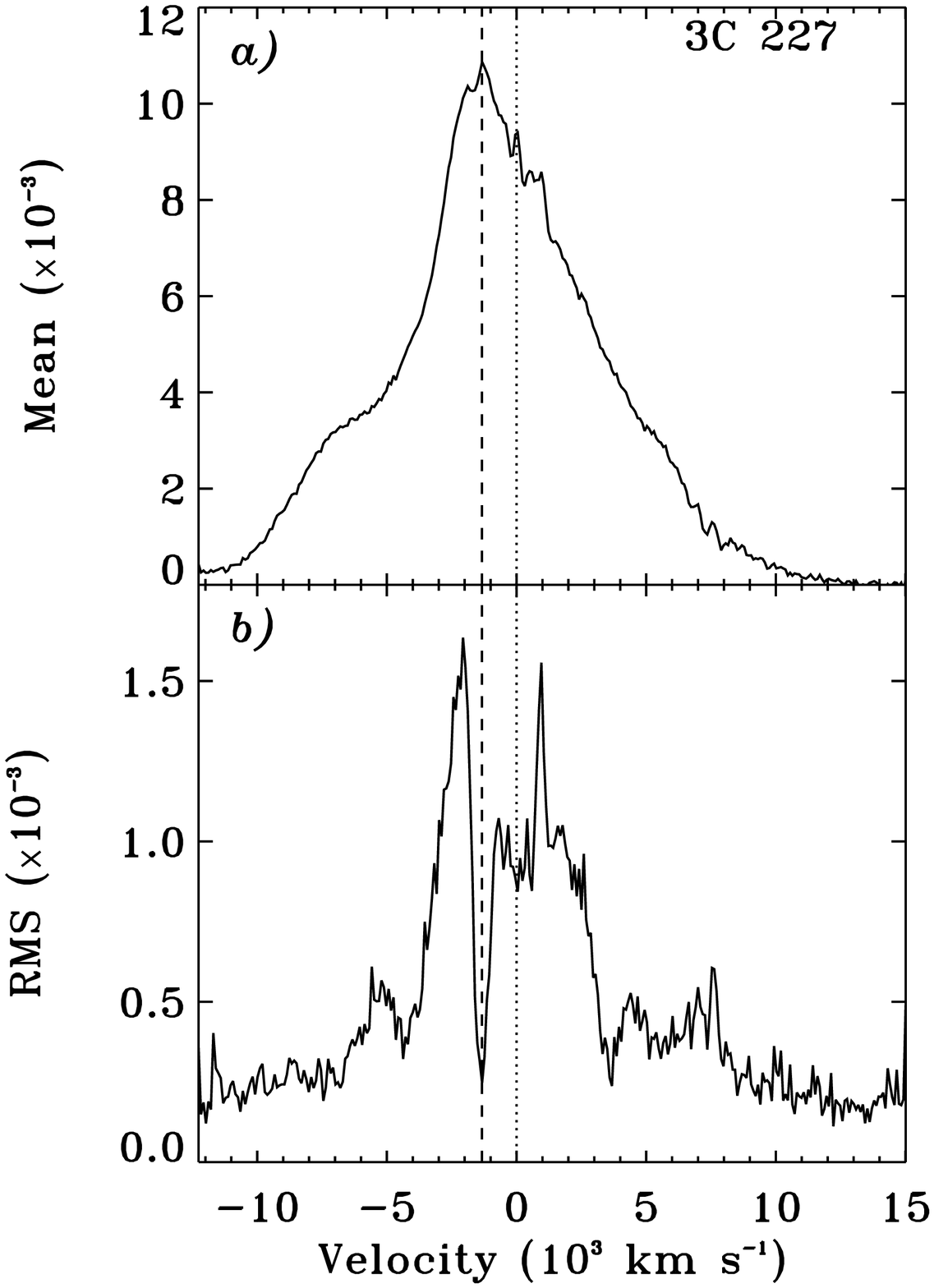}
\caption[3C 227: Mean and rms profile.]{3C 227: (a) Mean and (b) rms profile of the broad H$\alpha$ line in units of $f_{\nu}$ normalized by the total broad-line flux, with a dashed line indicating 
the velocity of the central blue peak of the mean profile, measured from the flux-weighted velocity
centroid of the top 10\% of the peak. \label{227:fig:meanrms}}
\end{center}
\end{figure}
 
\begin{figure} [tbp]\centering
\includegraphics[width=0.75\linewidth]{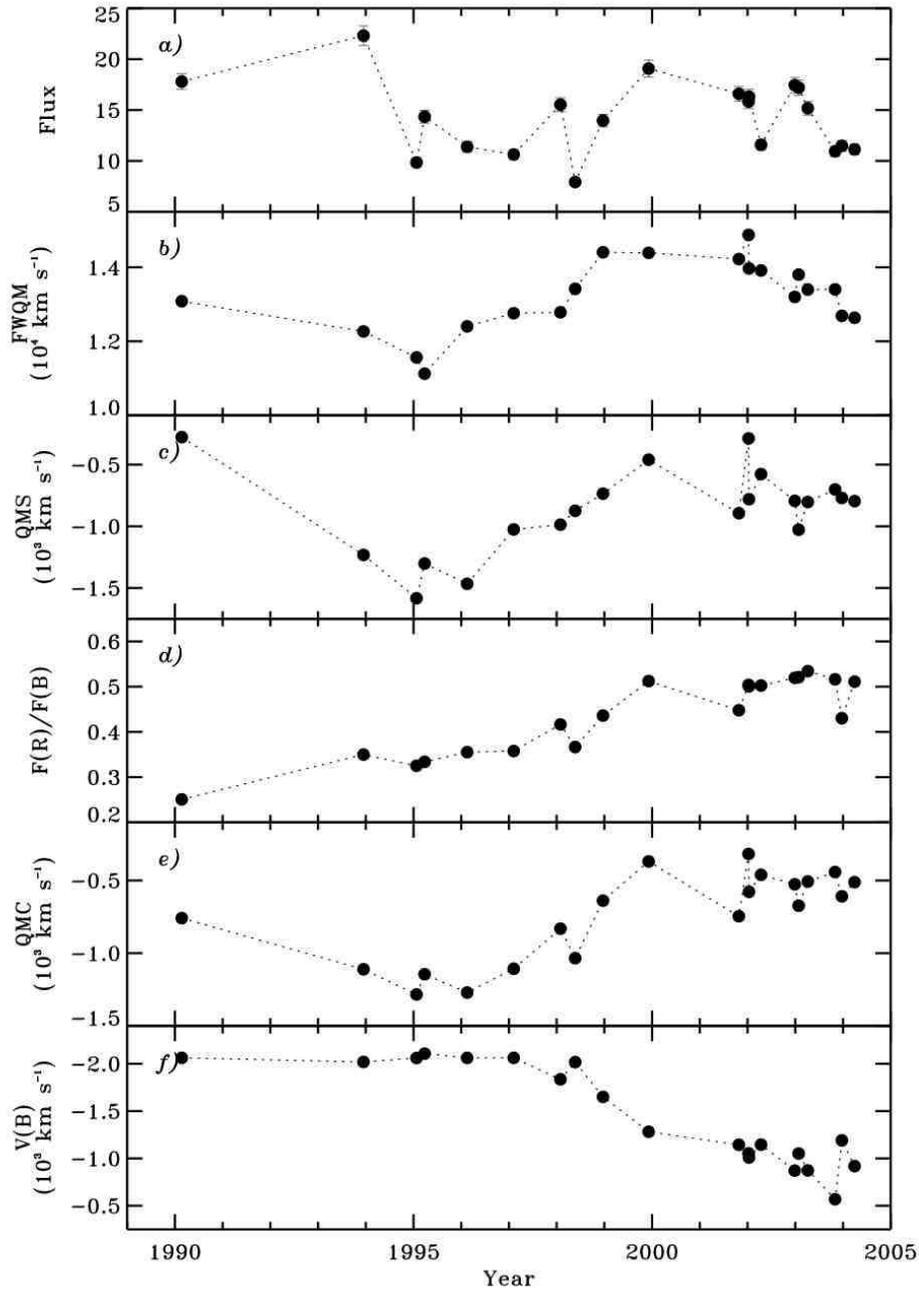}
\caption[3C 227: Characterization of the profiles.]{3C 227: 
(a) Total broad-line flux calibrated relative to the narrow H$\alpha$ line, (b) full width at quarter-maximum, (c) shift at quarter-maximum, 
(d) ratio of the integrated flux in fixed intervals in the red side (+1700 to +3400 \kms) and blue side ($-$1200 to $-$4400 \kms) of the profile, (e) flux-weighted centroid at quarter-maximum,
(f) flux-weighted velocity centroid of top 10\% of the central blue peak of the broad H$\alpha$ profiles over the entire duration of our monitoring program.
\label{227:fig:tp_227_3}}
\end{figure}

\begin{figure} [tbp]\centering
\includegraphics[width=0.9\linewidth]{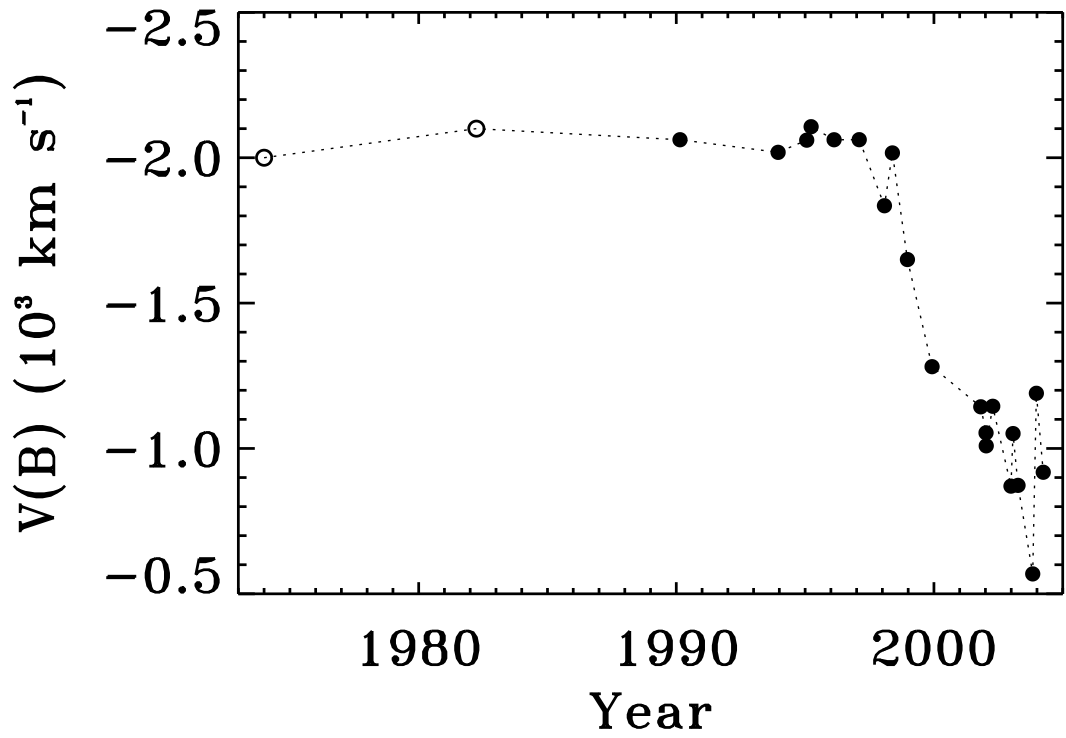}
\caption[3C 227: Historical velocity curve of blue peak.]{3C 227: Historical velocity curve of the blue peak of the 
broad H$\alpha$ line, with data in 1974 from Osterbrock, Koski, \& Phillips (1976), and in 1982 from Gaskell (1983) 
plotted with open circles, and data from our monitoring program plotted with filled circles.
\label{227:fig:vels}}
\end{figure}

\begin{figure} [tbp]\centering
\includegraphics[width=0.9\linewidth]{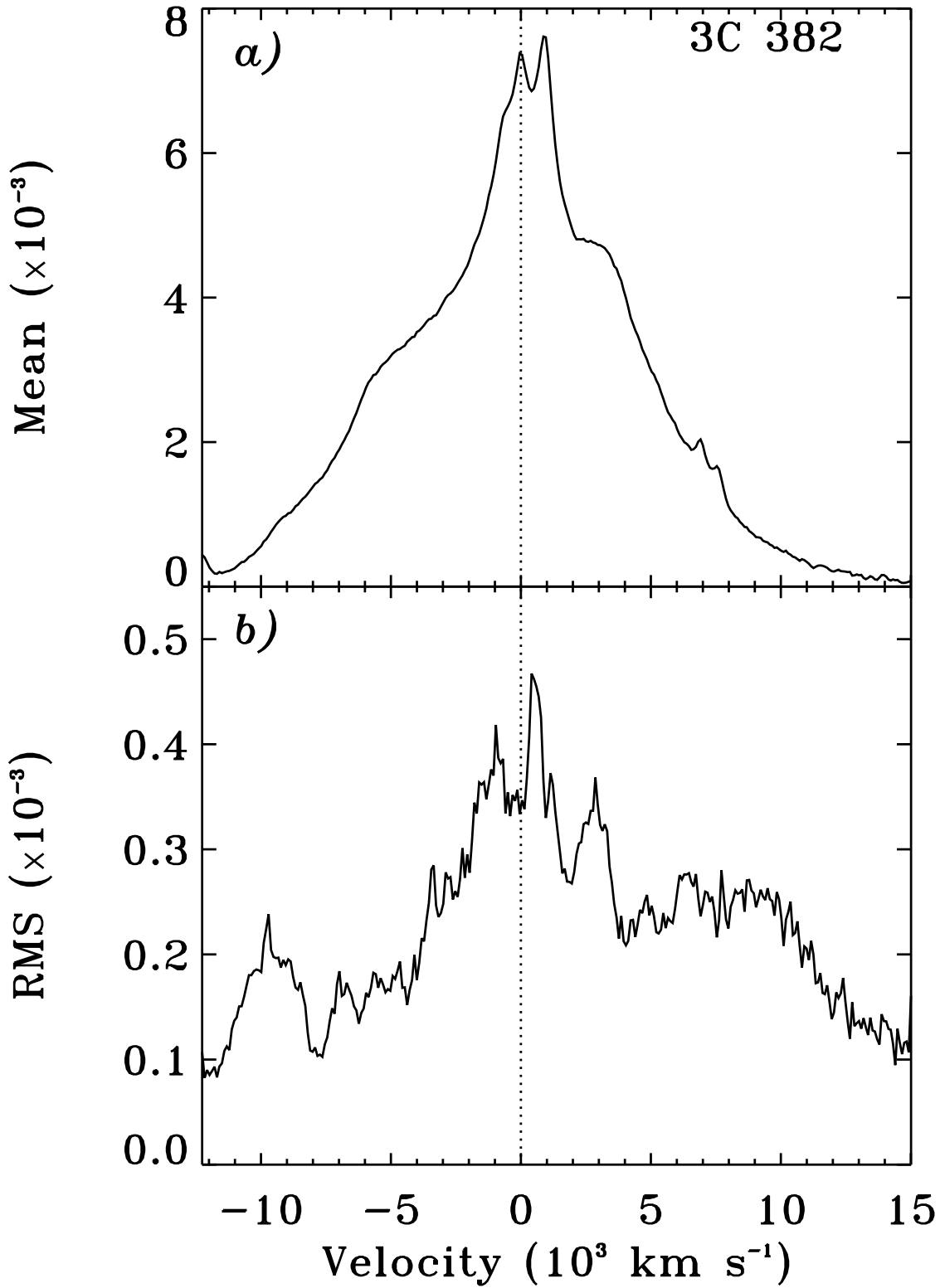}
\caption[3C 382: Mean and rms profile.]{3C 382:  (a) Mean and (b) rms profile of the broad H$\alpha$ line in units of $f_{\nu}$ normalized by the total broad-line flux in the regions of the profile not contaminated by the narrow emission lines. \label{382:fig:meanrms}}
\end{figure}

\begin{figure} [tbp]\centering
\includegraphics[width=0.9\linewidth]{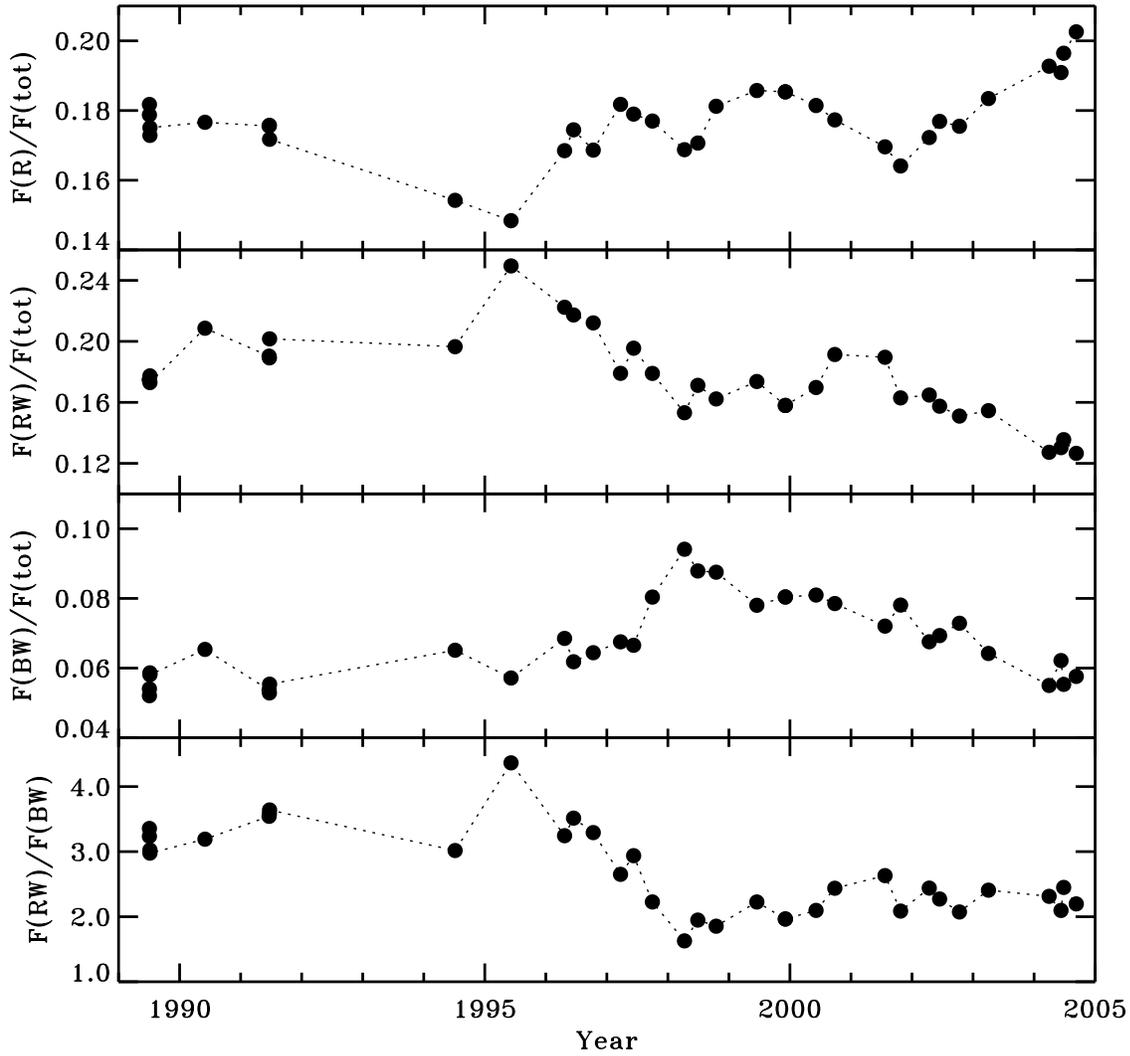}
\caption[3C 382: Normalized integrated flux in different parts of the profile.]{3C 382: Integrated flux in 
the red peak [$F$(R): +2100 to +3800 \kms], red wing [$F$(RW): +4400 to +6200, +8000 to +15,000 \kms], 
and far blue wing [$F$(BW): $-$7500 to $-$11,300 \kms] normalized by the total broad-line flux [$F$(tot)], 
and the $F$(RW)/$F$(BW) ratio of the broad H$\alpha$ profiles over the entire duration of our monitoring program.
\label{382:fig:tp_382_3}}
\end{figure}

\begin{figure} [tbp]\centering
\includegraphics[width=0.9\linewidth]{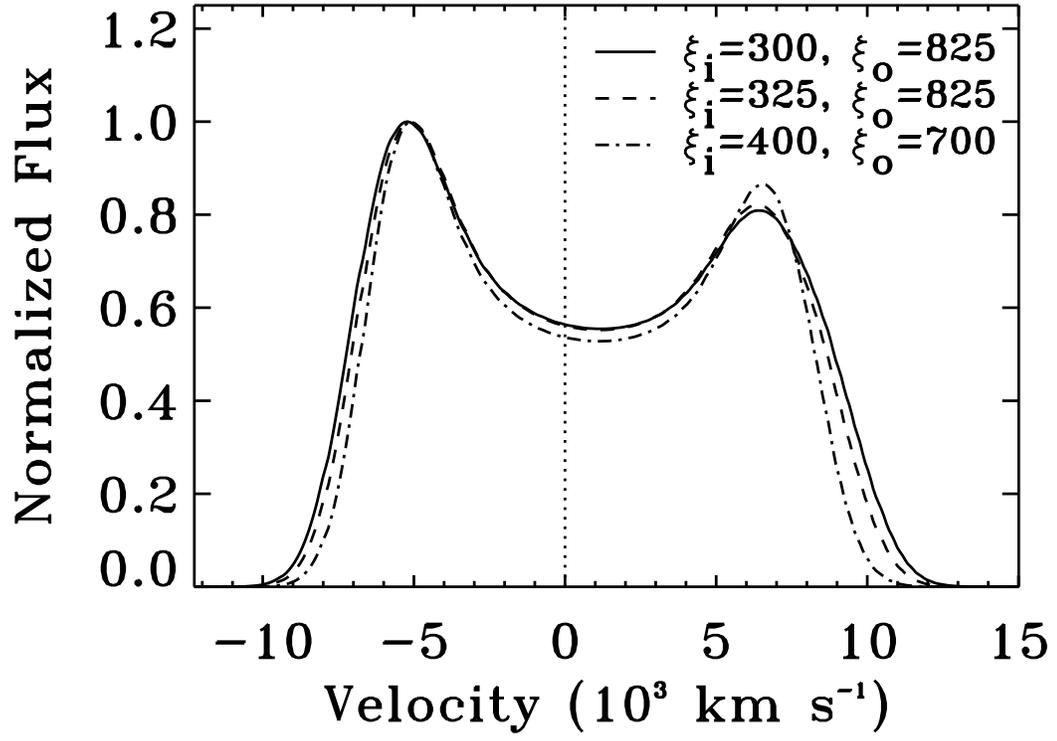}
\caption[Arp 102B: Circular disk models.]{Arp 102B: Circular disk models with changes in the
inner and outer radius of the disk ($\xi_{i}$ and $\xi_{o}$) that match the observed 
changes in FWQM of the profile, while
the other disk parameters remain fixed ($i=31\degr$,$q=3$, $\sigma=1050$ km s$^{-1}$).  The model parameters are defined in \S \ref{intro-sec2-ssec1}.
\label{arp:fig:modeldisks}}
\end{figure}

\begin{figure} [tbp]\centering
\includegraphics[width=0.9\linewidth]{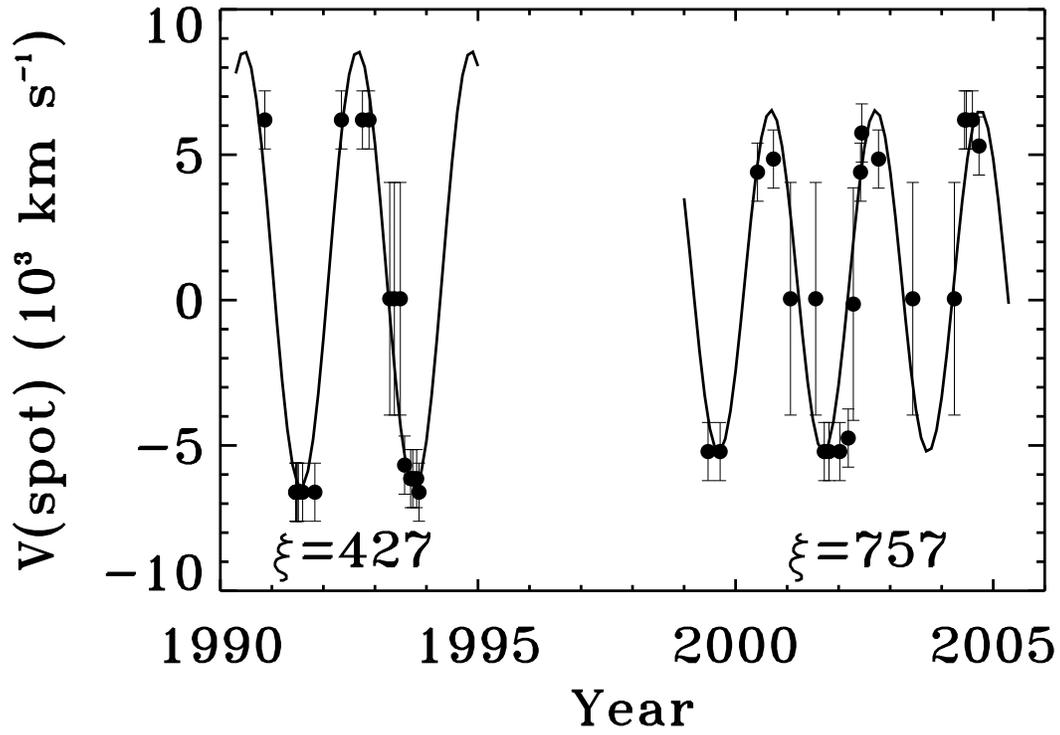}
\caption[Arp 102B: Velocity curve of the Gaussian excesses.]{Arp 102B: Velocity curve of the peaks of the model 
Gaussian excesses oscillating between 
the red and blue peaks.  Solid line shows the least-squares fit velocity curve of an orbiting bright spot for
the two episodes of oscillation.  Both episodes have a best-fit period of $\sim$ 2 yr, however the different amplitudes of their
velocity curves correspond to a bright spot orbiting at a radius of 427 \rg and 757 \rg, respectively.
\label{arp:fig:hotvel}}
\end{figure}

\begin{figure} [tbp]\centering
\includegraphics[width=0.9\linewidth]{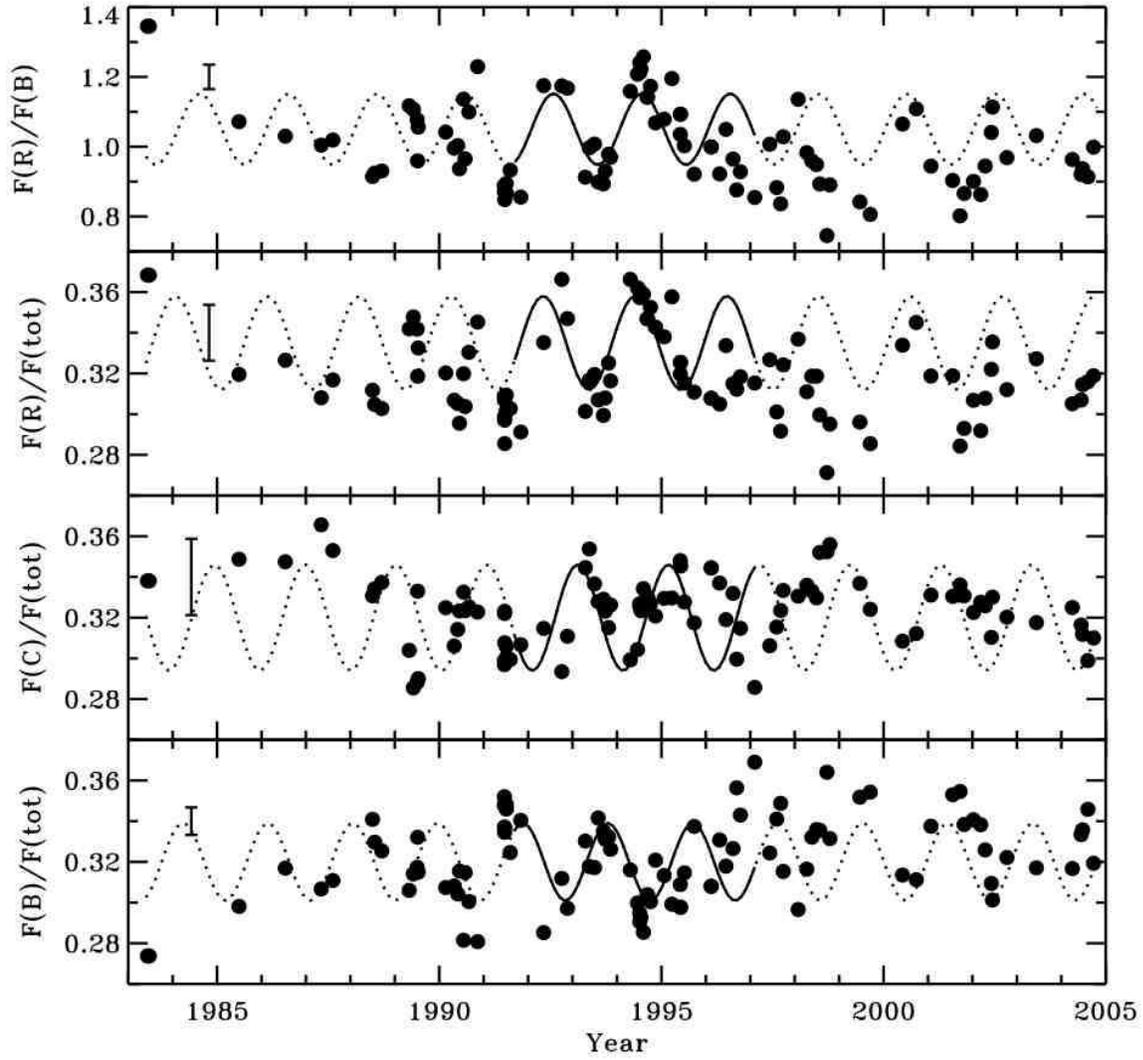}
\caption[Arp 102B: Integrated flux ratios defined by Sergeev \etal (2000).]{Arp 102B: Integrated flux ratios defined by Sergeev \etal (2000) measured for the broad H$\alpha$ profiles in our monitoring program.  
Error bars show the mean systematic error in the flux ratios due to the uncertainty in the narrow-line subtractions.  
Solid lines show the best-fit sine curves from Sergeev \etal (2000) over the time interval
from 1992 to 1995.  
Dotted lines show the sine curves
extended over the entire duration of our monitoring program.
\label{arp:fig:s00}}
\end{figure}

\begin{figure} [tbp]\centering
\includegraphics[width=0.9\linewidth]{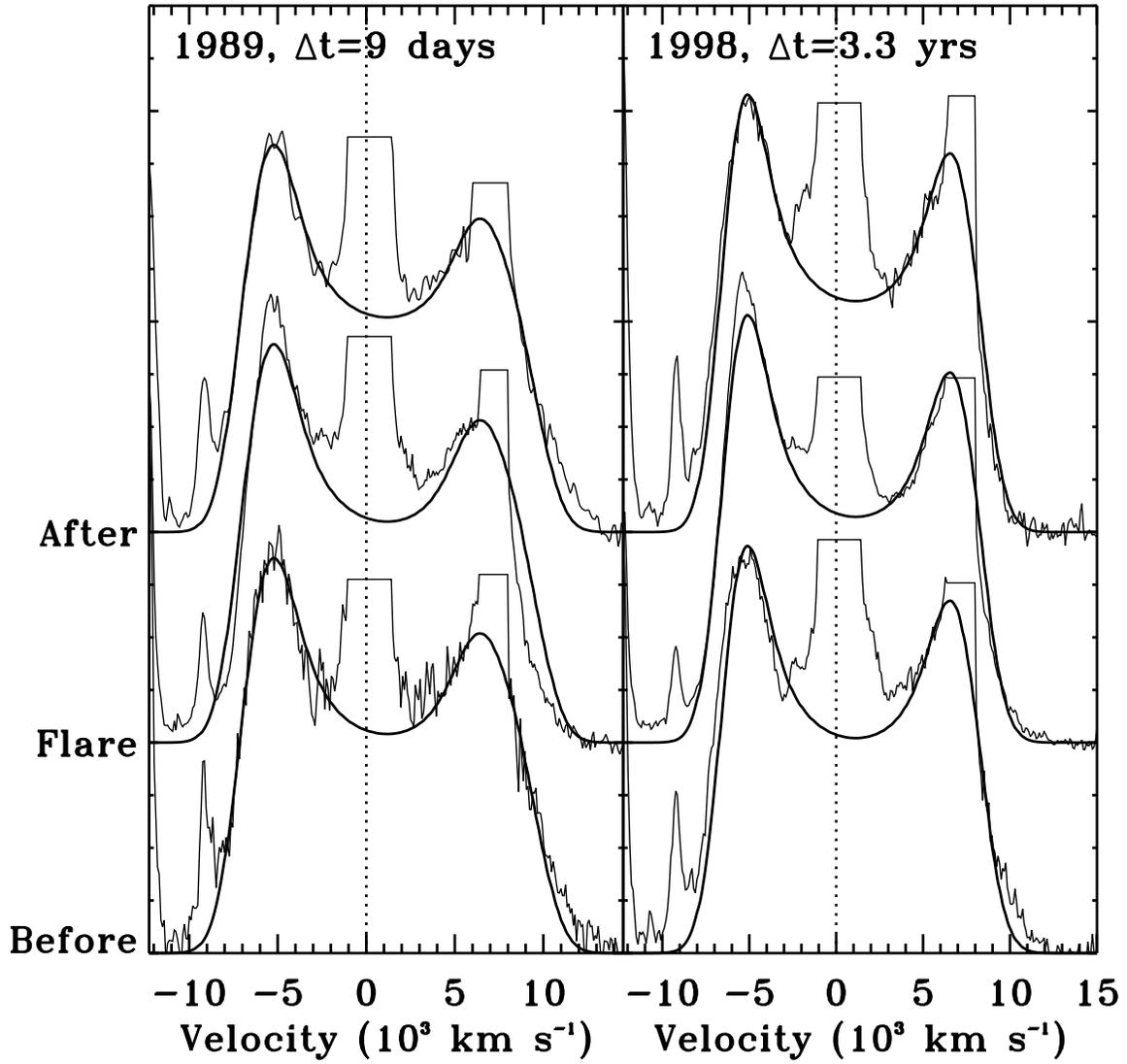}
\caption[Arp 102B: Change in the profile shape during a rapid flare in flux.]{Arp 102B: Change in shape of the profile 
during the rapid flare in the total broad-line flux in 1989 and a longer flare in 1998.  
Solid line shows the scaled disk profile fit to the broad H$\alpha$ line.\label{arp:fig:flare1}}
\end{figure}

\begin{figure} [tbp]\centering
\includegraphics[width=0.9\linewidth]{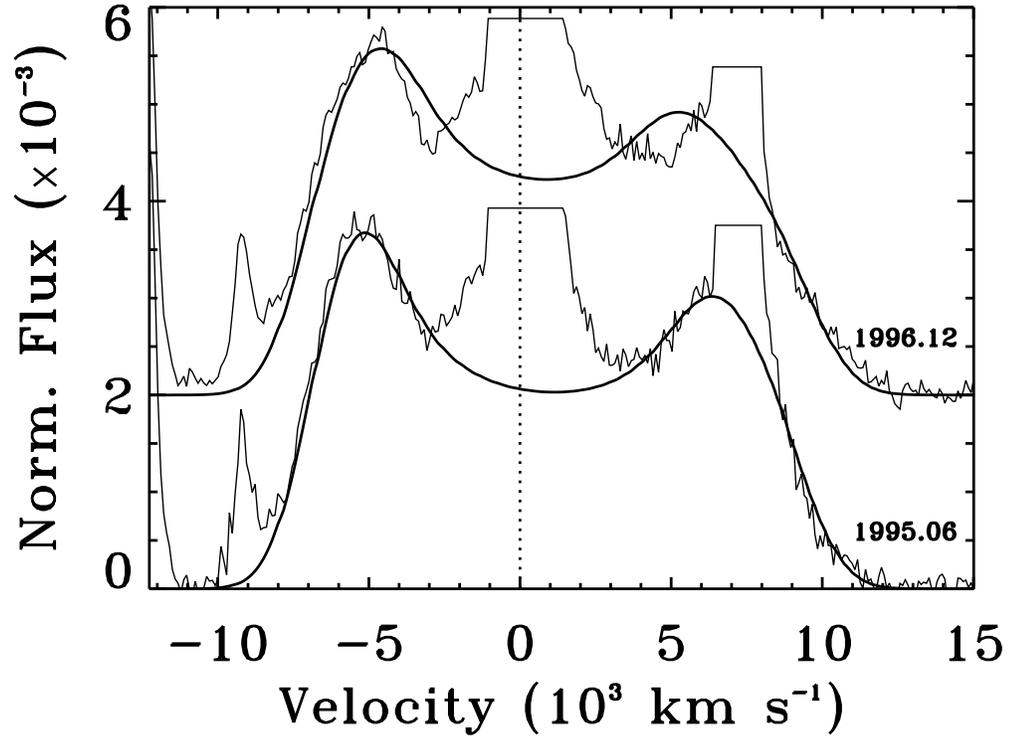}
\caption[Arp 102B: Changes in disk profile with an increase in outer radius.]{Arp 102B: Drift in the blue peak of the broad H$\alpha$ profile from 1995 to 1996 in comparison to the decrease in peak separation due to 
change of outer radius of the disk model from 825 \rg to 
1300 \rg shown with a thick solid line.  The drift
of the red peak and broadening of the peak widths in the disk model are not consistent with the spectra.
\label{arp:fig:veldrift2}}
\end{figure}

\begin{figure} [tbp]\centering
\includegraphics[width=0.9\linewidth]{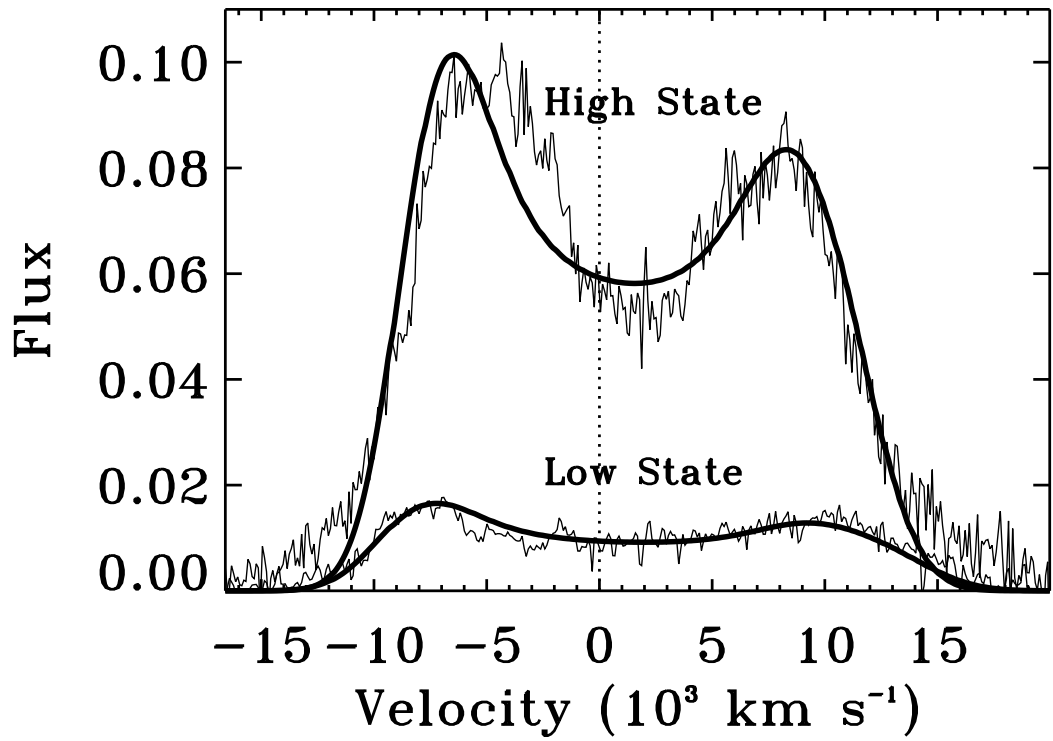}
\caption[3C 332: Model circular disk profiles.]{3C 332: Model circular disk profiles in 
comparison with the observed broad H$\alpha$ profiles, with disk parameters of $\xi_{i}$ = 190,
$\xi_{o}$ = 540 during the low state, and $\xi_{i}$ = 260, $\xi_{o}$ = 620 during the high state.
\label{332:fig:disks}}
\end{figure}

\begin{figure} [tbp]\centering
\includegraphics[width=0.9\linewidth]{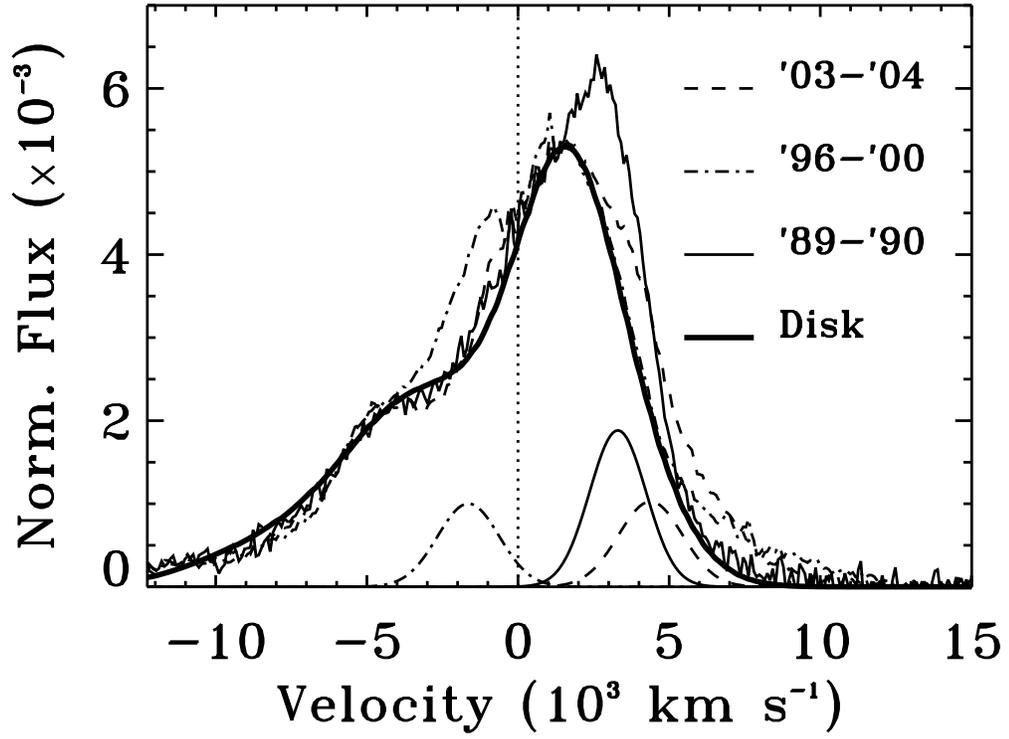}
\caption[Mkn 668: Dramatic changes in the red central peak.]{Mkn 668: Averaged broad H$\alpha$ profiles that demonstrate the dramatic changes in the red central peak.  The thick solid line
shows the eccentric disk fit to the non-varying portion of the profile.  The variations in the profile are modeled with
Gaussian excesses superposed on the eccentric disk profile, that drift from the red side to the blue side of the profile and back.  These are shown at the bottom of the figure, following the line style convention of the legend.  See also the discussion in \S \ref{ch2-sec7-ssec2-sssec2}.
\label{668:fig:spikes}}
\end{figure}

\clearpage
\begin{figure} [tbp]\centering
\includegraphics[width=0.9\linewidth]{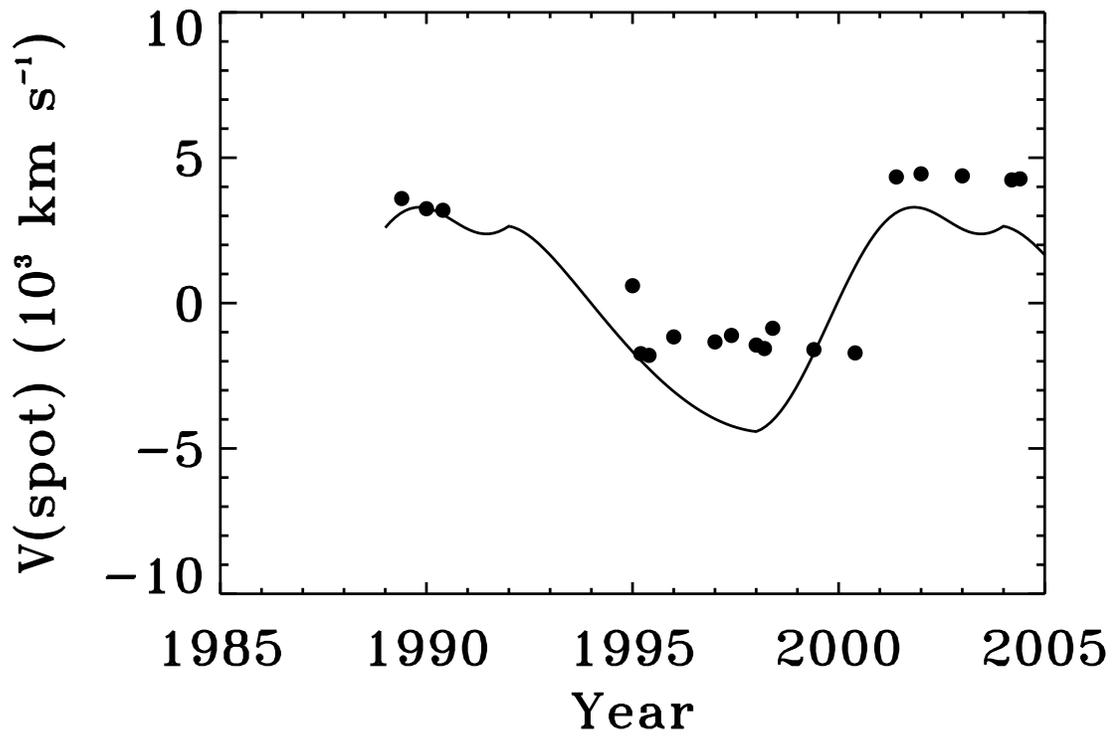}
\caption[Mkn 668: Velocity curve of Gaussian excesses.]{Mkn 668: Velocity curve of the Gaussian excesses in the eccentric disk profile, in comparison to the velocity curve
of a bright spot orbiting in the eccentric disk at a radius of 2125 \rg, and with an orbital period of 12 yr.
\label{668:fig:velspot}}
\end{figure}

\begin{figure}[tbp]\centering
\includegraphics[width=0.9\linewidth]{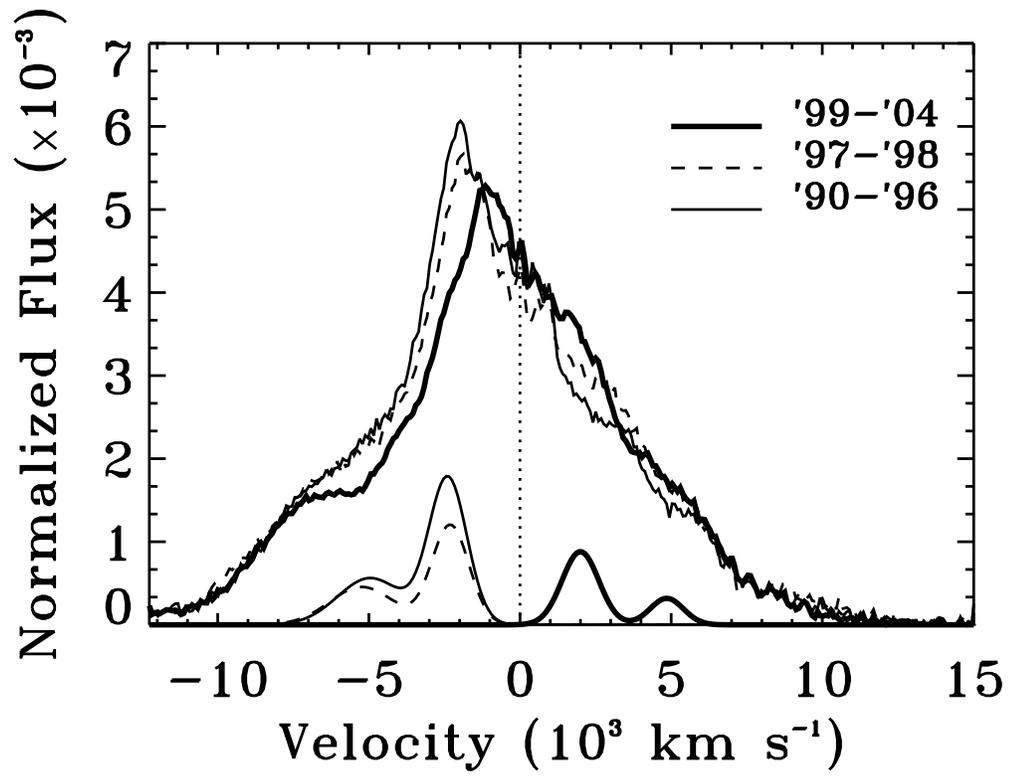}
\caption[3C 227: Averaged broad H$\alpha$ profiles that demonstrate the dramatic changes in the blue peak.]{3C 227: Averaged broad H$\alpha$ profiles that demonstrate the dramatic changes in the blue peak.  Gaussian fits to the excess that appears to traverse from the blue to the red side of the profile are also shown, following the line style convention of the legend.
\label{227:fig:spikes}}
\end{figure}

\clearpage

\appendix

In this Appendix we present the narrow-line flux calibrated spectra of the H$\alpha$ line profiles of the seven
broad-line radio galaxies in our spectroscopic monitoring program (Figures 33--52).  For all of the objects, except for 3C 382, the narrow
H$\alpha$ + [N II], [O I], and [S II] emission lines have been fitted and removed from the broad H$\alpha$ line profiles.
The subtracted narrow emission lines are shown with a dotted line, and a scaled model for the broad component of
the H$\alpha$ line is shown with a solid line.  Tick marks indicate the flux-weighted velocity centroids of the peaks
of the broad H$\alpha$ profiles.

\begin{figure} [tbp]\centering
\includegraphics[width=0.9\linewidth]{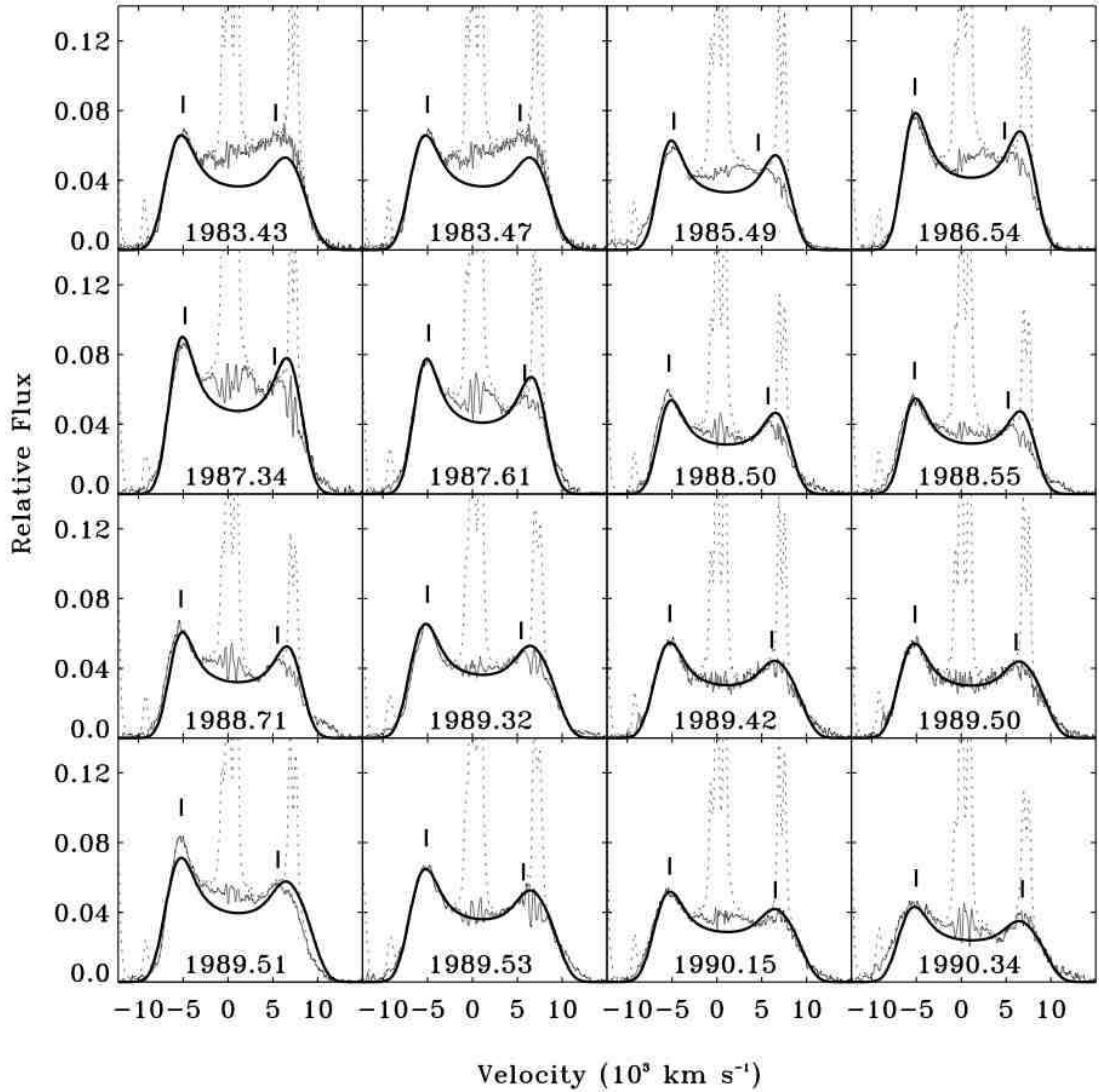}
\caption[Arp 102B: Broad H$\alpha$ profiles.]{Arp 102B: Broad H$\alpha$ profiles in units of $f_{\nu}$, flux calibrated relative to the narrow [O I] $\lambda$6300 line, with the 
subtracted narrow lines shown with dotted lines, and the best-fit scaled model disk profile plotted with a thick line.
Tick marks show the flux-weighted velocity centroid of the top 10\% of the red and blue peaks of the profile.
\label{arp:fig:pfits0}}
\end{figure}

\begin{figure} [tbp]\centering
\includegraphics[width=0.9\linewidth]{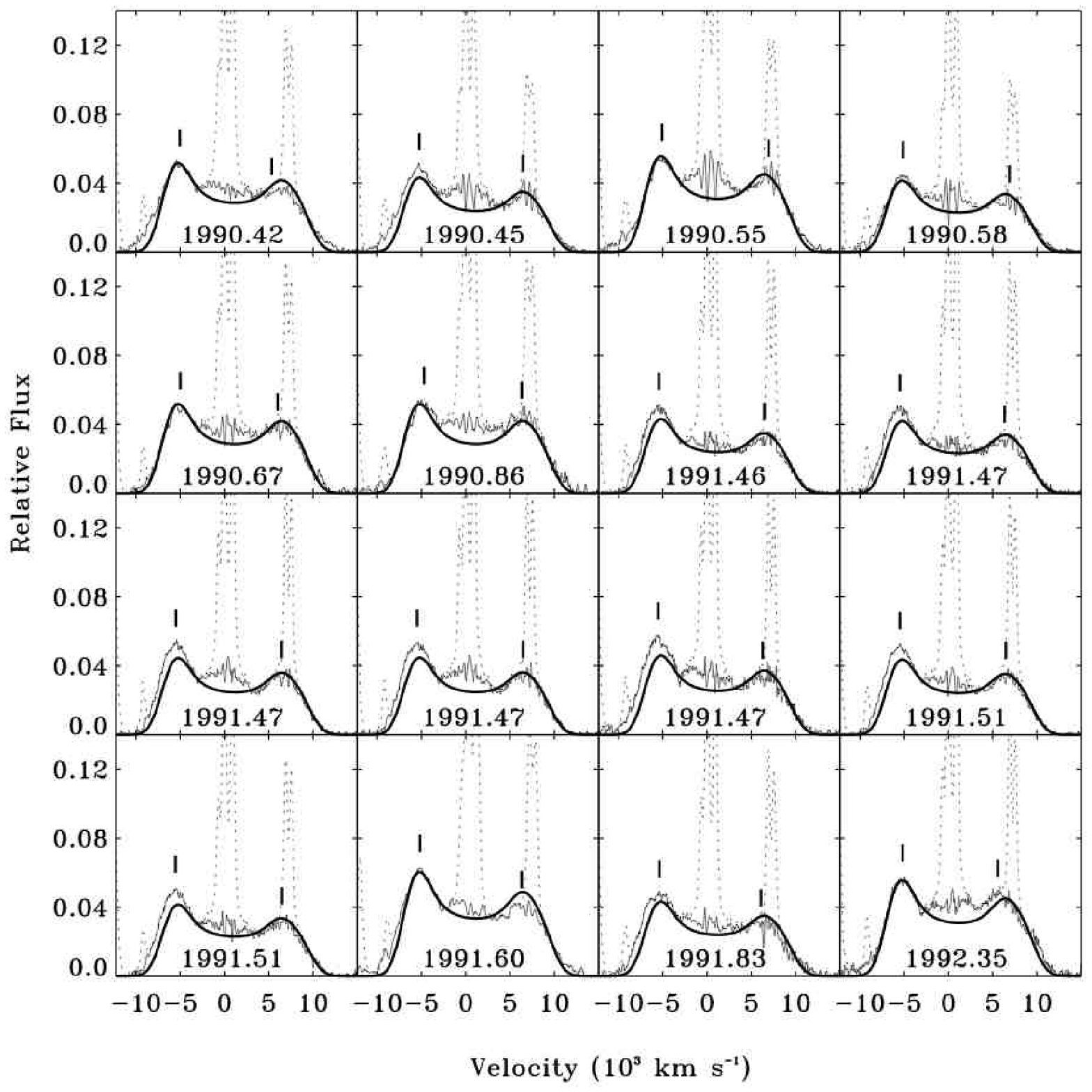}
\caption[Arp 102B: Broad H$\alpha$ profiles.]{Arp 102B: Broad H$\alpha$ profiles in units of $f_{\nu}$, flux calibrated relative to the narrow [O I] $\lambda$6300 line, with the 
subtracted narrow lines shown with dotted lines, and the best-fit scaled model circular disk profile plotted with a thick line.
Tick marks show the flux-weighted velocity centroid of the top 10\% of the red and blue peaks of the profile.
\label{arp:fig:pfits1}}
\end{figure}

\begin{figure} [tbp]\centering
\includegraphics[width=0.9\linewidth]{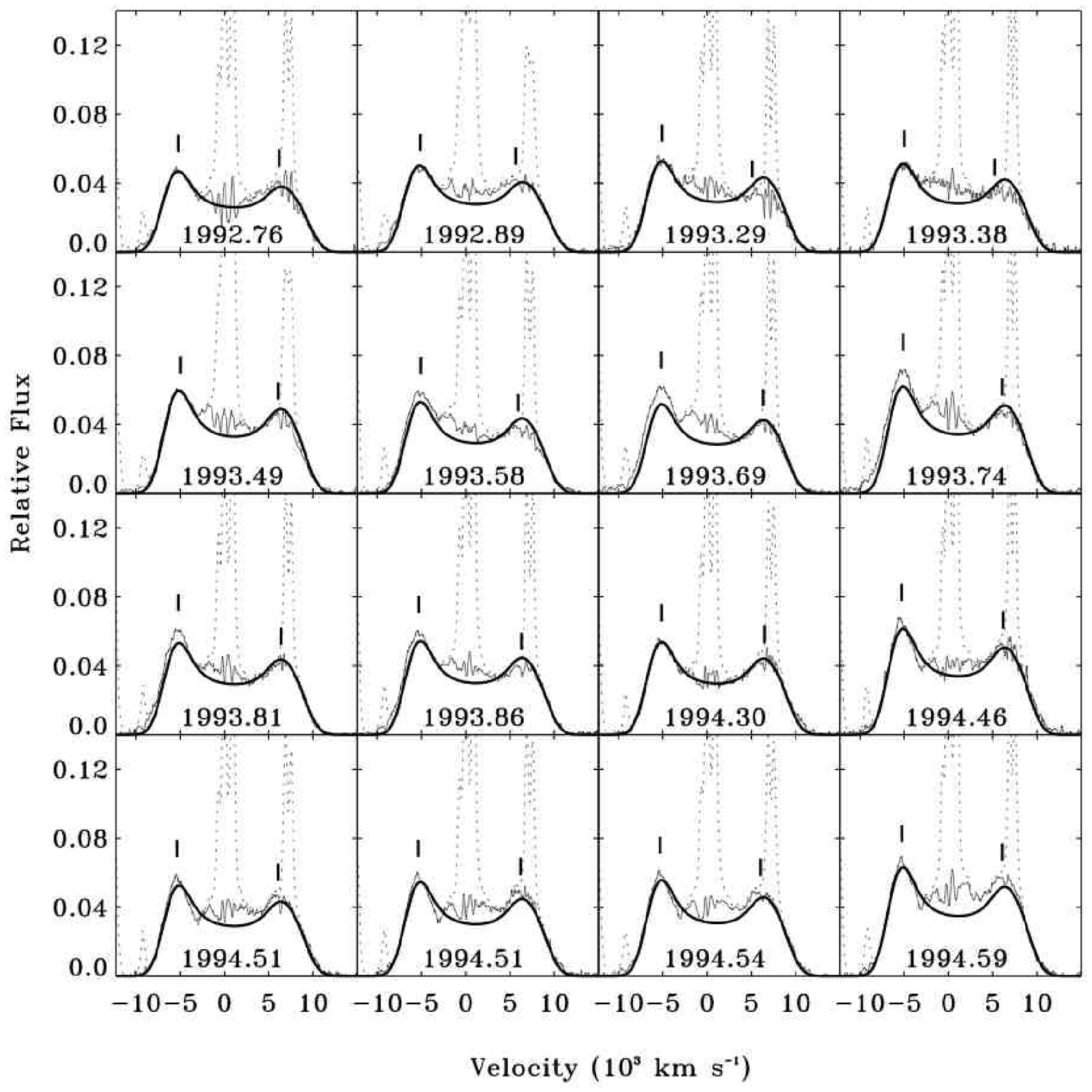}
\caption[Arp 102B: Broad H$\alpha$ profiles.]{Arp 102B: Broad H$\alpha$ profiles in units of $f_{\nu}$, flux calibrated relative to the narrow [O I] $\lambda$6300 line, with the 
subtracted narrow lines shown with dotted lines, and the best-fit scaled model disk profile plotted with a thick line.
Tick marks show the flux-weighted velocity centroid of the top 10\% of the red and blue peaks of the profile.
\label{arp:fig:pfits2}}
\end{figure}

\begin{figure} [tbp]\centering
\includegraphics[width=0.9\linewidth]{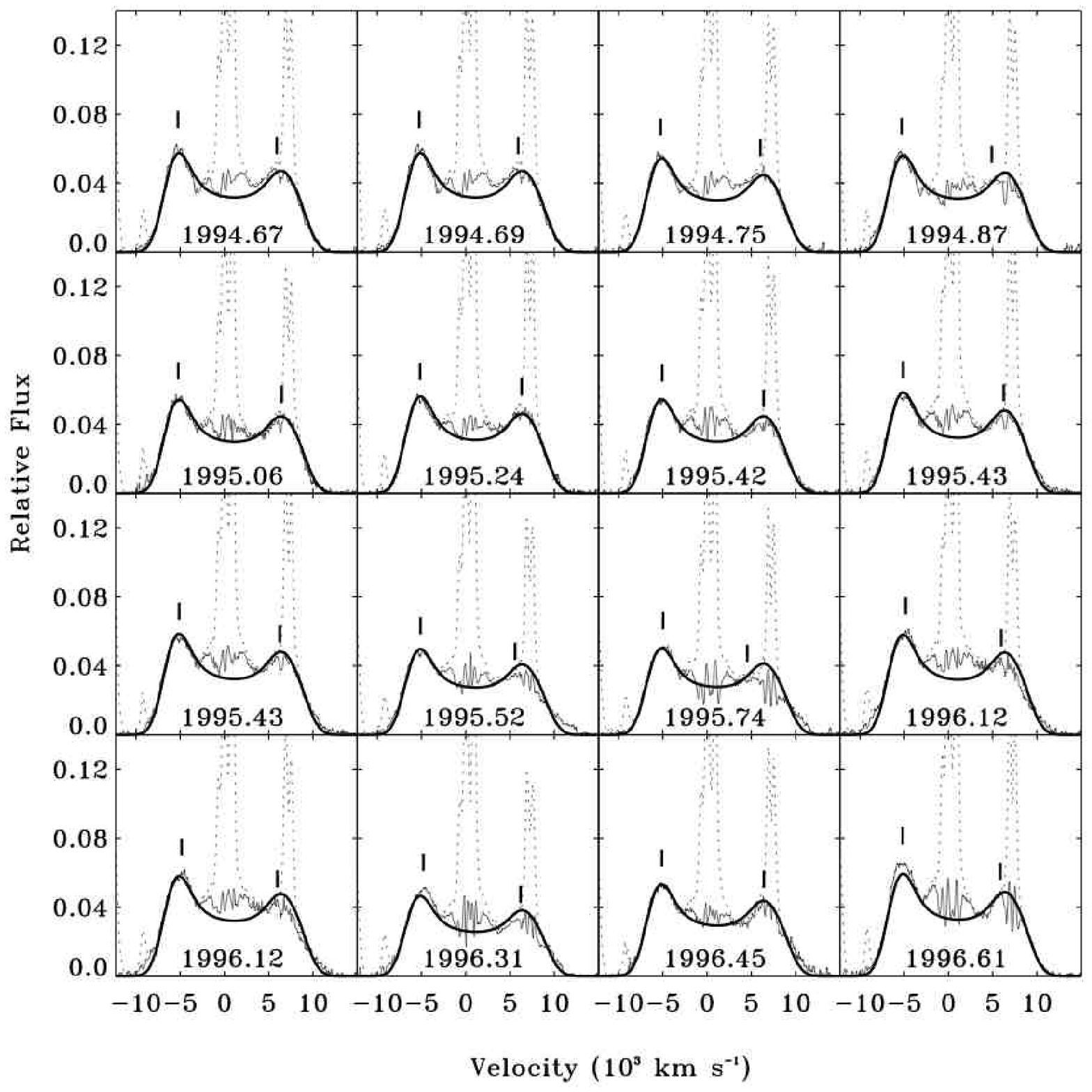}
\caption[Arp 102B: Broad H$\alpha$ profiles.]{Arp 102B: Broad H$\alpha$ profiles in units of $f_{\nu}$, flux calibrated relative to the narrow [O I] $\lambda$6300 line, with the 
subtracted narrow lines shown with dotted lines, and the best-fit scaled model disk profile plotted with a thick line.
Tick marks show the flux-weighted velocity centroid of the top 10\% of the red and blue peaks of the profile.
\label{arp:fig:pfits3}}
\end{figure}

\begin{figure} [tbp]\centering
\includegraphics[width=0.9\linewidth]{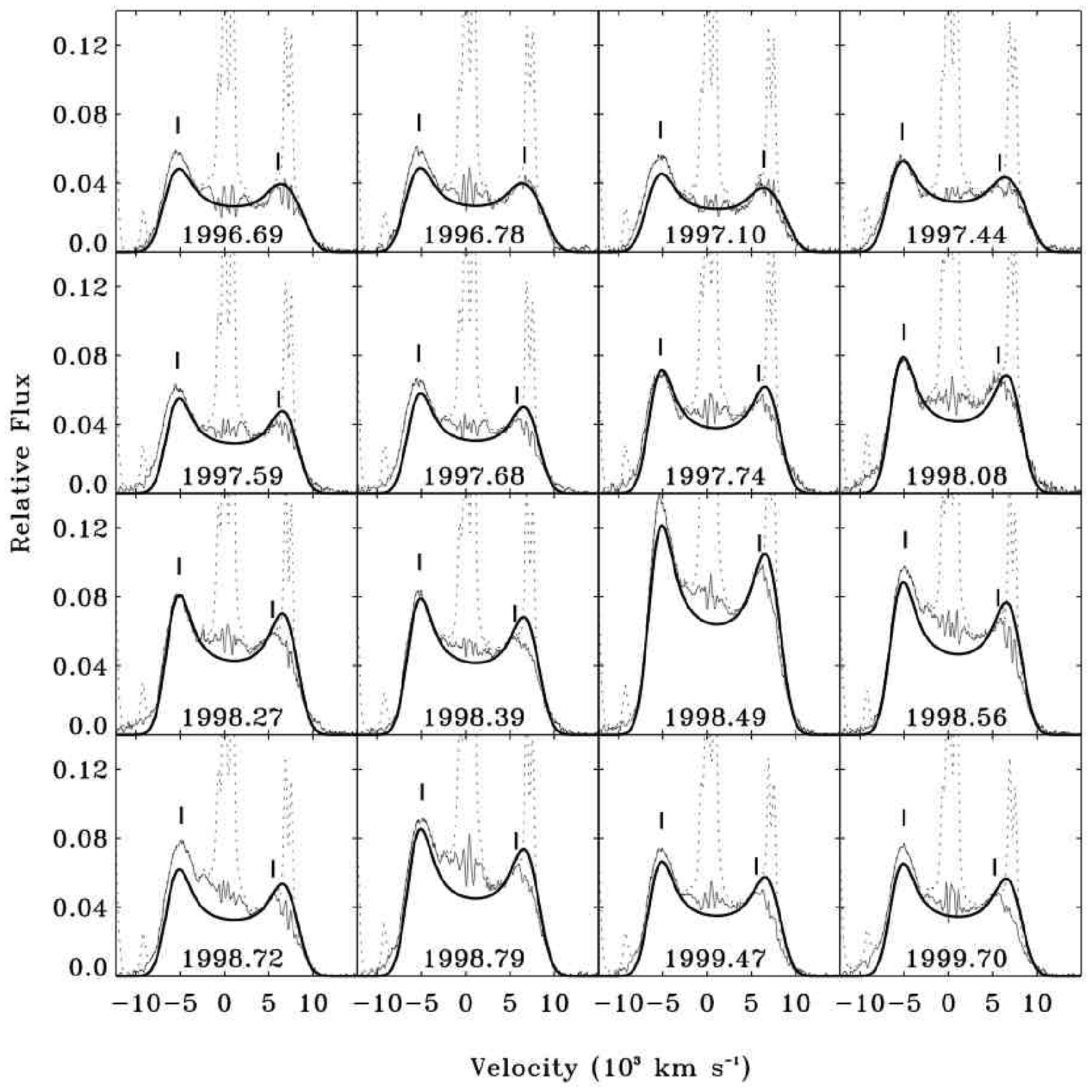}
\caption[Arp 102B: Broad H$\alpha$ profiles.]{Arp 102B: Broad H$\alpha$ profiles in units of $f_{\nu}$, flux calibrated relative to the narrow [O I] $\lambda$6300 line, with the 
subtracted narrow lines shown with dotted lines, and the best-fit scaled model disk profile plotted with a thick line.
Tick marks show the flux-weighted velocity centroid of the top 10\% of the red and blue peaks of the profile.
\label{arp:fig:pfits4}}
\end{figure}

\begin{figure} [tbp]\centering
\includegraphics[width=0.9\linewidth]{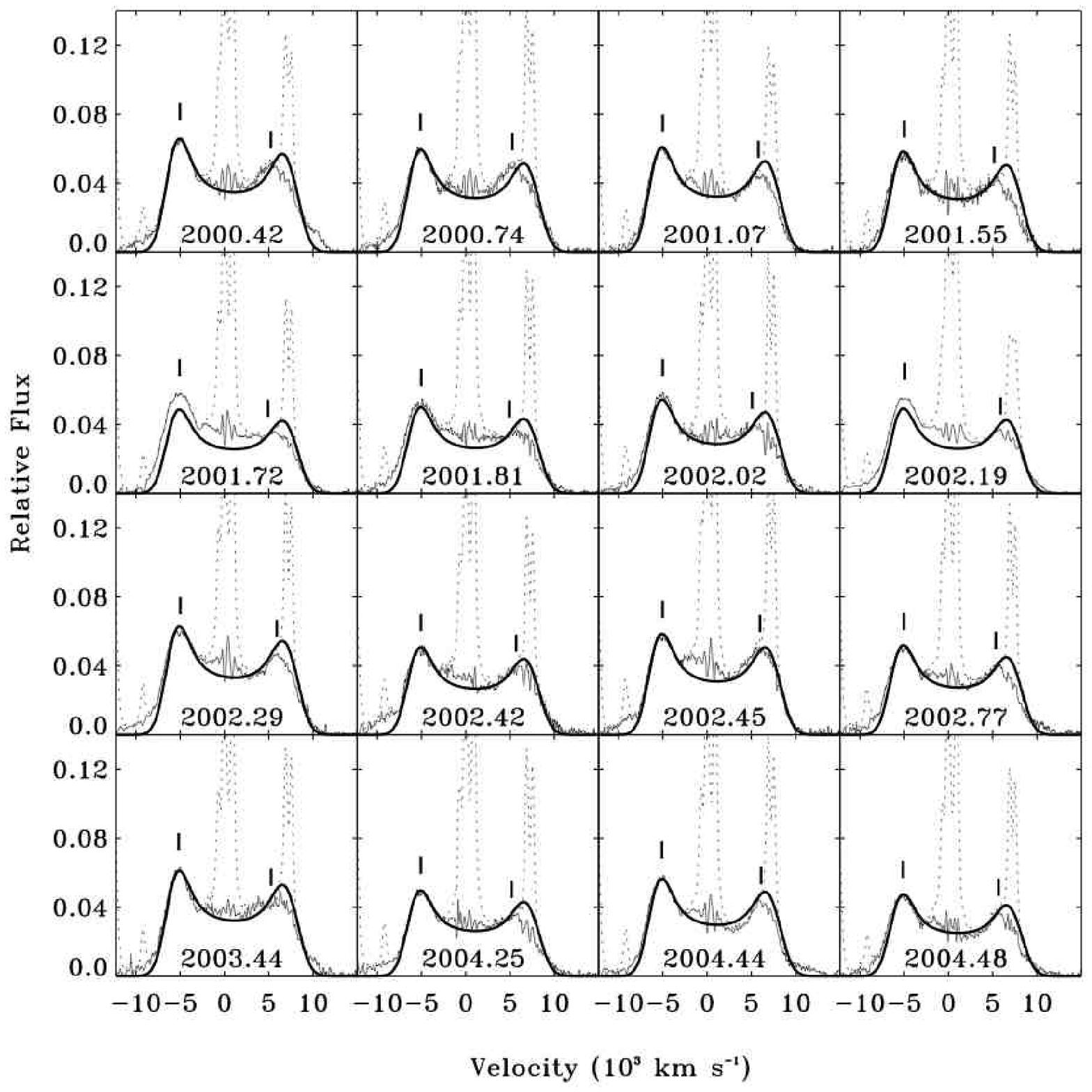}
\caption[Arp 102B: Broad H$\alpha$ profiles.]{Arp 102B: Broad H$\alpha$ profiles in units of $f_{\nu}$, flux calibrated relative to the narrow [O I] $\lambda$6300 line, with the 
subtracted narrow lines shown with dotted lines, and the best-fit scaled model disk profile plotted with a thick line.
Tick marks show the flux-weighted velocity centroid of the top 10\% of the red and blue peaks of the profile.
\label{arp:fig:pfits5}}
\end{figure}

\begin{figure} [tbp]\centering
\includegraphics[width=0.9\linewidth]{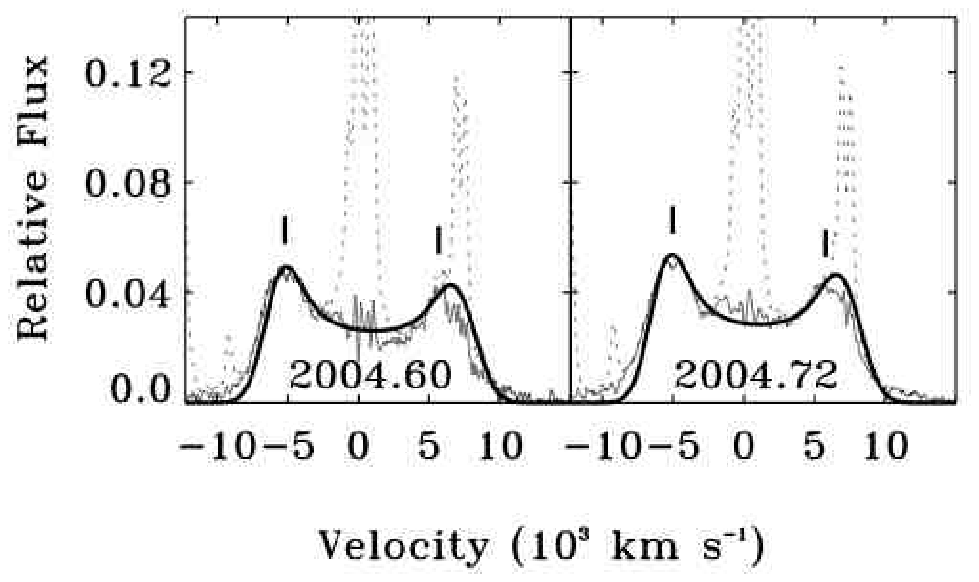}
\caption[Arp 102B: Broad H$\alpha$ profiles.]{Arp 102B: Broad H$\alpha$ profiles in units of $f_{\nu}$, flux calibrated relative to the narrow [O I] $\lambda$6300 line, with the 
subtracted narrow lines shown with dotted lines, and the best-fit scaled model disk profile plotted with a thick line.
Tick marks show the flux-weighted velocity centroid of the top 10\% of the red and blue peaks of the profile.
\label{arp:fig:pfits6}}
\end{figure}

\clearpage
 
\begin{figure} [tbp]\centering
\includegraphics[width=0.9\linewidth]{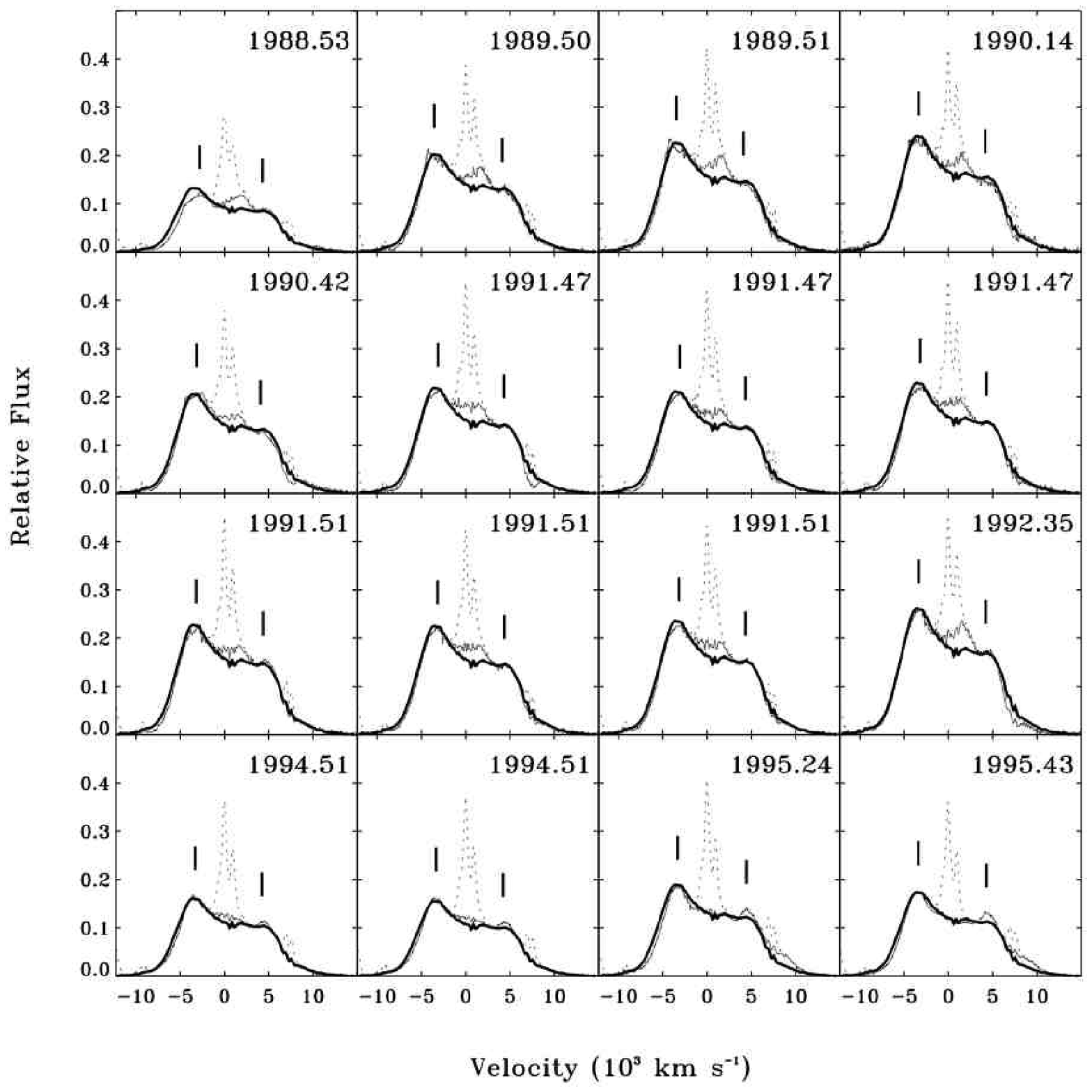}
\caption[3C 390.3: Broad H$\alpha$ profiles.]{3C 390.3: Broad H$\alpha$ profiles in units of $f_{\nu}$, flux calibrated relative to the narrow [O I] $\lambda$6300 line, with the 
subtracted narrow lines shown with dotted lines, and the scaled mean profile plotted with a thick line.
Tick marks show the flux-weighted velocity centroid of the top 10\% of the red and blue peaks of the profile.
\label{390:fig:pfits0}}
\end{figure}

\begin{figure} [tbp]\centering
\includegraphics[width=0.9\linewidth]{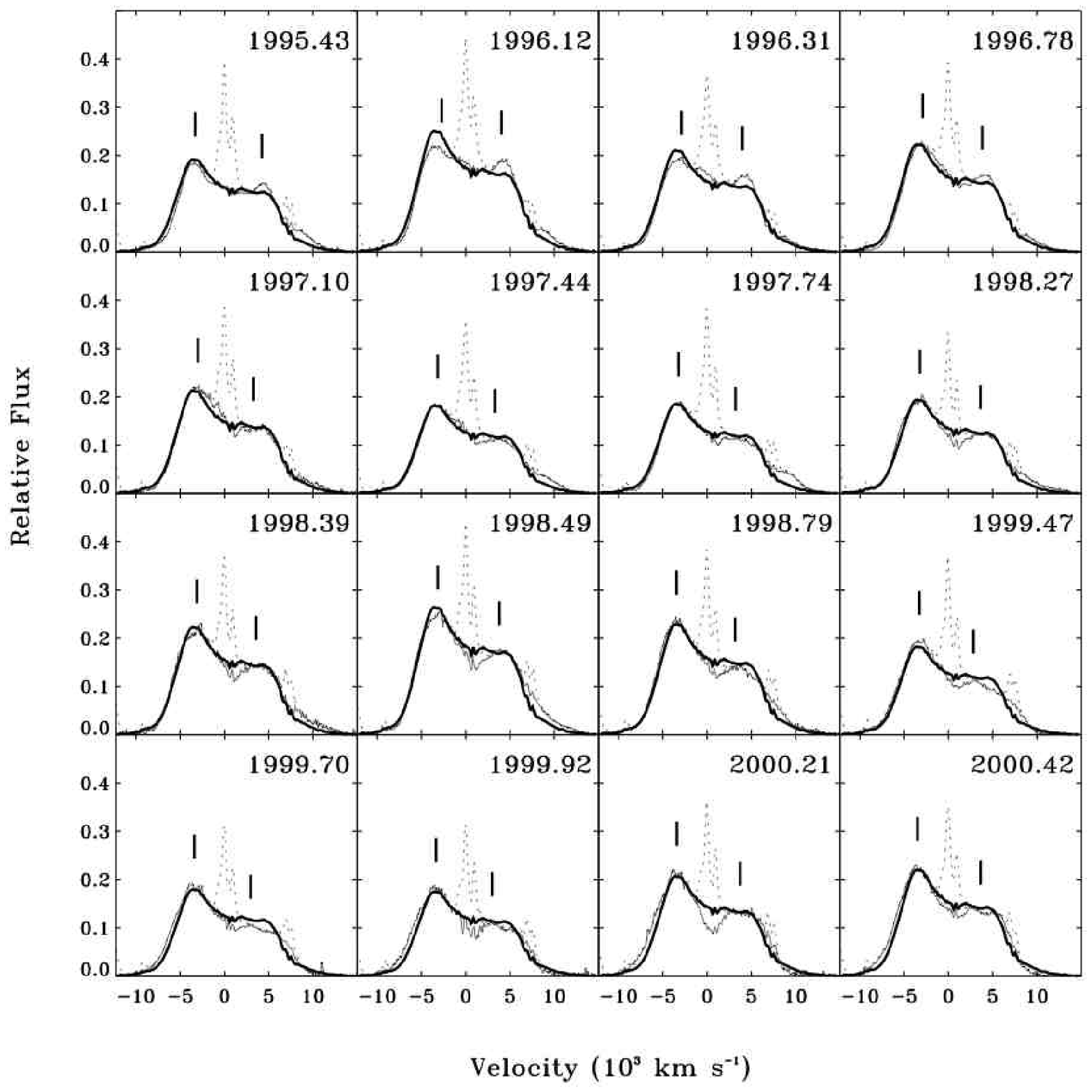}
\caption[3C 390.3: Broad H$\alpha$ profiles.]{3C 390.3: Broad H$\alpha$ profiles in units of $f_{\nu}$, flux calibrated relative to the narrow lines, with the 
subtracted narrow lines shown with dotted lines, and the scaled mean profile plotted with a thick line.
Tick marks show the flux-weighted velocity centroid of the top 10\% of the red and blue peaks of the profile.
\label{390:fig:pfits1}}
\end{figure}

\begin{figure} [tbp]\centering
\includegraphics[width=0.9\linewidth]{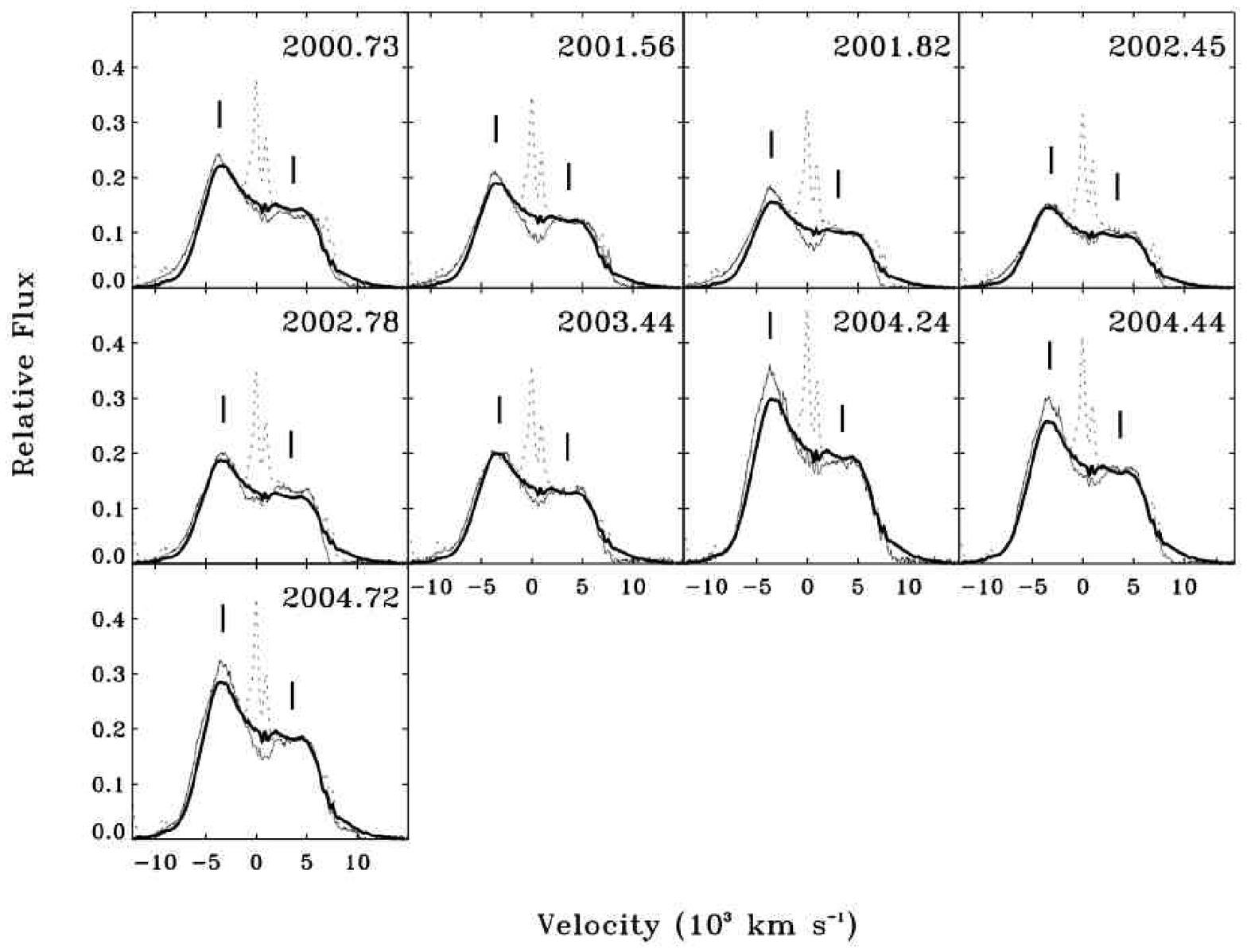}
\caption[3C 390.3: Broad H$\alpha$ profiles.]{3C 390.3: Broad H$\alpha$ profiles in units of $f_{\nu}$, flux calibrated relative to the narrow [O I] $\lambda$6300 line, with the 
subtracted narrow lines shown with dotted lines, and the scaled mean profile plotted with a thick line.
Tick marks show the flux-weighted velocity centroid of the top 10\% of the red and blue peaks of the profile.
\label{390:fig:pfits2}}
\end{figure}

\begin{figure} [tbp]\centering
\includegraphics[width=0.9\linewidth]{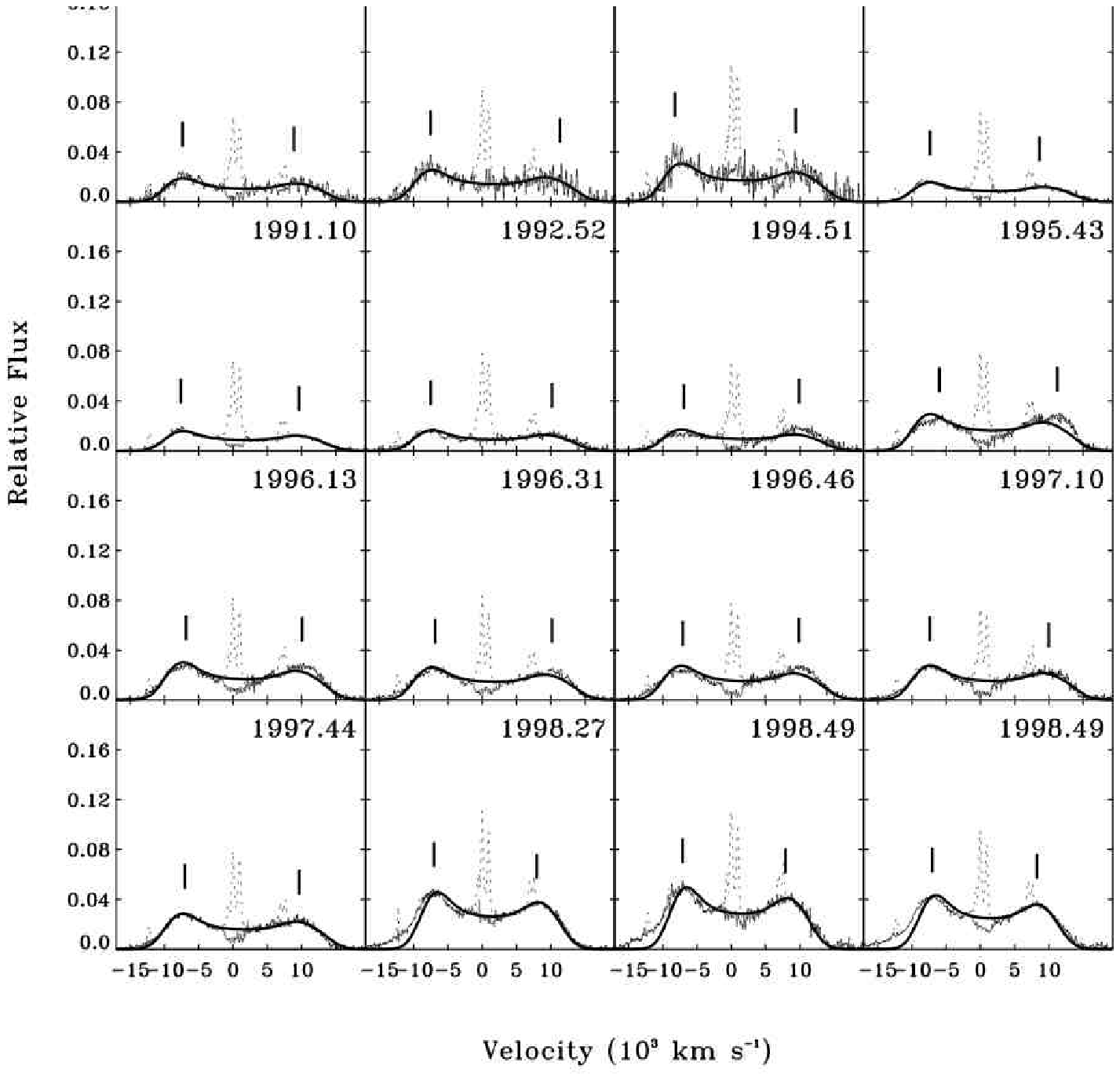}
\caption[3C 332: Broad H$\alpha$ profiles.]{3C 332: Broad H$\alpha$ profiles in units of $f_{\nu}$, flux calibrated relative to the narrow H$\alpha$ line, with the 
subtracted narrow lines shown with dotted lines, and the best-fit scaled model circular disk profile plotted with a thick line.
Tick marks show the flux-weighted velocity centroid of the top 10\% of the red and blue peaks of the profile.
\label{332:fig:pfits0}}
\end{figure}

\begin{figure} [tbp]\centering
\includegraphics[width=0.9\linewidth]{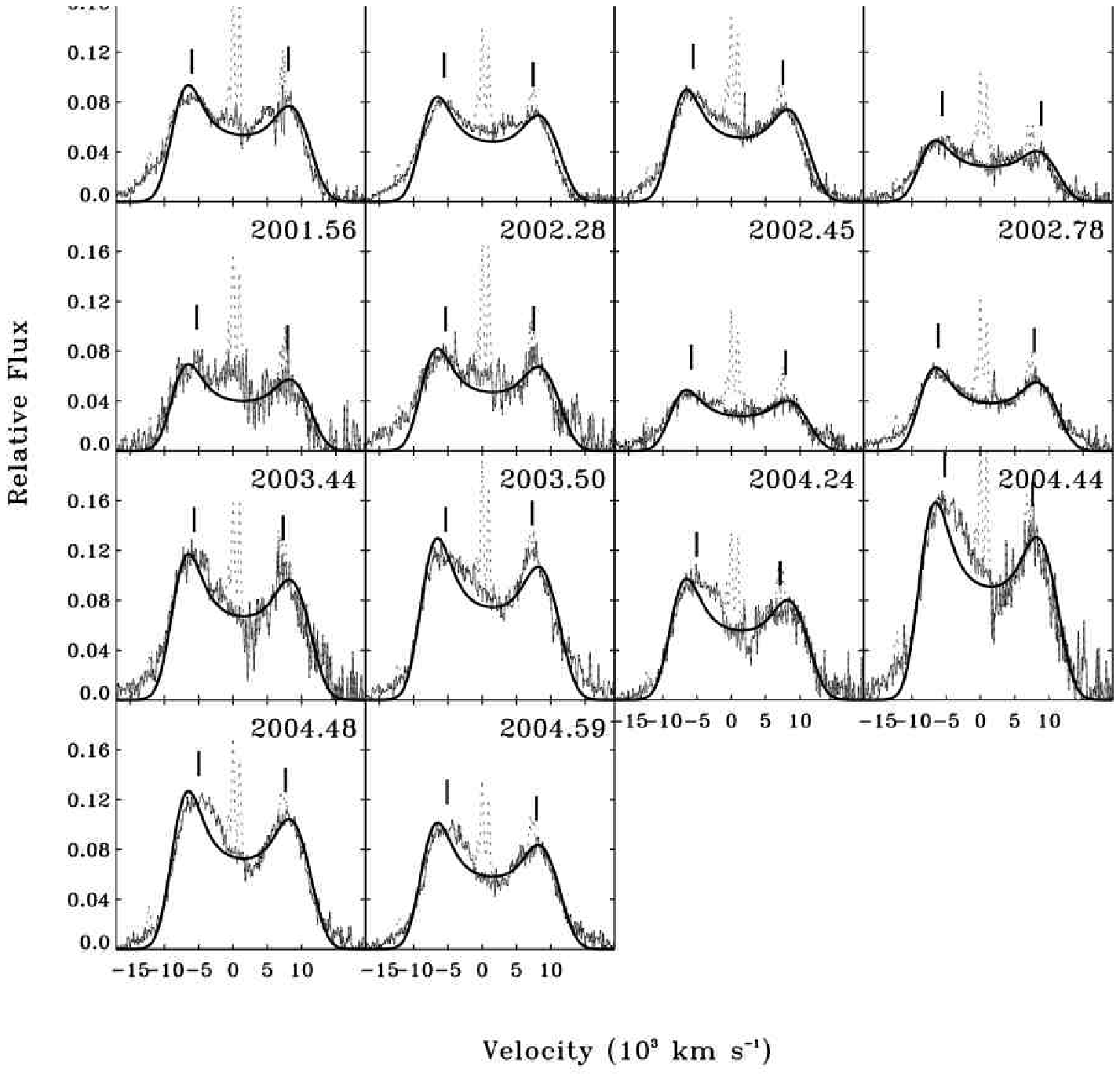}
\caption[3C 332: Broad H$\alpha$ profiles.]{3C 332: Broad H$\alpha$ profiles in units of $f_{\nu}$, flux calibrated relative to the narrow H$\alpha$ line, with the 
subtracted narrow lines shown with dotted lines, and the best-fit scaled model circular disk profile plotted with a thick line.
Tick marks show the flux-weighted velocity centroid of the top 10\% of the red and blue peaks of the profile.
\label{332:fig:pfits1}}
\end{figure}

\begin{figure} [tbp]\centering
\includegraphics[width=0.9\linewidth]{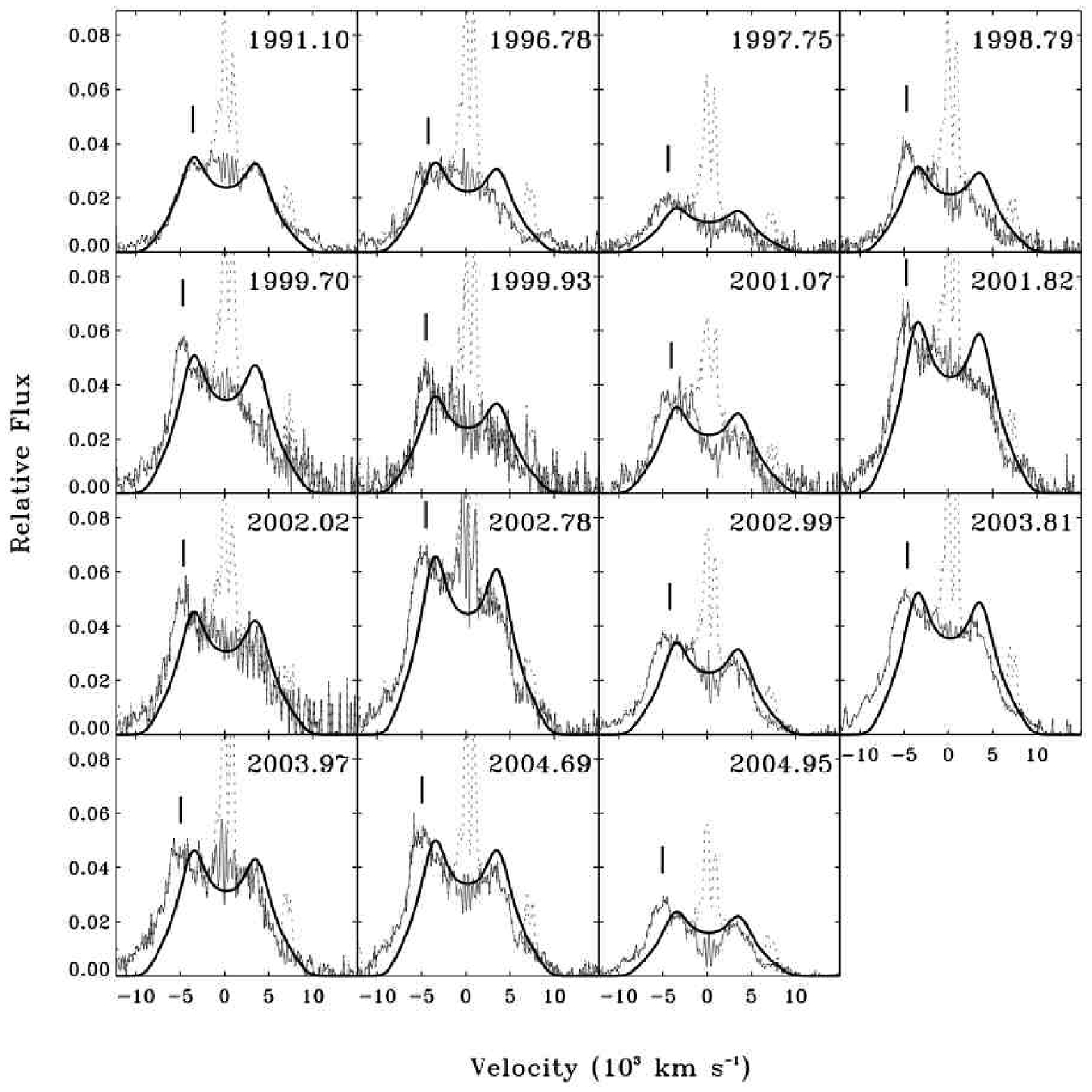}
\caption[PKS 0235+023: Broad H$\alpha$ profiles.]{PKS 0235+023: Broad H$\alpha$ profiles in units of $f_{\nu}$, flux calibrated relative to the narrow H$\alpha$ line, with the 
subtracted narrow lines shown with dotted lines, and the best-fit scaled model circular disk profile plotted with a thick line.
Tick marks show the flux-weighted velocity centroid of the top 10\% of the red and blue peaks of the profile.
\label{pks:fig:pfits_0}}
\end{figure}

\begin{figure} [tbp]\centering
\includegraphics[width=0.9\linewidth]{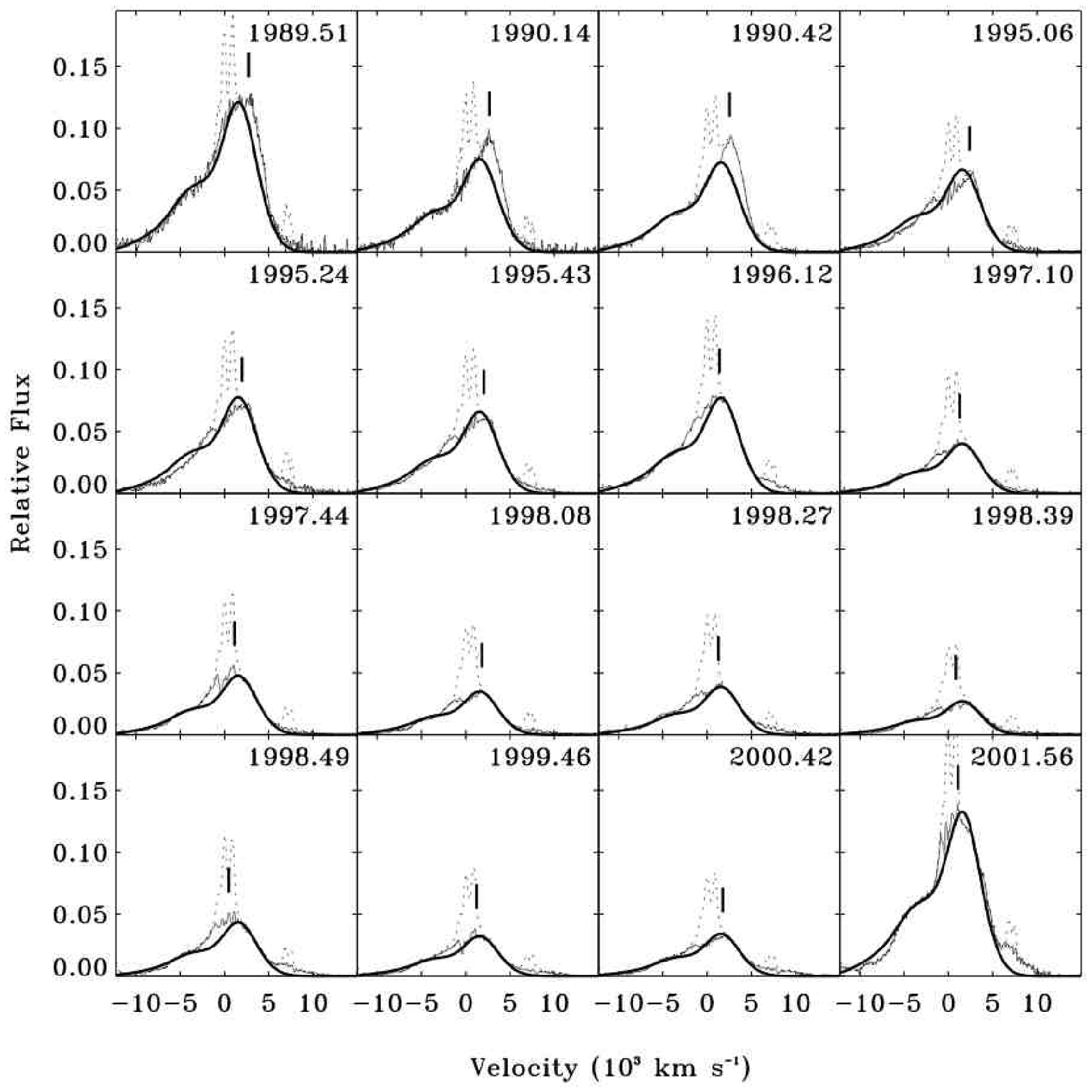}
\caption[Mkn 668: Broad H$\alpha$ profiles.]{Mkn 668: Broad H$\alpha$ profiles in units of $f_{\nu}$, flux calibrated relative to the narrow H$\alpha$ line, with the 
subtracted narrow lines shown with dotted lines, and the best-fit scaled model eccentric disk profile plotted with a thick line.
Tick marks show the flux-weighted velocity centroid of the top 10\% of the central red peak of the profile.
\label{668:fig:pfits_0}}
\end{figure}

\begin{figure} [tbp]\centering
\includegraphics[width=0.9\linewidth]{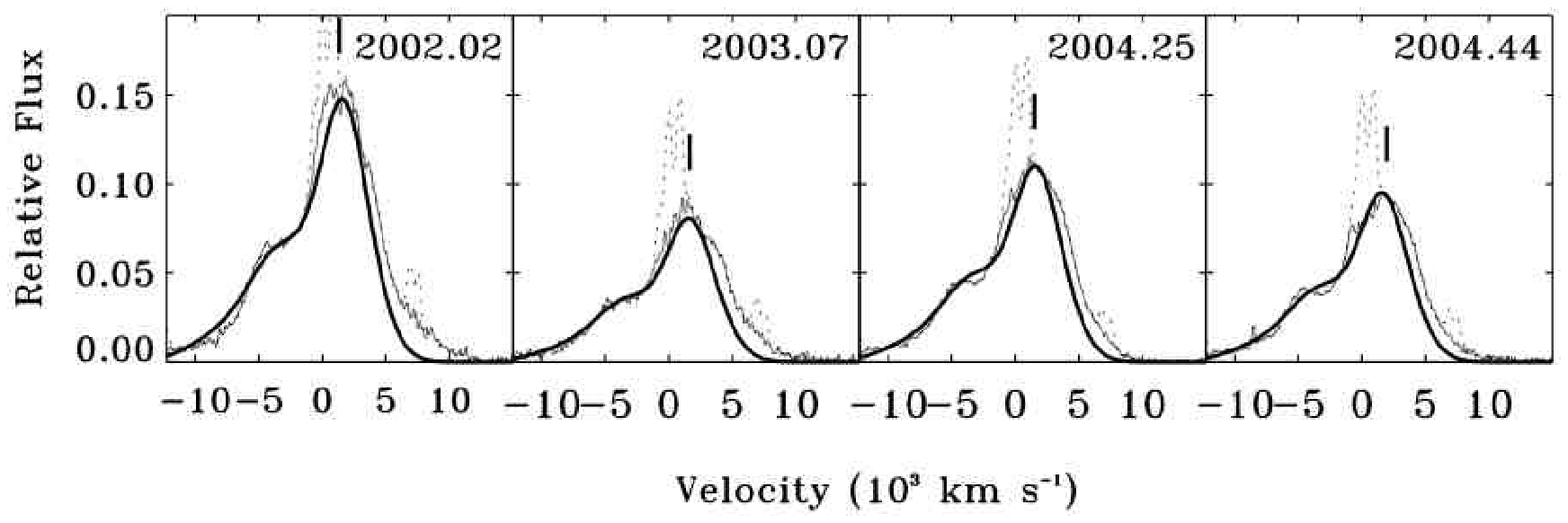}
\caption[Mkn 668: Broad H$\alpha$ profiles.]{Mkn 668: Broad H$\alpha$ profiles in units of $f_{\nu}$, flux calibrated relative to the narrow H$\alpha$ line, with the 
subtracted narrow lines shown with dotted lines, and the best-fit scaled model eccentric disk profile plotted with a thick line.
Tick marks show the flux-weighted velocity centroid of the top 10\% of the central red peak of the profile.
\label{668:fig:pfits_1}}
\end{figure}

\clearpage
 
\begin{figure} [tbp]\centering
\includegraphics[width=0.9\linewidth]{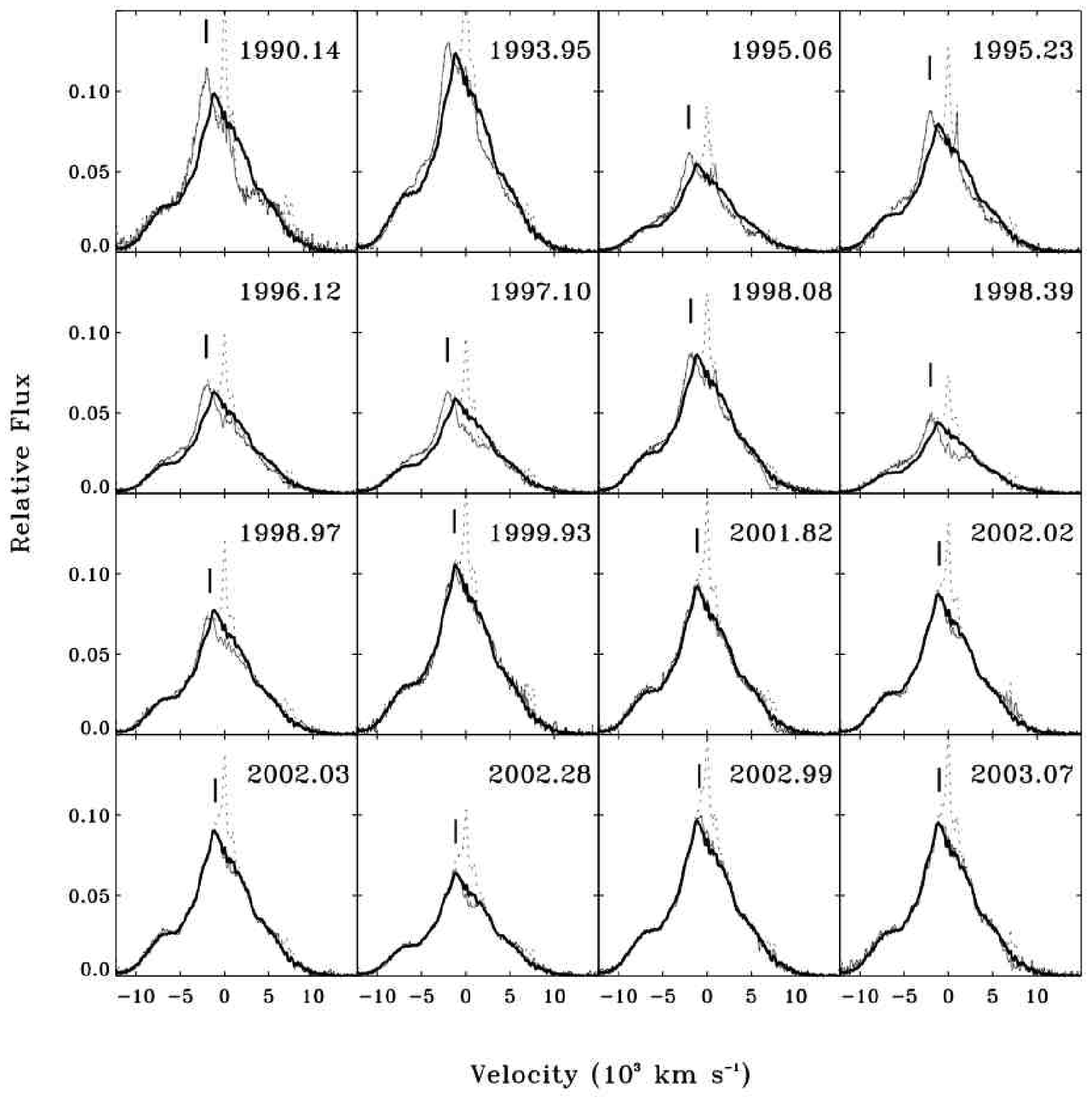}
\caption[3C 227: Broad H$\alpha$ profiles.]{3C 227: Broad H$\alpha$ profiles in units of $f_{\nu}$, flux calibrated relative to the narrow H$\alpha$ line, with the 
subtracted narrow lines shown with dotted lines, and the scaled mean profile from 1999-2004
plotted plotted with a thick line.
Tick marks show the flux-weighted velocity centroid of the top 10\% of the central blue peak of the profile.
\label{227:fig:pfits_0}}
\end{figure}

\begin{figure} [tbp]\centering
\includegraphics[width=0.9\linewidth]{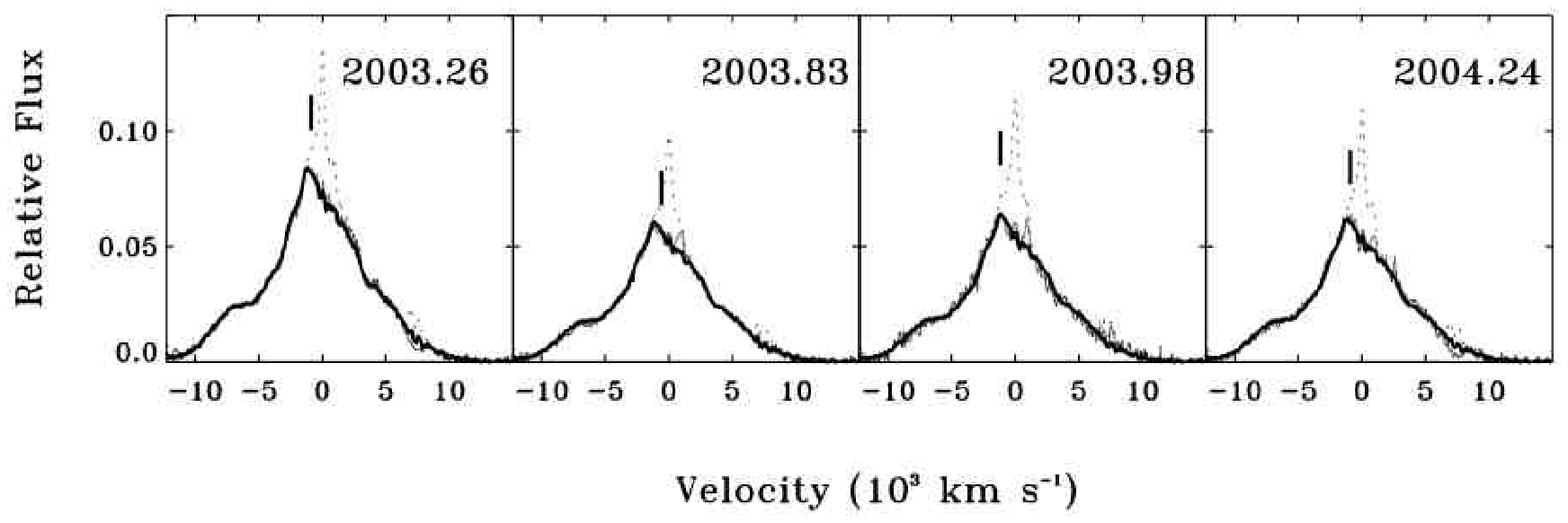}
\caption[3C 227: Broad H$\alpha$ profiles.]{3C 227: Broad H$\alpha$ profiles in units of $f_{\nu}$, flux calibrated relative to the narrow H$\alpha$ line, with the 
subtracted narrow lines shown with dotted lines, and the scaled mean profile from 1999-2004
plotted plotted with a thick line.
Tick marks show the flux-weighted velocity centroid of the top 10\% of the central blue peak of the profile.
\label{227:fig:pfits_1}}
\end{figure}

\begin{figure} [tbp]\centering
\includegraphics[width=0.9\linewidth]{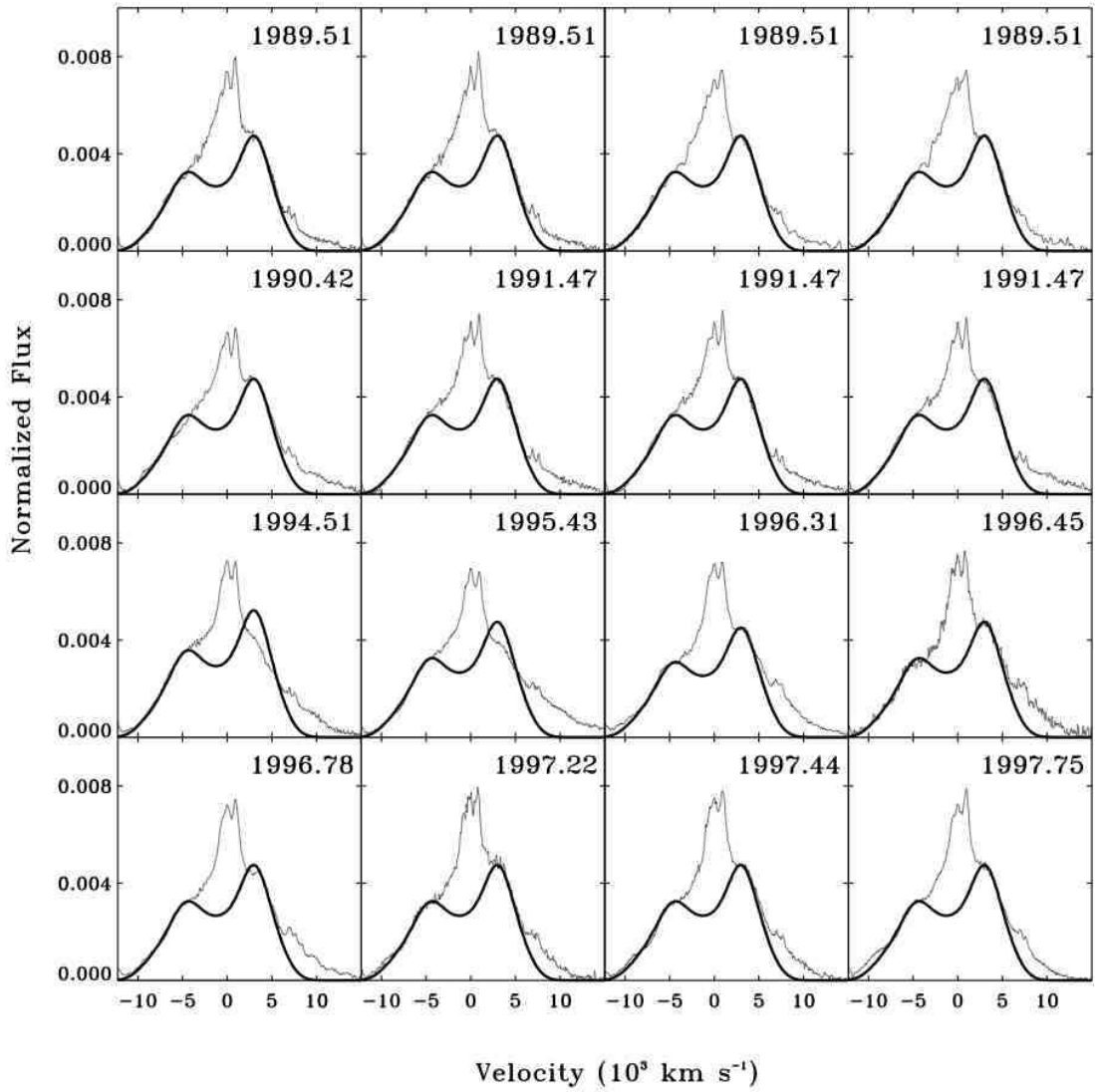}
\caption[3C 382: Broad H$\alpha$ profiles.]{3C 382: Broad H$\alpha$ profiles including the narrow lines in units of $f_{\nu}$, 
normalized by the total broad-line flux in the regions of the profile not contaminated by the narrow emission lines, and the best-fit scaled model disk profile plotted with a thick line.
\label{382:fig:pfits_0}}
\end{figure}

\begin{figure} [tbp]\centering
\includegraphics[width=0.9\linewidth]{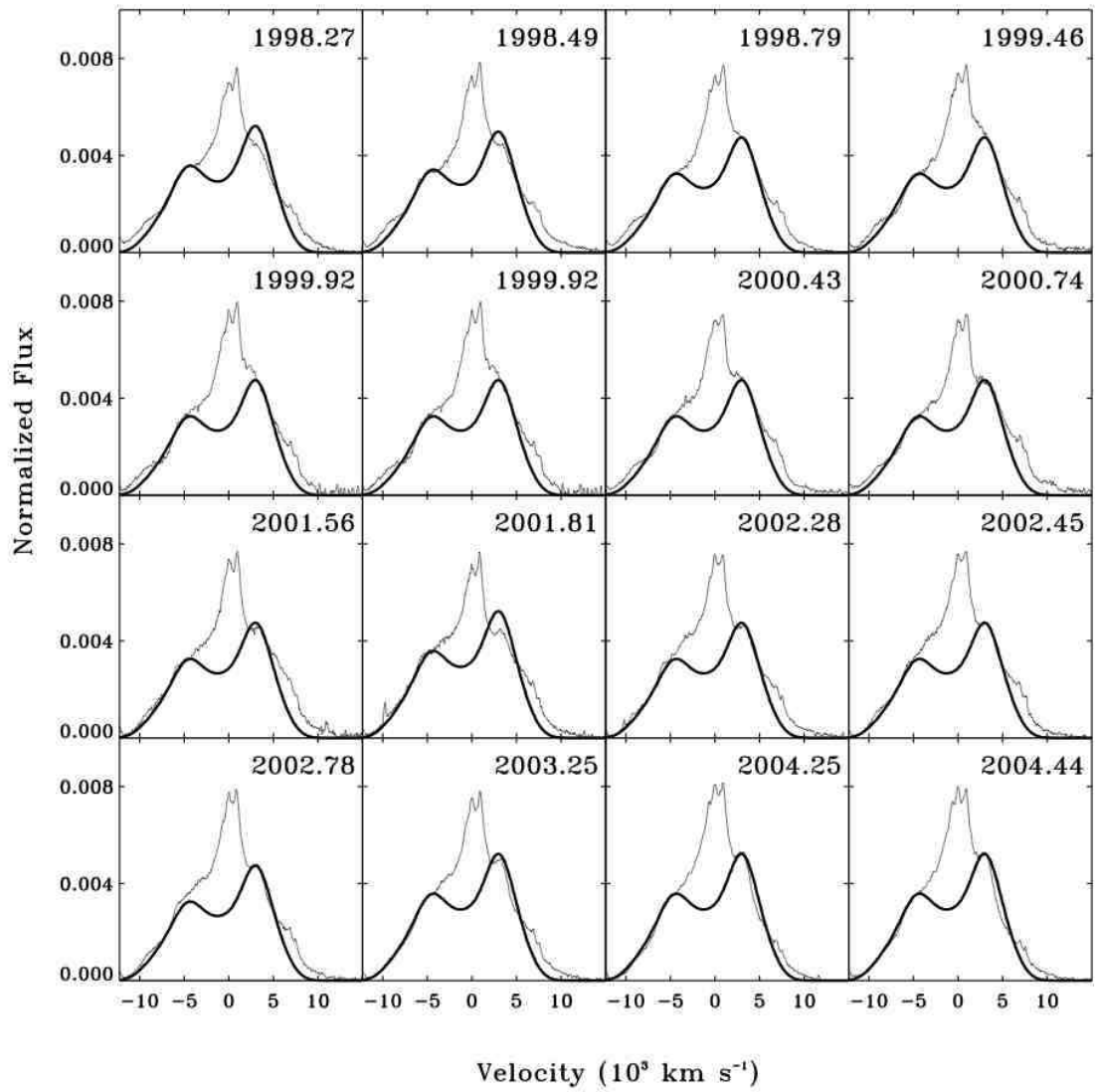}
\caption[3C 382: Broad H$\alpha$ profiles.]{3C 382: Broad H$\alpha$ profiles including the narrow lines, 
normalized by the broad-line flux, and the best-fit scaled model eccentric disk profile plotted with a thick line.
\label{382:fig:pfits_1}}
\end{figure}

\begin{figure} [tbp]\centering
\includegraphics[width=0.9\linewidth]{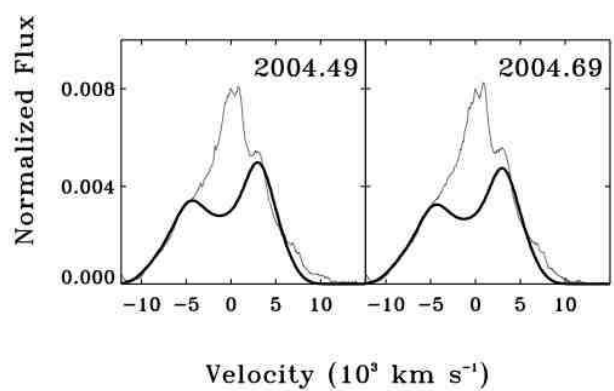}
\caption[3C 382: Broad H$\alpha$ profiles.]{3C 382: Broad H$\alpha$ profiles including the narrow lines, 
normalized by the broad-line flux, and the best-fit scaled model eccentric disk profile shown in blue.
\label{382:fig:pfits_2}}
\end{figure}

\end{document}